\documentclass[10pt]{article}

\usepackage{color}
\usepackage{makeidx}         
\usepackage{graphicx}
\usepackage{float}
\usepackage{amsfonts}
\usepackage{natbib}
\usepackage{fancyhdr}%
{%
}%
\usepackage{hyperref}

\definecolor{iblue}{RGB}{65,105,225}
\definecolor{ired}{RGB}{220,20,60}
\definecolor{igreen}{RGB}{50,205,50}
\definecolor{ipurple}{RGB}{75,0,130}
\definecolor{iochre}{RGB}{218,165,32}
\definecolor{iteal}{RGB}{51,204,204}
\definecolor{imauve}{RGB}{204,51,153}
\definecolor{RED}{RGB}{255,0,0}

\def\alertb#1{{\color{blue}#1}}
\def\alertr#1{{\color{red}#1}}
\def\alertr#1{{\color{RED}#1}}
\def\alertn#1{{\color{black}#1}}

\let\a=\alpha \let\b=\beta   \let\g=\gamma     \let\d=\delta  \let\e=\varepsilon
\let\z=\zeta  \let\h=\eta     \let\k=\kappa  \let\l=\lambda
\let\m=\mu         \let\x=\xi        \let\p=\pi     \let\r=\rho
\let\s=\sigma \let\t=\tau    \let\f=\varphi    \let\ch=\chi
\let\ch=\chi  \let\ps=\psi      \let\o=\omega  
 \let\D=\Delta       \let\X=\Xi
    \let\Si=\Sigma \let\F=\Phi       
\let\O=\Omega 
\def\V#1{{\bf#1}}\def\lhs{{\it l.h.s.}\ }
\def\*{\vskip 3mm}
\def\0{\noindent}
\def\be{\begin{equation}}
\def\ee{\end{equation}}
\def\bea{\begin{eqnarray}}
\def\eea{\end{eqnarray}}

\def\AA{{\mathcal A}}

\def\EE{{\cal E}}
\def\NN{{\cal N}}

\let\dpr=\partial\let\fra=\frac
\def\ie{{\it i.e.}\ }
\def\eg{{\it e.g.}\ }

\def\lis#1{\overline{#1}}
\def\defi{{\buildrel def\over=}}

\def\media#1{{\Blangle\,#1\,\Brangle}}

\def\otto{\,{\kern-1.truept\leftarrow\kern-5.truept\to\kern-1.truept}\,}

\def\tende#1{\,\vtop{\ialign{##\crcr\rightarrowfill\crcr
 \noalign{\kern-1pt\nointerlineskip} \hskip3.pt${\scriptstyle
 #1}$\hskip3.pt\crcr}}\,}
\def\ie{{\it i.e.\ }}\def\etc{{\it etc.\ }}

\def\tto{\Rightarrow}

\def\Ba   {{\mbox{\boldmath$ \alpha$}}}

\def\Brangle {{\mbox{\boldmath$ \rangle$}}}
\def\Blangle {{\mbox{\boldmath$ \langle$}}}

\def\({\left(}
\def\){\right)}

\def\iniz{\setcounter{equation}{0}}
\renewcommand{\theequation}{\arabic{section}.\arabic{equation}}

\newdimen\xshift \newdimen\xwidth \newdimen\yshift \newdimen\ywidth%
\def\ins#1#2#3{\vbox to0pt{\kern-#2pt\hbox{\kern#1pt #3}\vss}\nointerlineskip}

\def\eqfig#1#2#3#4#5{
\par\xwidth=#1pt \xshift=\hsize \advance\xshift
by-\xwidth \divide\xshift by 2
\yshift=#2pt \divide\yshift by 2%
{\hglue\xshift \vbox to #2pt{\vfil
#3 \includegraphics{#4.eps}
}\hfill\raise\yshift\hbox{#5}}}
\def\Eqfig#1#2#3#4#5#6{
\par\xwidth=#1pt \xshift=\hsize \advance\xshift
by-\xwidth \divide\xshift by 2
\yshift=#2pt \divide\yshift by 2%
{\hglue\xshift \vbox to #2pt{\vfil
#3 \includegraphics{#4.eps}\kern200pt\includegraphics{#5.eps}
}\hfill\raise\yshift\hbox{#6}}}

\def\ifnextchar#1#2#3{\let\tempe #1\def\tempa{#2}\def\tempb{#3}\futurelet
\tempc\ifnch}
\def\ifnch{\ifx\tempc\tempe\let\tempd\tempa\else\let\tempd\tempb\fi\tempd}
\def\gobble#1{}
\font\tengr=grreg10

\def\greekmode{%
\catcode`\<=13
\catcode`\>=13
\catcode`\'=11
\catcode`\`=11
\catcode`\~=11
\catcode`\"=11
\catcode`\|=11
\lccode`\<=`\<%
\lccode`\>=`\>%
\lccode`\'=`\'%
\lccode`\`=`\`%
\lccode`\~=`\~%
\lccode`\"=`\"%
\lccode`\|=`\|%
\tengr\def\bf{\tengrbf}
}
\newcount\vwl
\newcount\acct
\def\lt{<}

{
  \greekmode
  \gdef>{\ifnextchar `{\expandafter\smoothgrave\gobble}{\char\lq\>}}
  \gdef<{\ifnextchar `{\expandafter\roughgrave\gobble}{\char\lq\<}}
  \gdef\smoothgrave#1{\acct=\rq137 \vwl=\lq#1 \dobreathinggrave}
  \gdef\roughgrave#1{\acct=\rq103 \vwl=\lq#1 \dobreathinggrave}
  \gdef\dobreathinggrave{\ifnum\vwl\lt\rq140    
    \char\the\acct\char\the\vwl\else\expandafter\testiota\fi}
      \gdef\testiota{\ifnextchar |{\addiota\doaccent\gobble}{\doaccent}}
        \gdef\addiota{\ifnum\vwl=\lq a\vwl=\rq370
            \else\ifnum\vwl=\lq h\vwl=\rq371 \else\vwl=\rq372 \fi\fi}
              \gdef\doaccent{\accent\the\acct \char\the\vwl\relax}
              }

\newif\ifgreek\greekfalse

\def\begingreek{\bgroup\greektrue\greekmode}
\def\endgreek{\egroup}

\let\math=$
{\catcode`\$=13
}

\relax
\begingreek

\endgreek
\relax

\def\bgr{\begingreek}
\def\egr{\endgreek}

\def\eqalign#1{\null\,\vcenter{\openup\jot
  \ialign{\strut\hfil$\displaystyle{##}$&$\displaystyle{{}##}$\hfil
      \crcr#1\crcr}}\,}

\def\pagina{\vfill\eject}

\newdimen\u
\newcount\pdf
\author{\alertb{Giovanni Gallavotti}\footnote{
INFN-Roma1 \& Universit\`a ``La Sapienza'', Roma \& Rutgers
University, Math. Dept.\hfill\break
.\hskip.5cm email: {\tt giovanni.gallavotti@roma1.infn.it}}}
\title{\alertr{\bf Ergodicity: a historical perspective.\\
 Equilibrium and Nonequilibrium}}

\begin{document}
\maketitle
\kern-1cm
\begin{abstract}
A view on the physical meaning of the so called ergodic hypothesis: its
role on the foundations of equilibrium statistical mechanics in mid '1800,
its interpretations and hints at its relevance for modern nonequilibrium
statistical mechanics. Followed by appendices with
detailed comments on the original papers.
\end{abstract} \*
\0Keywords: {\small Ergodicity, Chaotic hypothesis, Gibbs
distributions, SRB distributions}
\*

\def\alertr{}\def\alertb{}

\setcounter{section}{-1}
\def\SEC{Contents}
\section{\SEC}
\label{sec:0}
\iniz
\lhead{\small \SEC}

\tableofcontents
\*\*

{\small Original Boltzmann papers, and their pages, are quoted by their
  number, and page numbers, in the {\it Wissenshaeftliche Abhandlungen} (WA).
  Appendices with
  comments to original papers usually follow a few partial translations.}

\vfill\eject

\0{\bf ``The entropy of the universe is always increasing'' {\it is not a
    very good statement of the second law}, \citep[Sec. 44.12]{Fe963}: this
  is Feynman.\footnote{Although I also think so it is safer to rely
    on authorities.}\\
  Clausius' formulation of the second law is ``{\it It is
  impossible to construct a device that, operating in a cycle will produce
  no effect other than the transfer of heat from a cooler to a hotter
  body}'', \citep[p.148]{Ze968}}.\\
  Existence of equilibrium entropy follows as a theorem: Clausius'
  ``fundamental theorem of the theory of heat'', here  ``heat
  theorem''. Is there an analogue of Ergodic Hypothesis and Entropy in
  nonequilibrium thermodynamics?
  \*
  
  \0{\it Some terminology:} 
The ``ergodic hypothesis'' in the sense of Boltzmann and Maxwell is that,
in a confined Hamiltonian system, ``a phase space point evolves in time and
eventually visits all other points with the same energy''
\footnote{\small Boltzmann and Maxwell considered systems in which
  the single, undisturbed motion, if pursued without limit in time, will
  finally traverse ``every phase point'' which is compatible with the given
  total energy, \citep[p.21]{EE911}: and Boltzmann called them
  ``ergodic''.}  (aside from obvious exceptions due to symmetries or to
integrability of the motions, like for harmonic chains): but this can make
sense only if the phase space is considered discrete and finite. 
\footnote{\small A ``quasi ergodic hypothesis'', \citep[p.90]{EE911},
  weaker than the original, has been introduced by requiring that, aside
  from trivial exceptions, the trajectories of the points should cover a
  dense set on the energy surfaces: this eliminates the contradiction
  forbidding that a single continuum trajectory visits the entire energy
  surface. In modern Mathematics a confined Hamiltonian system is called
  ``ergodic'' on an energy surface if all its initial data, but a set of
  zero Liouville measure, evolve spending in any measurable set a fraction
  of time proportional to its measure. }
I argue below that it is in this
sense that Boltzmann and Maxwell intended the hypothesis: furthermore in
Section\ref{sec:13} I discuss how the definition, so interpreted, could be
applied also to nonequilibrium situations, unifying conceptually
equilibrium and nonequilibrium theories of stationary states.

\setcounter{section}{0}
\def\SEC{Chronology}
\section{\SEC}
\iniz\label{sec:1}
\lhead{\small \SEC}


\0{\bf1866}: Boltzmann establishes a relation between the least action principle
and the entropy variation in a quasi static process assuming that atoms
move periodically and equilibrium states are identified with collections of
periodic orbits: this is a first version of the ``heat
theorem''. Determination of the probability distribution of phase space
points is not attempted, \citep[\#2]{Bo866}.  \*

\0{\bf1868}: In \citep[\#5]{Bo868} the distribution on phase space for a
system of $n$ atoms in equilibrium is defined as a probability density
which is positive on the region of phase space visited by a trajectory and,
at any point, it is identified as proportional to the fraction of the time
spent by the trajectory near that point.  The probability density is
supposed, without saying, smooth (possibly aside for discontinuities due to
hard cores or walls collisions) and its uniqueness is not (yet) challenged
by Boltzmann, see also comments in Appendix\ref{1868} below.  After several
increasingly complex examples he obtains what is now called the {\it
  microcanonical distribution} in the $6n-1$ dimensional energy surface. As
remarked by \citep{Ma879-c} the ergodic hypothesis (\ie the assumption that
a single trajectory visits the entire energy surface performing a periodic
motion in phase space) is implicit in the argument.%
\footnote{1868 seems to be the first time the hypothesis appeared: it has
  been again quite explicitly stated in \citep{Bo871-b}. It is well known
  that the strict interpretation of the hypothesis has been shown to be
  impossible: but the such an interpretation does not seem to have much in
  common with the work of the founding fathers, see below.  Neither
  Boltzmann nor Maxwell employ here the words ``ergodic hypothesis''.}  \*

\0{\bf1871}: Clausius obtains the 1866 result of Boltzmann: his work is clearer
but the conclusions are the same. The matter is clarified by the following
dispute on priority,\citep{Cl871,Cl872},\citep[\#17]{Bo871-0}, see also
comments
in Appendices\ref{1871},\ref{C1872} below.  \*

\0{\bf1871}: Boltzmann's 1871 ``trilogy''\\
{\bf(a)} In \citep[\#18]{Bo871-a} the distribution of the states of
individual $r$-atomic molecules in their center of mass is studied. Density
is assumed positive on the accessible regions of phase space which, because
of the collisions, is ``reasonably'' identified with the entire $6r-6$
dimensional phase space. The further assumptions are, first, negligible
collisions duration and, second, absence of multiple collisions. It is
shown, extending arguments of Maxwell for monoatomic gases, that the
distribution is the {\it canonical distribution}, as now it is
called. Having neglected multiple collisions leads Boltzmann to asking
wheter the distribution found is the unique stationary, see also comments
in Appendix\ref{1871-a} below.  \*
\0{\bf(b)} Derivation of the microcanonical distribution under the {\it
  ergodic hypothesis} in \citep[\#19]{Bo871-b}: formally set here for the
first time and applied to recover the results of (a) and, furthermore, to
show that any large subsystem is distributed with a {\it canonical
  distribution}. In this work the question is also raised on whether the
ergodic hypothesis is correct. A second aspect of this work is that for the
first time an ``ensemble'' is used as a device to compute the statistics of
the equilibria. It is remarkable that its introduction is not ``by an
axiom'': instead it is derived from (a), the preceding work. The gas of $n$
$r$-atomic molecules is considered as a collection of $n$ identical systems
of points moving independently and distributed (in their
$(6r-6)$-dimensional phase space) so as to keep the global phase space
density constant in time. Since (in (a)) the collision time is neglegible
and multiple collisions excluded, the trajectory changes due to collisions
are thought as interactions with an external warm body so that the system
of molecules becomes and ``ensemble''. Microscopic motion is still
periodic, because of the ergodic hypothesis, see also comments in
Appendix\ref{1871-b} below.\*
\0{\bf(c)} Boltzmann goes back to the 1866 heat theorem and shows that the
canonical distributions of a warm body, identified with the equilibrium
states (as in 1866 were the periodic orbits), verify, in a quasi static
process, the heat theorem, \citep[\#20]{Bo871-c}. {\it I.e.} it is
possible to define entropy, specific heat \etc bound by the appropriate
thermodynamic relations: this property will eventually be formalized as
``thermodynamic analogy'' or ``thermodynamics model'', see also comments
in Appendix\ref{1871-c} below.  \*

\0{\bf1872}: Boltzmann's evolution equation is established greatly
extending Maxwell's work where he had developed, in 1866
\citep{Ma867-b}, a set of equations equivalent to Boltzmann's equation
for the evolution of various key observables: it was a ``weak version'', to
use modern, mathematical, language. The extension is a major conceptual
advance, \citep[\#22]{Bo872}. It was a manifestly irreversible equation
and, via the H-theorem, allowed Boltzmann to develop a microscopic
interpretation of equilibrium entropy (for rarefied gases) and to extend it
to phenomena of evolution towards equilibrium. It ignited the discussion
over microscopic reversibility versus macroscopic irreversibility, even
though it did not lead to ``other'' advances with respect to Maxwell's work
(aside, of course, entropy formula and H-theorem).  \*

\0{\bf1877a}: Boltzmann gives a first detailed analysis of the relation
between microscopic reversibility and macroscopic irreversibility on
observable time scales, \citep[\#39]{Bo877a}. He also provides a more formal
definition of thermodynamic analogy (the term is, however, not yet
employed), with simple examples of the heat theorem in monocyclic systems,
\ie in systems in which all motions are periodic in phase space, see also
comments
in Appendix\ref{1877-a} below. Periodicity occurs in
isolated systems with superastronomical periods, hence unobservable, but it
is explained why irreversibility is nevertheless observable.  \*

\0{\bf1877b}: Detailed study, \citep[\#42]{Bo877b}, of phase space
represented as a discretization on regular (parallelepipedal) cells. The
count of configurations, somewhat extending the key 1868 work, leads 
Boltzmann to the well known expression of entropy, in equilibrium {\it as
  well as in systems approaching equilibrium}, as a measure of the
``permutability'' of the microscopic states. It achieves the identification
of equilibrium configurations with the ones which are the most numerous
among those which give the same values to a few observables of
thermodynamic relevance (\ie which realize the same ``macrostate'',
\citep{Le993}). The {\it essential use of regularity of the discretization
  cells} does not seem to have been stressed enough by commentators, see
also comments
in Appendix\ref{1877-b} and Sec.\ref{sec:8} below.  \*

\0{\bf1881}: Boltzmann, \citep[\#63]{Bo881}, presents Maxwell's analysis of
his 1868 work, \citep{Ma879-c}, attributing to Maxwell the ensembles view
of the equilibrium statistics (as it is now called) and stressing that his
own view (in the 1868 paper) rather considers the statistics defined by the
frequency of visit to phase space regions. Maxwell's paper states clearly
the ergodic hypothesis and does not mention that in \citep[\#19]{Bo871-b}
Boltzmann had already introduced and employed ensembles to describe
equilibrium statistics (see Appendix\ref{1871-b}). \*

\0{\bf1884}: Studies the Helmoltz' notion of thermodynamic analogy (which,
without using this name, he already used in several earlier papers),
\citep[\#73]{Bo884}.  The more formal set up leads Boltzmann to a general
definition of the statistical ensembles as collections of ``monodes'', \ie
probability distributions on phase space invariant under time evolution,
which are ``orthodes'', \ie generate a thermodynamic analogy. The main
examples that are described are: the ``holode'', the two parameters
collection (temperature and volume) of canonical distributions, and the
``ergode'', the two parameters (energy and volume) collection of
microcanonical distributions. The ensembles are connected via equivalence
properties and the ergodic hypothesis implies that the microcanonical
distributions describe the statistics controlling the physics of a system
modeled by given Hamiltonian microscopic equations. In spite of the
superastronomical (and unphysical) recurrence time, the hypothesis
establishes the connection between physics and the thermodynamic analogies,
see also comments
in Appendix\ref{1884}.  \*

\0{\bf1971$\to \infty$}: The ``Sinai-Ruelle-Bowen'', SRB, distributions are
introduced to describe chaotic stationary states,
\citep{Ru989,Ru995}. Phase space discretized on a regular lattice and the
ergodic hypothesis (hence periodic motions, still with superastronomical
periods) reappear: will it be possible to develop a unified theory of
equilibrium and stationary non equilibrium states? as well as of the
approach to those? (see below).
\*

\def\SEC{Least action \& periodicity: Boltzmann, Clausius.}
\section{\SEC}
\label{sec:2}
\iniz
\lhead{\small\thesection: \SEC}

In 1866 Boltzmann develops the idea that the second law simply reflects a
very general property, actually a theorem, of Hamiltonian mechanics: under
the ambitious title {\it On the mechanical meaning of the second
  fundamental theorem of heat theory}, \citep[\#2]{Bo866}.

After recalling that temperature should be identified as proportional to
the time-average of kinetic energy, independent of the particular atom of the
substance, (an already quite well known fact, \eg
see \citep{Kr856,Ma860-a}) Boltzmann tries  to obtain a theorem ``entirely
coincident'' with the form first discovered by Clausius, namely:
$$ \int \frac{dQ}T\le0 \eqno{(1)}$$
the integral being over a cyclic process in which ``{\it actions and
  reactions are equal to each other, so that in the interior of the body
  either thermal equilibrium or a stationary heat flow will always be
  found}'', \citep[\#2,p.24]{Bo866}.

Boltzmann makes the point, in Sec.IV, that Eq.(1) is an extension of the
least action principle: the latter compares very close motions which
develop in a given time interval at {\it fixed initial and final
  positions}, concluding that if a motion satisfies Newton's equations then
the variation of the action at it vanishes. The extension considered by
Boltzmann compares instead very close {\it periodic motions} under the
assumption that {\it both} satisfy Newton's equations: however technically
the analytical procedure employed to perform the comparisons, \ie the
calculus of variations, is identical in the two cases, see
p.\pageref{actionpr} below.

The basic assumption, \citep[\#2,p.24]{Bo866}, is that:
\* 
\0''{\it An arbitrarily selected atom moves, whatever is the state of the
  system, in a suitable time interval (no matter if very long), of which
  the instants $t_1$ and $t_2$ are the initial and final times, at the end
  of which the speed and the direction come back to the original value in
  the same location, describing a closed curve and repeating, from this
  instant on, their motion. In this case the equality holds in \rm
  Eq.(1).}''  \*

In fact on the same page, Boltzmann mentions that, sometimes, motion might
be not periodic: immediatey after stating the recurrence assumptions he
comments that recurrence may occur ``possily not exactly, nevertheless so
similar ...''.%
\footnote{And later in the paper, p.30, he acknowledges again that
    the motion might be aperiodic and develops an argument (rather
    involved) that, even so, it could possibly be regarded as periodic with
    infinite period,
\\
  ``{\it''... this explanation is nothing more than the mathematical
  formulation of the statement according to which paths that do not close
  themselves in any finite time can be regarded as closed in an infinite
  time}''.\phantomsection\label{infinitetime} This seems to point at the
remark that a stationary motion, confined in phase space, is generically
bound to recur (as later formalized by Poincar\'e's recurrence theorem). An
attempt to interpret this statement in a physical application is in
\citep[I-6 \& Appendix B]{Ga013b} or \citep[Appendix 9.A3]{Ga000}.}

Notice that the above assumption only demands that each atom goes through
a periodic orbit. %
 The mathematical analysis is carried (see Sec.\ref{sec:3} below for a
 version of the proof) in the case of a system consisting of a single atom
 or of a system of $N$ atoms whose stationary motion is periodic in the $6N$
 dimensional phase space. And it is extended to more general cases in which
 the motion can be quasi periodic involving many atoms each moving
 periodically (and, less formally, even more general).

It is remarkable that an analogous hypothesis, independently formulated,
can be found in the successive paper  \citep[l.8, p.438]{Cl871}:
and indicates that, at the time, the idea of recurrence must have been
quite common.

Clausius firstly imagines that the system consists of a single atom moving
periodically; the proof is outlined in Sec.\ref{sec:3} below (as
in Boltzmann's work, it would apply to the case of an arbitrary system moving
periodically on a path in the $6N$ dimensional phase space).
\footnote{\alertr{However Clausius, and Boltzmann earlier, really
    discuss the motion of a single atom and choose not to stress that the
    proof would also work if the considered motion is that of the
    point representing the system in phase space.}} The position on the path is
determined by the ``phase'', \ie by the time, in units of the period, taken
by the motion to return to its position and velocity from an arbitrarily
fixed initial point; and at each time many phases are occupied (each by a
different atom).

But Clausius, as already Boltzmann, is not strict about the recurrence.
The assumption evolves eventually to one in which the atoms are supposed to
be in different groups in which each atom follows a closed path but the
period might depend on the group to which it belongs, and both authors
go on to even more general cases (see Sec\ref{sec:6} below).

Today any periodicity assumption, even in its evolved forms discussed in
the following, is still often strongly criticized: social pressure induces
to think that chaotic motions are non periodic and ubiquitous: and their
images fill both scientific and popular magazines.  It is however
interesting and conceptually important that the ideas and methods developed
or based on the primitive forms of periodic recurrence have laid the
foundations of equilibrium statistical mechanics and have even been the
basis of the chaotic conception of motion and of the possibility of
reaching some understanding of it. {\it It will be argued below that
  periodicity can be a guide to the modern developments of nonequilibrium
  statistical mechanics.}

\def\SEC{Heat theorem}
\section{\SEC}
\label{sec:3}
\iniz

The proofs of Boltzmann and Clausius can be read in the original papers,
and compared; Clausius' version of the heat theorem is more general: for a
comparison between the two proofs see Appendices \ref{1871}, \ref{C1872}.

The analysis are based on what the Authors consider a version for periodic
motions of the action principle in Maupertuis' form (at fixed extremes)
$\d\int \frac{m}2c ds=0$ with $c$ the speed and $ds$ the arc length of the
path followed by the motion $x$, rather than in the equivalent form $\d
\int_{t_1}^{t_2} (\frac{m}2c^2-V(x))\frac{dt}{t_2-t_1}=0$ (at fixed extremes
in $x$ and $t$).  It will be summarized here following Clausius notations
which are more clear, in the case in which the motion of the whole system
is periodic.

If $t\to x(t)$ is a periodic motion developing with kinetic energy $K(x)$
and under the action of internal forces with potential energy $V(x)$ and of
external forces with potential $W$,%
\footnote{An example for a $n$ points system $x=(q_1,\ldots,q_n)$ enclosed
  in a container with boundary $\dpr\O$ could be
  $V(x)=\sum_{i<j}v(q_i-q_j)+\sum_i v'(q_i)$, $W(x)= \sum_i
  w(d(q_i,\dpr\O))$ with $w(r)$ a function of the distance $r=d(q,\dpr\O)$
  of $q$ to $\dpr\O$, steeply diverging at $r=0$ (to model the container
  wall) and $v,v'$ some potentials; if some part of $\dpr\O$ is moved, as
  in a piston, the external potential changes by $\d_e W$; of course if $\O$
  is fixed $W$ can be thought as included in $V$.}
then the average action of $x$ is defined, if its period is $\t$,
\footnote{The original notation of Clausius for the period was $i$, which
  today might seem unusual, while for Boltzmann was $t_2-t_1$: here the
  period will be denoted $\t$, for simplicity}
by ${\cal
  A}(x)=\frac1\t\int_0^\t \big(\fra{m}2\dot x(t)^2 -V(x(t)-W(x(t)))\big)\,dt$.
\\ We are interested in variations $x'$ of $x$ which are periodic: they
will be represented as
$$ \eqalign{
  x'(t)&=x'(\t'\f)\defi\x'(\f), \quad t\in [0,\t'],\cr
  x(t)&=x(\t\f)\defi\x(\f), \quad t\in [0,\t],\cr
  \d \t&=\t'-\t\cr}
$$
where $i',i$ are the periods, and $\x',\x$ are periodic functions in
$\f\in[0,1]$; the $\f$ is the {\it phase}, as introduced in
\citep{Cl871}, and corresponds to the variable called {\it average anomaly}
in astronomy, see Fig.1.

\def\CAP1{{\small Fig.1: A periodic path and another one close to
    it: the paths can be imagined to be two curves in phase space of
    equations $\f\to\x(\f)$ or $\f\to\x'(\f)$ periodic in $\f\in [0,1]$; if
    the periods of the two motions are $i$ and $i'$ the position at time
    $t$ of each will be $\x(\frac{t}\t)$ or $\x'(\frac{t}{\t'})$. Hence
    Clausius' phases correspond to the astronomical average anomalies.}}

\ifnum\pdf=1
.\kern3cm\hbox{\includegraphics[width=140pt]{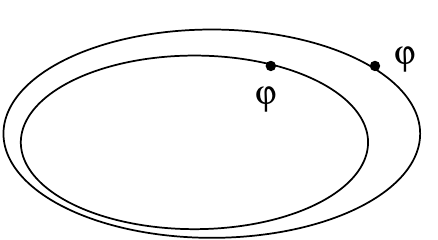}}\kern2cm%
\raise45pt\hbox{\small fig.1}
\fi
\ifnum\pdf=0
\eqfig{136}{70}{
}
{Fig1}{fig1}
\fi

\0\CAP1
\*

The role of $\f$ is simply to establish a correspondence between points on
the initial trajectory $x$ and on the varied one $x'$: it is manifestly an
arbitrarily chosen correspondence, which could be defined differently
without affecting the final result, but convenient in order to follow the
algebraic steps. It should be noted that in \citep[\#2]{Bo866} the phase is
not introduced and this makes the computations difficult to
follow.\footnote{\alertr{ In the sense that once realized which is the
    final result, it is easier to reproduce it rather than to follow his
    calculations.}}

The two motions to be compared have to be thought as characterizing the
system state at two successive steps of a reversible process: the duration
of each step of the process is very long, enough to cover many full periods
so that each observable has acquired a well defined new average value.

Set $\lis{F}(x)=\t^{-1}\int_0^\t F(x(t))dt=\int_0^1 F(\x(\f))\,d\f $ for a
generic observable $F(x(t))$; for instance $F$ could be the difference
between kinetic energy and potential energy
$\AA(x(t))=K(x(t))-V(x(t)-W(x(t)))$ and its average would be the average
action $\lis \AA(x)$.

Then the new form of the ``action principle for
periodic motions'' is, calling $\d_eW$ the variation of external potential
(with no loss of generality $W=0$ could be supposed to hold at the beginning
of the process considerably simplifying the following formulae)
$$\d\lis\AA\ \defi\ \d \lis{ K- V-W}=-2 \lis K \d\log \t+\d_e\lis{W}
\eqno{(*)}$$
allowing here {\it also} for a variation $\d_e W$ of the potential of the
external forces (think of a volume change and of the work by the pressure)
driving the new motion $x'$ of period $\t'$ (which is itself a variation of
the motion $x$ of period $\t$): Eq.(*) yields the correction to the
expression $\d\lis\AA=\d\lis{K-V-W}=0$ for the variation at
{\it fixed} temporal extremes and fixed potentials, familiar from the
action principle.%
\footnote{\alertr{ To derive Eq.(*) represent a periodic motion of period
    $i$ as $x(t)=\x(\f)$, with $t=\t\f$ and $\f\to\x(\f)$ periodic with
    period $1$; and, likewise, represent a close motion with period $\t'$ as
    $x'(t)=\x(\f)+\h(\f)$ with $t=\t'\f$: computing to first order the
    $\frac1{\t'}\int_0^{\t'} dt (\frac{m}2\dot x'(t)^2 -V(x'(t))-
    W(x'(t))-\d_e W(x'(t)))$} and {\it applying the equations of motion}
  for a motion $x$ developing under the influence of a force with potential
  $V+W$, or for $x'$ developing under a potential $V+W+\d_e W$: $\ddot
  x'(t)=-\dpr_x (V(x'(t)+W(x'(t))+\d_eW(x'(t))))$ in the process changing
  $x$ into $x'$.\\ Remark that if $\d Q=0,\d_e W=0$ Eq.(*) becomes $\d
  (\t\lis{ K})=0$, \ie $\d\int_0^\t K dt=2\d\oint c ds$ which can be seen as
  a periodic version of the least action principle applied to either the
  initial motion or to the varied motion.}

In deriving Eq.(*), which is the authors' common conclusion, there is {\it
  no stationarity condition} to consider: it holds if and only if the
variation changes a periodic motion $x$ into a nearby periodic one $x'$
developing under forces varied by the variation $\d_e W$ of the external forces
potential, as it is imagined in the theory considered ({\it but in
Boltzmann it is supposed $\d_e W=0$}, see below).  Furthermore {\it the
variation is not arbitrary} (unlike the case of the usual least action
principle) but occurs between given motions which are both periodic
solutions of the corresponding equations of motion.

The connection with the heat theorem derives from the remark that in the
infinitesimal variation of the orbit from $x$ to $x'$
its new total energy is $\lis U'\defi
\lis K'+\lis V'+\lis{W'}$ and changes from the old $\lis K +\lis V+\lis W$ by
$\d\lis{ K+ V+W})$. 

Since $\d_e\lis{W}$ is interpreted as the external work %
the quantity $\d Q\defi\d\lis U-\d_e\lis {W}\equiv\d(\lis K+\lis
V+\lis W)$ is interpreted as the heat $\d Q$ received by the system and the
variational relation (*) above can be rewritten:
$$\eqalign{
  &-\d Q+2\d\lis K+2\lis K\d\log \t\equiv 
  -\d Q+2\lis K \d \log(\lis K \t)=0\cr  &
  \quad\tto\quad\frac{\d Q}{\lis K}
= 2\d \log(\lis K\, \t)\cr
}\eqno{(**)}$$
with the last $=$ signs holding if the system can be considered to satisfy
the equations of motion during the transformation, which means if the
variation $\d$ is ``quasi static''. Hence in a quasi static transformation
$\lis K^{-1}$ is an integrating factor for $\d Q$: and the primitive
function (\ie the entropy) is twice the logarithm of the ordinary action
$S=2\log \t\lis K$, up to an additive constant.  \*

About Eq.(*) Boltzmann says:
\*
\0''{\it It is easily seen that our conclusion on the meaning of the
  quantities that intervene here is totally independent from the theory of
  heat, and therefore the second fundamental theorem is related to a
  theorem of pure mechanics to which it corresponds just as the kinetic
  energy principle corresponds to the first principle; and, as it
  immediately follows from our considerations, it is related to the least
  action principle, in a somewhat generalized form.}'',
\citep[\#2,sec.IV]{Bo866}  \*

Both Boltzmann and Clausius call Eq.(*) a ``generalization of the action
principle'': it should be noted that the latter principle uniquely
determines a motion, {\it i.e.} it determines its equations; Eq.(*),
instead, does not determine a motion but it only establishes a relation
between the variation of average kinetic and potential energies of close
periodic motions under the assumption that they satisfy the equations of
motion; and it does not establish a variational property.

Also Eq.(*) does not establish the general inequality
$\oint\frac{dQ}T\le0$: which in Boltzmann is obtained via a further
argument at the end of the paper, while Clausius does not even mention the
inequality, {\it leaving it, implicitly and to the reader, as consequence}
of having allowed in his formulation variable external forces (\ie $\d_e
W\ne0$) so that the result applies to a general cycle in which heat and
work are involved: remarkably his paper contains {\it no signs $\ge$ or
  $\le$, but only equalities}\ !\ %
\footnote{\alertr{ This is an important point: Eq.(**) does not give
    to the periodic orbits describing the state of the system any
    variational property (of minimum or maximum): the consequence is that
    it does not imply $\oint \frac{\d Q}T\le 0$ in the general case of a
    cycle but only $\oint \frac{\d Q}T=0$ in the (considered) cases
    of reversible cycles. In Clausius' derivation, thanks to the inclusion
    of extrenal forces, the equality proves existence of entropy,
    therefore, by earlier classical arguments (again due to Clausius), also
    the inequality follows for non reversible cycles.}}

In the case in which the motion of the atoms is not periodic in phase space
and the atoms can be divided in groups with different periods the heat
theorem can be extended: both Boltzmann and Clausius, in modern notations,
say the following. Let the internal interaction potential be
$V(x)=\frac12\sum_{r\ne s} V(x_r-x_s)$ and suppose that particle $s$ moves
with period $i_s$: then the variational analysis leading to Eq.(*) above
yields (for simplicity take $W=0$ in the first motion)
$$\d{(\lis K_r-\frac12\sum_{s\ne r} \lis V(x_r-x_s))}=
-2\lis K_r\d\log\t_r+\d_e\lis W_r$$
hence, as above,
$$\eqalign{
  &\d Q_s\defi \d(\lis K_s+\frac12 \sum_{s\ne r} \lis V(x_r-x_s))=
  2\lis K_s+2\lis K_s
\d\log \t_s,\qquad {\rm summing:}\cr
&\d Q=\sum_s\d Q_s=\d (\lis K +\lis V)=\sum_s (2\d\lis K_s +2\lis K_s\d\log
\t_s)\cr}$$
and {\it since $\lis K_s$ is $s$-independent} by the earlier equipartition
result (of Maxwell) this shows that $\frac{\d Q}{\lis K}$ is exact.

Hence if there are several periods, so that the motion is quasi periodic,
the heat theorem still holds (because in the quasi static process the time
that elapses in an infinitesimal step of a process is ``quasi infinite'' so
that all averages are reached). However the simplest assumption is to
suppose that the motion is periodic in phase space: which {\it
  appropriately formulated} will become the ergodic hypothesis.

\def\SEC{An example}
\section{\SEC}
\label{sec:4}
\iniz

It is instructive to illustrate via a simple explicit example, the
construction of the function $S$. The example in the form below is
extracted from the later work \citep[\#39,p.127-148]{Bo877a}. It
explains why and how a concept that will play an important role in the
development of the theory of the ensembles, namely the ``thermodynamic
analogy''\footnote{ Original Boltzmann's term.} or ``model of
thermodynamics'' already arises in very simple systems in which a general
analysis can be performed.

The example is built on a case in which all motions are really periodic,
namely a one-dimensional system with potential $\f(x)$ such that
$|\f'(x)|>0$ for $|x|>0$, $\f''(0)>0$ and
$\f(x)\tende{x\to\infty}+\infty$. All motions are periodic (systems with
this property are called {\it monocyclic},
\citep[\#73]{Bo884},\citep[p.205]{Ga013b},\citep[p.45]{Ga000}. Suppose
  that the potential $\f(x)$ depends on a parameter $V$, thus denote it
  $\f_V$.

Define {\it a state} to be a motion with given energy $U$ and given
$V$ (\ie a periodic motion). And call:
\*

\halign{#\ $=$\ & #\hfill\cr
$U$ & total energy of the system $\equiv K+\f$\cr
$T$ & time average of the kinetic energy $K$\cr
$V$ & the parameter on which $\f$ is supposed to depend\cr
$p$ & $-$ average of $\dpr_V \f$.\cr}
\*
\0A state (\ie a periodic motion) is therefore parameterized by $U,V$ and
if such parameters change by $dU, dV$, respectively, let
$$
dW=-p dV, \qquad dQ=dU+p dV,\qquad \lis K= T.$$
Then, if $i=i(U,V)$ is the period, the heat theorem is, in this case,
\citep[\#6]{Bo868b}, \citep[\#39]{Bo877a},\citep{He884b}: : \*
\0{\sl Theorem} {\it The differential
  $({dU+pdV})/{T}$ is exact and equal to the differential of ``entropy''
  $S\defi2\log (i T)$, see Eq.(**) Sec.\ref{sec:3}.}  \*

In fact let $x_\pm(U,V)$ be the extremes of the oscillations of the motion
with given $U,V$ and define $S$ as:

$$ S=2\log 2\int_{x_-(U,V)}^{x_+(U,V)}\kern-3mm \sqrt{K(x;U,V)}dx=2\log
\int_{x_-(U,V)}^{x_+(U,V)}\kern-3mm2 \sqrt{U-\f(x)}dx$$
so that $dS=\fra{\int \big(dU-\dpr_V\f(x) dV\big)\, \fra{dx}{\sqrt{K}}}{ \int
  K\fra{dx}{\sqrt{K}}}\equiv \frac{d Q}T$, and $S=2\log i\lis K$ if
$\fra{dx}{\sqrt K} =\sqrt{\fra2m} dt$ is used to express the period $i$ and
the time averages via integrations with respect to $\fra{dx}{\sqrt K}$.

Therefore Eq.(**) of Sec.\ref{sec:3} is checked in this case.  This
completes the discussion of the case in which motions are periodic.  For an
interpretation of the above proof in a general monocyclic system see
\citep[\#73]{Bo884}.

The case of Keplerian motions, actually treated in \citep[\#39]{Bo877a},
is more delicate and interesting because all confined motions are periodic
but there is an extra constant of motion,
\citep[\#73]{Bo884},\citep[p.45]{Ga000} and \citep[Appendix D]{Ga013b}.

\def\SEC{Boltzmann vs. Clausius: on priority}
\section{\SEC}
\label{sec:4}
\iniz

Since both Boltzmann and Clausius reach a similar conclusion a priority
discussion arose leading to a detailed comparison of the two
works. Boltzmann started with:
\* 
\0''{\it .... I believe I can assert that the fourth Section of my paper
  published four years earlier is, in large part, identical to the quoted
  publication of Mr. Clausius. Apparently, therefore, my work is entirely
  ignored,...}'', \citep[\#17]{Bo871-0}.
\*
\0Clausius reacted politely but firmly. He objected to
Boltzmann the obscurity of his derivation obliging the reader to interpret
suitably formulae that in reality might be not explained (and ambiguous),
\citep[p.267]{Cl872}: \*
\0''{\it... the first of these Boltzmann's equations will be identical to my
  Eq.(I), if the value assigned to $\e$ can be that of my equation.}''
\*
\0but the values of $\e$, according to Clausius, could not be compared
because of the different definition of averages and variations.

Such first difference between the two works, as Clausius points out, is due
to some ambiguity in the definition of the averages because it is not clear
which is the correspondence, between the times in the motion $x$ and in the
varied motion $x+\d x$, that should be considered in comparing the averages;
Clausius is very careful about this point and he says that, because
of this difference, the two variations appearing in the definition of $\e$
above cannot be really compared without a proper interpretation of the
symbols.\footnote{\alertr{Actually the correspondence is mentioned quite
  explicitly in the formulation of the action
  principle in \citep[\#2,p.32]{Bo866}.}}

The main difference, however, is that in Boltzmann's discussion the motions
whose actions are compared are imagined as end products of an infinitesimal
step of a quasi static thermodynamic process in which {\it the external
  potential does not change}: therefore the class of transformations to
which the result applies is very restricted, \eg in the case of a gas it would
restrict the theory to the isovolumic transformations, as stressed by
Clausius, see Appendix\ref{C1872}: who points out that this is a key
difference between his analysis and Boltzmann's which otherwise would be
equivalent.

Boltzmann had realized the point, and admitted it, already in his priority
claim paper: while promising to take variable external potential into
account in his later works, he had also written that disregarding the
external potential variability had been a choice done to simplify the
argument since inclusion of external potential variations would have been
easy without affecting his deduction.

After Clausius reply (and strong disagreement) he did not insist on the
matter and did not profit from the fact that his formulae remain the same,
under further interpretations, even if the external potential changes; this
was concluded by Clausius himself at the end of his exegesis of Boltzmann's
work, see \hbox{p.\pageref{deltacomment} below}, by the sentence ``{\it
  ... but I want to stress that such solution appears well simpler when it
  is found than when it is searched}''.

Still Boltzmann did really deal with the promise about taking into account
the external forces in the same year, \citep[\#20]{Bo871-c}, in great detail
and adopting a new viewpoint in which periodic motion, as a representative
of a state, was replaced by stationary probability distributions,
considering in particular the cases of the canonical and microcanonical
distributions (and referring (on p.302 of \citep{Bo871-c}) to the connection
with his old work without mentioning again Clausius): and developed key
examples of what later would be called {\it thermodynamic analogy}.

It should also be said that Boltzmann's analysis, even if restricted to the
simpler completely periodic case, is quite difficult to follow: but as
Clausius says, if interpreted correctly, it is right; on the other hand
Clausius' is very clear (at least when treating the same case).

Of course with Boltzmann remains, besides the proposal of a relation
between the microscopic Hamiltonian laws of motion and the second law of
(equilibrium) thermodynamics, the ``novel idea'' that the states of the
system are to be identified with the average values of physical quantities
computed on a (periodic) trajectory thus building, at least, a
``thermodynamic analogy'', see Sec.\ref{sec:10} below: the 1866 work is an
early stage of the developing concept of state as a stationary
distribution. An equilibrium state is identified with the average values
that the observables have in it: in the language of modern measure theory
this is a special probability distribution, with the property of being
invariant under time evolution.

It should finally be kept in mind that, by the time Clausius' work appeared,
and Boltzmann was upset, Boltzmann had gone {\it far beyond} his work of
1866, see also Appendix\ref{1868} below: but the priority claim discussion
dealt solely with the subject of the two papers on the least action
principle which were, in a sense, {\it already quite outdated} with respect
to the early breakthrough work \citep[\#2]{Bo866}, mainly because of the
new major breakthrough by Boltzmann himself in \citep[\#5]{Bo868}.

For more detailed comments see Appendices\ref{1871},\ref{C1872} below.

\def\SEC{On periodicity}
\section{\SEC}
\label{sec:6}
\iniz

Both Boltzmann and Clausius were not completely comfortable with the
periodicity of atomic motions.

In Sec.\ref{sec:3} Boltzmann has been quoted as assuming periodicity and
then saying that strict periodicity is not necessary and even presented an
elaborate argument to conclude that, at least for his purposes, in the
cases in which the paths did not close nevertheless they could be ``{\it
  considered closed in an infinite time}'', see p.\pageref{infinitetime}
above.

Clausius worries about such a restriction more than Boltzmann does; he is
led to think the system as consisting of many groups of points which
closely follow an essentially periodic motion, \citep[Sec.13,p.452]{Cl871}:
\*
\0''{\it ..temporarily, for the sake of simplicity we shall assume, as
  already before, that all points describe closed trajectories. For all
  considered points, that move in a similar manner, we suppose, more
  specifically, that they go through equal paths with equal period,
  although other points may run through other paths with other periods.  If
  the initial stationary motion is changed into another, hence on a
  different path and with different period, nevertheless these will be
  still closed paths each run through by a large number of points.}''
\*
\0each atom or small group of atoms undergoes a periodic motion and the
statistical uniformity follows from the large number of evolving units.
\footnote{\alertr{ The assumption can be seen as an assumption that
    all motions are quasi periodic and that the system is integrable: it is
    a view that {\it mutatis mutandis} resisted until recent times both in
    celestial mechanics, in spite of Poincar\'e's work, and in turbulence
    theory as in the first few editions of Landau-Lifschitz' treatise on
    fluid mechanics, \citep{LL971}.}}

Then he dedicates the entire last three sections (13--15) of his work to
weakening further the periodicity assumption:

\* \0{\it ''... In the present work we have supposed until now that all points
  move on closed paths. We now want to set aside also this assumption and
  concentrate on the hypothesis that the motion is stationary.
\\
For motions that do not run over closed trajectories the notion of
recurrence is no longer usable in a literal sense, therefore it is
necessary to talk about them in another sense. Consider therefore right
away motions that have a given component in a given direction, for instance
the $x$ direction in our coordinates system.  It is then clear that motions
go back and forth alternatively, and also for their elongation, speed and
return time, as it is proper for stationary motion, the same form of motion
is realized.  The time interval within which each group of points which,
approximately, behave in the same way admits an average value.''}\*

\0without assuming the {\it one dimensional} motions of the three coordinates
separately periodic with the same period (which would mean that the motions
are periodic) he supposes ``only'' that the variation $\d i$ of the periods
are the same for the three coordinates when the motions of the 
particles are varied. The particles can be collected into groups with the same
$\d i$ and the same kinetic energy:
\*
\0{\it To proceed further to treat such equations a difficulty arises
because the velocity $v$, as well as the return time interval $i$, may be
different from group to group ...}
\*

\0an argument follows to justify that the {\it average} $\lis {v^2}$ can be
regarded at each step of a quasi static process as constant in a
stationary state%
\footnote{\alertr{This is because each periodic motion is supposed run by
    a large number of particles of a given group with different phases and
    the independence (\ie equipartition), from the group itself, of the
    average kinetic energy of the motions in each group is due to the
    stationarity of the state of the system.}}  for
all groups of particles and concludes that the analysis leading to Eq.(**),
in Sec.\ref{sec:3} above, can be repeated and the entropy variation can be
expressed as proportional to $\sum \d\log(T i)$ (with $T$ a constant
proportional to the average kinetic energy but with $i$ possibly variable,
from group to group and particle to particle appearing in the sum).

Summarizing, a simplified picture is: particles move periodically and can
be collected in groups with the same average kinetic energy $T$ per
particle, whose equipartition is implied by stationarity, and $\frac{\d
  Q}T$ is exact and the entropy is proportional to $\sum \log (T i)$, where
the sum runs over the different particles: \ie the variation of the entropy
is the sum of the variations of the entropies of each group.

Finally it should be mentioned, as I argue explicitly in
Sec.\ref{sec:8},\ref{sec:9} below, that it cannot be said, that Boltzmann
abandoned the periodic orbits and the related Ergodic Hypothesis to switch
to the theory of ensembles.
%
\def\SEC{Ergodic hypothesis: Boltzmann \& Maxwell}
\section{\SEC}
\label{sec:7}
\iniz

The work \citep[\#5]{Bo868}:%
\footnote{\alertr{ Initially motivated to reproduce Maxwell's result ``{\it
      because the exposition of Maxwell in its broad lines is difficult to
      understand}''.  The strong assumption, set at the beginning, of
    restricting the result to the case of a very rarefied gas (to neglect
    the time of interaction versus the time of free flight of an atom) is,
    later in the paper, removed and periodicity can be imagined to have
    been {\it implicitly} reintroduced, see Maxwell comment below, through
    the identification between probabilities and frequencies of visits.}}
is remarkable because in it Boltzmann derives the probability distribution,
today called {\it microcanonical}, for the phase space points that
represent an entire ``warm body''. But the key assumption in the derivation
is not proposed explicitly and has to be extracted from the text because,
although used, it is not mentioned as such. It was later recognized and
most vividly commented in \citep[p.734]{Ma879-c}, as:
  \*
\0''{\it The only assumption which is necessary for the direct proof
  {\rm [of the microcanonical distribution by Boltzmann]}
is that the system, if left to itself in its actual state of motion,
will, sooner or later, pass through every phase which is consistent with
the equation of energy.  Now it is manifest that there are cases in which
this does not take place
\\...
\\
But if we suppose that the material particles, or some of them,
occasionally encounter a fixed obstacle such as the sides of a vessel
containing the particles, then, except for special forms of the surface of
this obstacle, each encounter will introduce a disturbance into the motion
of the system, so that it will pass from one undisturbed path into
another....}''\phantomsection{\label{maxcit}}
\* \0It might take a long time to do the travel but eventually it will be
repeated.%
\footnote{\alertr{ Here Maxwell refers to the proof of the microcanonical
    distribution in \citep[\#5]{Bo868}, where it had
    been derived after releasing, in Sec.II,III, the strong assumption made
    in Sec.I, of a very rarefied gas (to neglect the time of interaction
    versus the time of free flight of an atom). In the paper the
    probability is identified and defined as proportional to the time of
    visit to small rectangular cells filling the entire available phase
    space: and this is simply realized if the phase space point
    representing the microscopic configurations visits all the discrete
    cells (since the cells may contain several particles one cannot really
    imagine other possibilities if the size of the cells is left arbitrary,
    so Maxwell's interpretation seems the only possible once a discrete
    phase space is accepted).  Maxwell's remark is interesting as it
    formulates again the hypothesis in \citep{Bo868,Bo871-b}
    in the strong form which, if interpreted as several critics did, would
    be mathematically incorrect. In \citep{In015} is analyzed the importance
    of this work of Maxwell for the foundations of the theory of the
    ensembles as representing a single invariant distribution on phase
    space via a collection of independent copies of the same system.  }}
Maxwell interprets Boltzmann without invoking external random actions, and
if intepreted in this way (see the full quote of Maxwell below,
p.\pageref{Max1879}) this is essentially the {\it Ergodic Hypothesis}, see
below.

In \citep[\#5]{Bo868}, besides a new proof of
Maxwell's distribution, the microcanonical distribution (so called
since Gibbs) is obtained not only without explicit reference to periodicity but,
partitioning phase space into cells, via the combinatorial count of the
number of ways to divide energy (just kinetic in Sec.II and total in
Sec.III of \citep[\#5]{Bo868}) into a sum of energies, keeping the total sum
fixed.

The combinatorial argument in Sec.II, dealing with the case of a rarefied
system, in which the interaction time can be neglected so that kinetic
energy can be considered constant, and aiming at obtaining Maxwell's
velocity distribution, supposes that each among the discrete kinetic energy
values probability is proportional to the time spent by the system in the
cell.  After checking that the {\it velocity space} volume element remains the
same at every pair collision this is done by inserting in the energy levels
one particle at a time and counting how many possibilities one has at each
time: this will be the number of different microscopic configurations with
a given total kinetic energy: it can be identified with their probability
if the different microscopic configurations are cyclically permuted by the
time evolution as it becomes more clear in Sec.III.

The argument is extended in Sec.III to cover the case of {\it non
  negligible time of interaction between the particles}.  The phase space
volume elements conservation is checked {\it without assuming only pair
  collisions} (an assumption impossible in presence of non negligible range
of interaction) and in the full phase space%
\footnote{This follows via the by now usual argument employed to prove
  Liouville's theorem, \citep[\#5,p.93-95]{Bo868}, 1868. It will be again
  proved later in \citep[\#18]{Bo871-a}. It is unclear why Maxwell, in
  commenting the latter work in \citep{Ma879-c}, while attributing to
  Boltzmann the theorem, develops a proof quite close to the one in
  \citep[\#18]{Bo871-a} but refers to the 1868 paper only. Also unclear is
  why neither author refers to Liouville's theorem of 1838, while proving
  it over and over again and still in 1881,
  \citep{Bo868,Bo868b,Bo871-b,Bo877b,Bo872,Ma879-c,Bo881}, so that one has to
  conclude that the theorem was not well known, yet, at the date. It should
  also be stressed that nowhere the authors mention that the volume
  elements are deformed by the evolution, so that it has to be imagined
  that they are really thought as points, see Sec.\ref{sec:8},\ref{sec:9}
  below.} The kinetic energy of the $n$-particles configurations with the
$3n$-position coordinates $q=(q_1,\ldots,q_n)$ in a given position space
cell is $n\k-\chi(q)$, if $\ch(q)$ is potential energy, is partitioned in all
possible ways among the kinetic energy levels. The microcanonical
distribution is obtained by essentially the same combinatorial argument of
Sec.II but using also a discretization of positions: thus assuming that the
phase space trajectory visits all cells in the {\it available} phase space,
as pointed out by Maxwell.%
\footnote{\alertr{The result is given only after integrating
    it over the velocities, \citep[\#5,p.95,l.-6]{Bo868}: for details see
    Appendix \ref{1868}.}}

The above mentioned hypothesis may seem stronger than the ergodic
hypothesis because it deals with phase space and does not mention the
exceptions that are allowed in the mathematical version of the hypothesis:
but it deals with motion discretized in finitely many cells and therefore
it is simply different from the mathematical version, and the usual
impossibility argument cannot be used.%
\footnote{\alertr{From a mathematical point of view it is obviously
    impossible that a periodic trajectory covers entirely a region of space
    and the assumption needs intepretation (because, out of respect for the
    founding fathers, it cannot be just dimissed).  For instance, if the
    forces are supposed smooth, consider a point in a small enough volume
    $\D$ in phase space: the velocity (of the phase space point
    representing the system configuration) will be non zero and data
    initially in $\D$ will go out of it without being able to return before
    a time $\t$ necessarily positive and independent of the initial
    location within $\D$. Therefore the initial point can return to $\D$
    only finitely many times during the period if its motion; each time it
    will cover a smooth segment in $\D$; but a finite number of smooth
    segments cannot cover the continuum of points in $\D$.}}

The periodicity property remains in the background and the possibility
arises that it could even be forgotten and replaced by a direct assumption that
the probability of configurations in phase space is simply obtained by
counting the number of ways of realizing them in phase space.  This may
seem (a new assumption and)  an easy way out of the unphysically long time
scales in which periodicity would become manifest.

But Boltzmann always identified probabilities with visits frequencies and
did not really abandon the periodic conception of the systems motions:
eventually developed arguments to show that periodicity plaid the ``role of
a symmetry property'' which implied, for very small or very large systems,
properties that become observable in ``human time scales'', as predicted by
Boltzmann's equation for dilute gases in \citep[\#22]{Bo872}, for the deep
reason that they hold for {\it most} system configurations,
\citep[\#39]{Bo877a} and \citep[\#73]{Bo884}.

\def\SEC{Discrete phase space, ergodicity \& regular cells.}
\section{\SEC}
\label{sec:8}
\iniz

The hypothesis that the statistics of the equilibrium states are given by
probability distributions over all states compatible with energy
conservation, through which the phase space point representing the state
cycles periodically, had been used in the 1868 work of Boltzmann, as
recognized by Maxwell, see Sec.\ref{sec:7} above and Appendix \ref{1868}
below. It has to be properly interpreted and it is the ``Ergodic
Hypothesis''.

It is important {\it not} to identify this with the similar statement
mentioned by Boltzmann around 1870 for the molecules in a rarefied gas, see
p.237,l.8, in \citep[\#18]{Bo871-a}, about the internal motion of
atoms in single molecules of a gas: ``{\it it is clear that the various gas
  molecules will go through all possible states of motion}''. This property
is attributed as due to collisions between different molecules: viewed as
non periodic, stochastic, events which make the state of a single molecule
assume all the possible configurations (on a time scale far shorter than
any global recurrence). As shown in the same key work
\citep{Bo871-a}, the latter property implies the canonical distribution for
the molecules states, while the ergodic hypothesis implies the
microcanonical distribution for the whole ``warm body''.

Boltzmann, shortly afterwards and closer to Maxwell's interpretation
(Sec.\ref{sec:7} above, quoted in full in Appendix\ref{1868},
p.\pageref{Max1879} below) of the hypothesis of his own 1868 paper, gives a
more precise formulation, \citep[\#19,p.284]{Bo871-b} (see also
Appendix\ref{1871-b} below):, \*

\0{\it Finally from the equations derived above we can, under an assumption
  which it does not seem to me of unlikely application to a warm body,
  directly access to the thermal equilibrium of a polyatomic molecule, and
  more generally of a given molecule interacting with a mass of gas. The
  great chaoticity of the thermal motion and the variability of the force
  that the body feels from the outside makes it probable that the atoms
  [{\rm of the molecules}] take in the motion, that we call heat, all
  possible positions and velocities compatible with the equation of the
  kinetic energy {[\rm conservation of energy]}, and even that the atoms of a
  warm body can take all positions and velocities compatible with the last
  equations considered.}%
\footnote{\alertr{ The ``variability of the force that the body feels from
    the outside'' and molecular collisions appear essential for the
    statement to hold for the atoms of a single molecule, see also
    \citep{Br003}: however the variability {\it should not} be intended as
    ``random'' when stating ``...even that the atoms of a warm body
    ...''. In \citep{Ma879-c} this ``variability'' is exemplified
    by referring to the roughness (and not the randomness) of the
    interactions with a container: so the statement is still about
    deterministic, time independent, interactions. Only in cases in which
    the system admits constants of motion due to symmetries that are not
    generic, the system may avoid visiting all possible positions and
    velocities compatible with the equation of the kinetic energy,
    \citep[\#5,p.96]{Bo868}, see also the comment in the recent
    collection \citep[p.xxviii]{Fl000}.}. }  \*

The last sentence here has to be particularly remarked as it jumps from the
properties of {\it atoms inside molecules} immersed in a gas to properties
of {\it atoms in a warm body} as demanded by the ergodic hypothesis.

In the \citep[\#5]{Bo868} work the assumption is used to imply that
since the motion visits the whole phase space then the frequencies of visit
to regions must be described by a density function which must be invariant
under time evolution (because of the phase space volume conservation),
leading to the microcanonical distribution for the whole system, see
\citep[\#5,p.95,l.-6]{Bo868}, and to a canonical distribution for the
individual molecules.

The view that the equilibrium distribution should be representable by a
density on phase space which ought to be $>0$ for all configurations
compatible with energy conservation did not really demand the intervention
of random external forces. This is made clear by the immediately following
work, \citep[\#6]{Bo868b}, where an example is discussed in detail: however
the example shows that Boltzmann did not {\it yet} conceive the possible
existence of other probability distributions {\it with positive density} on
the region of phase space allowed by the values of other constants of
motion beyond the energy, see comments in Appendix\ref{1868},\ref{1868b}
below.\footnote{ By other constants of motion one should understand smooth
  constants. Boltzmann did not really consider functions which are not
  smooth (\ie just measurable) and which are essential in modern
  mathematical problems in ergodic theory.}

His later works, starting with \citep[\#18]{Bo871-a}, show that he later
realized this possibility and was very concerned about it: density is still
taken for granted but its dependence on the only energy is openly
questioned, and absence of multiple collisions or of other constants of
motion have to be added as supporting assumptions.

Most important, in the three 1871 papers \citep{Bo871-a,Bo871-b,Bo871-c}, as
already in the earlier \citep[\#5]{Bo868} work, equipartition may
appear as obtained without using the equations of motion, except for the
pair collisions conservation laws or via a discrete representation of the
states of the atoms, classified through discrete values of their possible
kinetic energy and positions, determined by a partition of all the
available phase space into parallelepipedal cells (as in the above quoted
hypothesis).%
\footnote{\alertr{The initial restriction that the density be so low
    that the time of free flight be overwhelmingly large compared to the
    collision time {\it is lifted} in Sec.III of \citep[\#5]{Bo868}. }}
This allowed to set up the determination of the probability of the states
(as points in phase space) via a combinatorial argument.

Nevertheless dynamics is deeply linked to the derivations because the
probability of finding a microscopic configuration is {\it identified with
  the fraction of time, in its evolution, it spends inside a given cell}
and is implied by the assumption that the phase space point representing
the system visits cyclically all {\it discrete cells} into which phase
space is divided. And this is the complete {\it interpretation of the
  Ergodic Hypothesis} (distinct from its modern interpretation in the
ergodic theory of dynamical systems).

Discretization of phase space into cells is done systematically in
\citep[\#42]{Bo877b} going well beyond the \citep[\#5]{Bo868}, in
the sense that the program of computing probabilities by counting the
number of ways of realizing the microscopic configurations in a discretized
phase space is presented as a general method and used to provide strength
to the analysis in the earlier paper on Loschmidt's irreversibility
question, where the ``sea of Boltzmann'' (as Uhlenbeck,\citep{Ul968}, called
it) was introduced, \citep[\#39]{Bo877a}, see
Appendices\ref{1877-a},\ref{1877-b}.

It should be noted and stressed that in the discretisation used the phase
space cells are parallelepipeds:\footnote{\alertr{Cubes in space times
    cubes in momentum. At least this is so in \citep[\#42]{Bo877b}: while in
    \citep[\#5]{Bo868} the cells in momentum space were simply cells in
    which energy $\frac12m c^2$ was between $e$ and $e+de$, with
    $de=\frac1p$ ($p$ very large) and had a multiplicity proportional to
    $\sqrt{e} de$ in $3$-dimensional space, which is appropriate for {\it
      cubic cells} of side $\frac1p$.}}
{\it this is an essential point not really stressed in the
  literature}. Boltzmann does not comment on that: but he wants to think,
as he repeatedly wrote since realizing that his ideas were misunderstood,
that derivatives and integrals are just ``approximations'' of certain
ratios and sums, \citep[p.25-29]{Du959}.

{\it This means, if ratios and sums are to be interpreted as the
usual derivatives and integrals, that the discretization of phase space and
time must be done via a regular lattice}.\phantomsection{\label{regularcells}}

The laws of nature can be formulated {\it via discrete difference equations
or equivalently by the usual corresponding differential equations} only
if the discretization is imagined on a {\it regular lattice}, \eg a
parallelepipedal one (the parallelepipeds being $3N$ dimensional cubes in
positions space times $3N$ dimensional cubes in momentum space). If
discretizations were performed on discrete approximations of fractals, \ie
irregular (even though scaling symmetric) sets of points, currently
considered in various fields, the laws of nature would be very different
(maybe needing fractional derivatives in their continuum versions) and the
Gibbs distributions, for instance, could be quite different.

This becomes a key point in the modern approaches to nonequilibrium in
which the analogue, for instance, of the microcanonical distribution in
general dissipative systems is determined,\footnote{\alertr{ This is
    the SRB distribution, \citep{Ru989}.}} and in the conceptual unification
between equilibrium and nonequilibrium, see \citep{Ga013b}.

\def\SEC{Boltzmann on discreteness}
\section{\SEC}
\label{sec:9}
\iniz

The discrete view of phase space makes it mathematically possible to say
that a point visits all phase space cells: but this statement was
attributed to {Boltzmann} and criticized over and over again (see
\citep[p.505]{Br003}) even by Physicists, including in the influential book,
\citep{EE911}, although enlightened mathematicians could see better, see
p.385 in \citep{Br976}).

\* Therefore the question: {\it can it be said that the not mathematically
  formalized ergodic hypothesis of Boltzmann, Clausius and Maxwell by
  chance led them to statistical mechanics}, just on the grounds that its
mathematical interpretation ``cannot be true'' as proved in several
Mathematics and Physics works?  should be answered simply ``no''. But not
simply because formalizing a statement into one that via Mathematics can be
proved right or wrong is ambiguous.%
\footnote{\alertr{ Many famous results in Mathematics became correct
    after the very formulations of the problems originating them was
    slightly changed, as the KAM theorem story reminds.}}

The point is that for {Boltzmann}, for instance, phase space was discrete
and points in phase space had finite size, that I will call $h^3$ as
customary, and were located on a {\it regular lattice}. And time evolution
was a permutation of them: ergodicity meant therefore that the permutation
was just a {\it one cycle permutation}. This is very cearly discussed by
Dugas in \citep[p.25-29]{Du959}, as followed here.

It does not seem that in the original viewpoint Boltzmann's systems of
particles were really thought of as susceptible of assuming a $6N$ {\it
  dimensional continuum of states}.

The relation between the apparent continuum of reality, as we perceive it
and input it in most of our models or theories for its interpretation and
understanding, and the possibly intrinsic and deep discrete nature of
reality and/or of our own thinking is exemplified explicitly in many among
Boltzmann's writings, for instance, \citep[p.169]{Bo974}: \*

{\it Therefore if we wish to get a picture of the continuum in words,
we first have to imagine a large, but finite number of particles with
certain properties and investigate the behavior of the ensemble of
such particles. Certain properties of the ensemble may approach a
definite limit as we allow the number of particles ever more to
increase and their size ever more to decrease. Of these properties one
can then assert that they apply to a continuum, and in my opinion this
is the only non-contradictory definition of a continuum with certain
properties}
\*

\0Or also, \citep[p.56]{Bo974}.:

\*
``{\it The concepts of differential and integral calculus separated from
any atomistic idea are truly metaphysical, if by this we mean, following
an appropriate definition of Mach, that we have forgotten how we acquired
them}''
\*
\0And, \citep[p.55]{Bo974}:
\*
``{\it Through the symbols manipulations of integral calculus, which have
become common practice, one can temporarily forget the need to start
from a finite number of elements, that is at the basis of the creation
of the concept, but one cannot avoid it}''.
\*
\0or, \citep[p.227]{Bo974}:
\*
``{\it Differential equations require, just as atomism does, an
initial idea of a large finite number of numerical values and points
...... Only afterwards it is maintained that the picture never
represents phenomena exactly but merely approximates them more and
more the greater the number of these points and the smaller the
distance between them. Yet here again it seems to me that so far we
cannot exclude the possibility that for a certain very large number of
points the picture will best represent phenomena and that for greater
numbers it will become again less accurate, so that atoms do exist in
large but finite number}
\*
\0and, \citep[p.55-56]{Bo974}:
\*
``{\it This naturally does not exclude that, after we got used once and
for all to the abstraction of the volume elements and of the other
symbols {\rm[of calculus]} and once one has studied the way to operate
with them, it could look handy and luring, in deriving certain
formulae that Volkmann calls formulae for the coarse phenomena, to
forget completely the atomistic significance of such
abstractions. They provide a general model for all cases in which one
can think to deal with $10^{10}$ or $10^{10^{10}}$ elements in a cubic
millimeter or even with billions of times more; hence they are
particularly invaluable in the frame of Geometry, which must equally
well adapt to deal with the most diverse physical cases in which the
number of the elements can be widely different. Often in the use of
all such models, created in this way, it is necessary to
put aside the basic concept, from which they have overgrown, and perhaps
to forget it entirely, at least temporarily. But I think that it would
be a mistake to think that one could become free of it
entirely.}''
\*

The latter sentence also reminds us that the evaluation of the corrections
is of course a harder problem, which it would be a mistake to set aside. In
fact the corrections are quite important and somehow even more important
than the continuum models theselves, which will remain as a poor
idealization of far more interesting cooperative phenomena (see, for
instance, the microscopic theory of phase separation in the Ising model,
\citep{Ga000}).

The discrete conceptions of Boltzmann, perfectly meaningful mathematically,
were apparently completely misunderstood by his critics: yet it was clearly
stated for instance in the analysis of the Loschmidt remarks on
irreversibility \citep[\#39]{Bo877a}, or in one of the replies to
  Zermelo, \citep[\#119,1896]{Bo896}, and in the book on gases,
\citep[1896]{Bo896a}, see also \citep{Ga995a}.

But the seed had been sown and the discrete phase space and the ergodic
hypothesis became unseparable.

\def\SEC{Thermodynamic analogies. Ensembles,}
\section{\SEC}
\label{sec:10}
\iniz

The notion of thermodynamic analogy, or model of thermodynamics, was
formulated and introduced by Helmoltz for general systems admitting only
periodic motions (called {\it monocyclic}), \citep{He884a,He884b}.%
\footnote{\alertr{ However the notion had been developed earlier although
    not formalized, by Boltzmann already in 1866, and by Clausius in
    1871. It is clearly used, again without a formal definitition, in
    \citep[\#20]{Bo871-c} and again developed in \citep[\#39]{Bo877a},
    where also an explicit example is discussed.\\
    It is possible that Boltzmann did not claim priority over Helmoltz'
    work not only because he had become older, hence wiser, but also
    because he wrote this paper while he was in contact with Helmoltz who
    was inviting him to accept a professorship in Berlin: I conjecture that
    he had started the 1884 work to establish a closer scientific contact
    with Helmoltz but, while thinking to it, he realized that he could go
    far beyond the monocyclic systems and formalized and unified into a
    fully developed theory of ensembles his older theory of the
    microcanonical ensemble, built on the combinatorial count of the
    microscopic configurations in \citep[\#5]{Bo868} and
    \citep[\#42]{Bo877b}, and of the canonical and microcanonical
    ensembles built on a form of the ergodic hypothesis in
    \citep[\#20]{Bo871-c}. and the other ``trilogy'' papers.}}

\* \0{\bf Thermodynamic analogy:} {\it Given a Hamiltonian system and a
  collection $\EE$ of invariant probability distributions on phase space
  depending on parameters $\Ba=(\a_1,\a_2,\ldots)$ denoted $\m_{\Ba}$ let
  $U(\Ba),V(\Ba),p(\Ba),T(\Ba)$ be $\m_{\Ba}$-average values of suitable
  observables. If $\Ba$ is varied by $d\Ba$ let $d U,d V,d p,d T$ be the
  corresponding variations of $U(\Ba),V(\Ba),p(\Ba),T(\Ba)$. Then the
  ``ensemble'' of the distributions $\m_\Ba$ and the functions
  $U(\Ba),V(\Ba),p(\Ba),T(\Ba)$ describe a thermodynamic analogy if
$$\frac{dU+pdV}T\,=\,\hbox{exact differential}\eqno(@)$$
\ie if there is a function $S(\Ba)$ such that $d S=  \frac{dU+pdV}T$.}
\*

An example, actually illustrating the early works of Boltzmann and
Clausius, is in Sec.\ref{sec:4} above.

The wide generality of the notion provided a new perspective, and grounds,
to make popular the theory of the ensembles, and to show that the
thermodynamic relations would hold in {\it most mechanical systems}, from
the small and simple to the large and complex. 

In the first cases the relations would be trivial identities of no or
little interest, just {\it thermodynamic analogies}, but in the large
systems they would become nontrivial and interesting being relations of
general validity. In other words they would be a kind of symmetry property
of Hamiltonian Mechanics.

Then in the fundamental paper \citep[\#73]{Bo884}, following and
inspired by the quoted works of {Helmoltz}, {Boltzmann} was able to achieve
what I would call the completion of his program of deducing the second law,
Eq.(@), from Mechanics. In simple monomolecular systems: \*

\0(1) the {\it absolute temperature} $T$ is identified with the average
kinetic energy over the periodic motion following the initial datum
$(\V p,\V q)$ of a macroscopic collection of $N$ identical particles
interacting with a quite {\it arbitrary} pair interaction, and

\0(2) the {\it energy} $U$ is $H(\V p,\V q)$ sum of kinetic and of a
potential energy,

\0(3) the {\it volume} $V$ is the volume of the region where the positions
$\V q$ are confined (typically by a $V$-dependent hard wall potential),

\0(4) the potential energy of interaction is $\f_V$, usually dependent on
the volume $V$, and the {\it pressure} $p$ is the average of $-\dpr_V \f_V$
(with the interpretation of force exercized on the walls by
    the colliding particles),
\*

\0then, from the assumption that each phase space point would evolve
periodically visiting every other point on the energy surface ({\it i.e.}
assuming that the system could be regarded as ergodic in the sense of
Sec.\ref{sec:8} or monocyclic in Helmoltz' sense) it would follow that the
quantity $p$ could be identified with the $\media{-\partial_V\f_V}$, time
average of $-\partial_V \f_V$, and Eq.(@) would follow as a {\it heat
  theorem}.

The heat theorem would therefore be a consequence of the general properties
of monocyclic systems, which already in the 1866 work of Boltzmann had been
shown to generate a thermodynamic analogy.

This led Boltzmann to realize, in the same paper, that there were a
large number of mechanical models of thermodynamics: the macroscopic states
could be identified with regions whose points would contribute to the
average values of quantities with thermodynamic interpretation ({\it
i.e.} $p,V,U,T$) with a weight (hence a probability) also invariant
under time evolution.

Hence imagining the weights as a density function one would see the
evolution as a motion of phase space points, each representing possible
atomic configuration, leaving the density fixed. Such distributions on phase
space were called {\it monodic}, because they keep their identity with time
or, as we say, are invariant.

And in \citep[\#73]{Bo884} several {\it collections} $\EE$ of weights
or {\it monodes} were introduced: today we call them invariant
distributions on phase space or {\it ensembles}. Among the ensembles $\EE$,
{\it i.e.}  collections of monodes, Boltzmann singled out the ensembles
called {\it orthodes} (``behaving correctly''): they were the families
$\EE$ of probability distributions on phase space depending on a few
parameters (normally $2$ for simple one component systems) such that the
corresponding averages $p,V,U,T$, defined in (1-4) above, would vary when
the parameters were varied. And the variations $dU,dV$ of average energy
and volume would be such that the {\it r.h.s} of Eq.(@) would be an exact
differential, thereby defining the {\it entropy} $S$ as a function of
state, see \citep{Ga995a,Ga000}.

The ergodic hypothesis yields the ``orthodicity'' of the ensemble $\EE$,
that today we call {\it microcanonical} (in \citep[\#73]{Bo884} it was
named {\it ergode}): but ergodicity, {\it i.e.} the dynamical property that
evolution would make every phase space point visit every other, was not
necessary to the orthodicity proof of the ergode: {\it but it is necessary
  for its physical interpretation} of a distribution that describes an
equilibrium state with given macroscopic parameters.

In fact in \citep[\#73]{Bo884} the relation Eq.(@) is proved directly
without recourse to dynamical properties (just as we do today, see
\citep{Fi964,Ru968,Ga000} and Appendix\ref{1884} below); and in the same way
the orthodicity of the {\it canonical ensemble} (called {\it holode} in
\citep[\#73]{Bo884}) was obtained and shown to generate a Thermodynamic
analogy which is equivalent (for large enough systems) to the one
associated with the microcanonical ensemble.%
\footnote{\alertr{ Still today a different interpretation of the word
    ``ensemble'' is widely used: the above is based on what Boltzmann calls
    ``{\it Gattung von Monoden}'', see p.132,l.14 of
    \citep[\#73]{Bo884}: unfortunately he is not really consistent in
    the use of the name ``monode'' because, for instance in p.134 of the
    same reference, he clearly calls ``monode'' a collection of invariant
    distributions rather than a single one; further confusion is generated
    by a typo on p.132,l.22, where the word ``ergode'' is used instead of
    ``holode'' while the ``ergode'' is defined only on p.134. It seems
    beyond doubt that ``holode'' and ``ergode'' were intended by Boltzmann
    to be {\it collections} $\EE$ of invariant distributions (parameterized
    respectively by $U,V$ or by $(k_B T)^{-1},V$, in modern notations):
    Gibbs instead called ``ensemble'' each single invariant distribution
    (\ie a ``monode'', as a family of identical non interacting systems
    filling phase space with an invariant density), or at least that is
    what is often stated. Although the original names proposed by Boltzmann
    may be more appropriate, of course, we must accept calling
    ``microcanonical ensembles'' the elements of ergode and ``canonical
    ensembles'' those of the holode, see \citep{Ga000}, but I would prefer
    replacing the word ensemble by distribution.}}

In the end in \citep[1868-71]{Bo868,Bo871-b,Bo877a} and, in final form, in
    \citep[\#73]{Bo884} the theory of ensembles and of their
    equivalence was born without need of the ergodic property: however its
    still important role was to guarantee that the quantities $p,V,U,T,S$
    defined by orthodic averages with respect to invariant distributions on
    phase space had the physical meaning implied by their names.  This was
    true for the microcanonical ensemble (or ``ergode'') by the ergodic
    hypothesis, and followed for the other ensembles by the equivalence.

Unfortunately the paper \citep[\#73]{Bo884} has been overlooked until
quite recently by many, actually by most, physicists.%
\footnote{{Possibly because it starts, in discussing the
    thermodynamic analogy, by giving the Saturn rings as an ``example'': a
    brilliant one, certainly but perhaps discouraging for the suspicious
    readers of this deep and original paper on Thermodynamics.}}
 See p.242 and p.368 in \citep{Br976} for an exception, perhaps the first.

Of course conceiving phase space as discrete is essential to formulate the
ergodicity property in a mathematically and physically acceptable way: it
does not, however, make it easier to prove it, even in the discrete sense
just mentioned (nor in the sense acquired later when it was formulated
mathematically for systems with continuous phase space).

 It is in fact very difficult to be {\it a priori\ }\ sure that the
 dynamics is an evolution which has only one cycle. Actually this is very
 doubtful: as one realizes if one attempts a numerical simulation of an
 equation of motion which generates motions which are ergodic in the
 mathematical sense, like motions in convex billiards, \citep{Si977}.

And the difficulty is already manifest in the simpler problem of simulating
differential equations in a way which rigorously respects the uniqueness
theorem. In computers the microscopic states are rigorously realized as
regular cells (because points are described by integers, so that the cells
sizes are limited by the hardware performance and software precision) and
phase space is finite. By construction, simulation programs map a cell into
another: but it is extremely difficult, and possible only in very special
cases (among which the only nontrivial that I know is \citep{LV993}) without
dedicating an inordinate computing time to insure a $1-1$ correspondence
between the cells.

Nevertheless the idea that phase space is discrete and motion is a
permutation of its points is very appealing because it gives a privileged
role to the {\it uniform distribution} on the phase space region in
which the motion develops, {\it i.e.\ } the energy surface, if the ergodic
hypothesis holds, assuming it discretized on a regular lattice.

Furthermore the hypothesis has to be supplemented with the realization,
made possible by the discretization of phase space which allows a detailed
combinatorial analysis, that its predictions can be checked on
short time scales because they are properties of most microscopic
configurations, making the unobservable recurrence time irrelevant.

The success of the ergodic hypothesis has several aspects. One that
will not be considered further is that it is not necessary: this is
quite clear as in the end we want to find the relations between a very
limited number of observables and we do not need for that an
assumption which tells us the values of all possible averages, most of
which concern ``wild'' observables (like the position of a tagged
particle). The consequence is that the ergodic hypothesis is intended
in the sense that confined Hamiltonian systems ``can be regarded as
ergodic for the purpose of studying their equilibrium properties''.
\*

\0{\it Remark:} The above historical analysis is concentrated on the
ergodic hypothesis and only touches the theory of the ensembles in the
years following 1884: hence the role of Gibbs, Planck, Einstein ..., as
well as the natural role played in quantum mechanics by Boltzmann's initial
view on microscopic motion periodicity, \citep{dB995}, is not even mentioned
here. A very detailed study mostly centered on the development of the
theory of the ensembles has recently appeared, \citep{In015}.

\def\SEC{From equilibrium to stationary monequilibrium}
\section{\SEC}
\iniz\label{sec:11}

What is, perhaps, the most interesting aspect of the Ergodic Hypothesis
formulated imagining space-time and phase space as a {\it regular array of
points} is that it can hold for systems of any size and lead to relations
which are essentially size independent as well as model independent and
which become interesting properties when considered for macroscopic
systems.  \* \0{\it Is it possible to follow the same path in studying
  nonequilibrium phenomena?}  \*

The simplest such phenomena arise in stationary states of systems subject
to the action of nonconservative forces and of suitable heat-removing
forces (\ie ``thermostats'', whose function is to forbid indefinite build
up of energy in the system).

Such states are realized in Physics with great accuracy for very long
times, in most cases longer than the available observation times. For
instance it is possible to keep a current circulating in a wire
subject to an electromotive force for a very long time, provided a
suitable cooling device is attached to the wire.

As in equilibrium, the stationary states of a system will be described by a
collection $\EE$ of probability distributions $\m$ on phase space,
invariant with respect to the dynamics, which I continue to call {\it
  ensemble}: the distributions $\m\in\EE$ will be parameterized by a few
parameters $U,V,E_1,E_2,\ldots$ which have a physical interpretation of
(say) average energy, volume, intensity of the nonconservative forces
acting on the system (that will be called ``external parameters'').

Each distribution $\m$ will describe a macroscopic
state in which the averages of the observables will be their integrals
with respect to $\m$.  The equations of motion will
be symbolically written as
$$\dot{\V x}=\V f(\V x)$$
and we shall assume that $\V f$ is smooth, that it depends on the
external parameters and that the phase space visited by trajectories
is bounded (at fixed external parameters and initial data).

Since we imagine that the system is subject to nonconservative forces
the phase space volume (or any measure with density with respect to
the volume) will not be preserved by the evolution and the divergence
$$\s(\V x)=-\sum_{i}\partial_{x_i} f_i(\V x)$$
will be {\it different} from $0$.
We expect that, in interesting cases, the time average $\s_+$ of
$\s$ will be positive:\citep{Ru996,Ru997}
$$\s_+\defi \lim_{\t\to\infty}\frac1\t \int_0^\t
\s(S_t\V x)\,dt\,>\,0\,.$$ 
and, with ``few'' exceptions, $\V x$--independent.

This means that there cannot be invariant distributions that can be
represented by a density with respect to the volume. And the problem to
find even a single invariant distribution is nontrivial except possibly for
some concentrated on periodic orbits.

Occasionally an argument is found whereby, in equilibrium, motion can be
regarded as a permutation of cells ``because of the volume conservation due
to Liouville's theorem''. But this cannot be a sensible argument due to the
chaoticity of motion: it cannot be ignored (whether in equilibrium or in
nonequilibrium states) that any volume element will be deformed under
evolution and stretched along certain directions while it will be
compressed along others. Therefore the points of the discretized phase
space should not be thought as small volume elements, ``cells'' with
positive volume, but precisely as individual ($0$ volume) points which the
evolution permutes.

The problem can be attacked, possibly, by following  again Boltzmann's
view of dynamics as discrete, (``{\it die Zahl der lebendigen Kr\"aft ist
eine diskrete}'', see p.167 in \citep[\#42]{Bo877b}).

\def\SEC{Nonequilibrium and discrete phase space}
\section{\SEC}
\iniz\label{sec:12}

In a discrete conception, points in phase space {\it are not
  extended cells} but they evolve as points without dimension. {\it This
  has become a common way of imagining dynamical systems, particularly
  because of the development of simulations}.

Simulations have played a key role in the recent studies on
nonequilibrium. And simulations operate on computers to perform solutions
of equations in phase space: therefore phase space points are given a
digital representation which might be very precise but rarely goes beyond
$32$ bits per coordinate. If the system contains a total of $N$ particles
each of which needs $4$ coordinates to be identified (in the simplest
$2$--dimensional models, $6$ otherwise) this gives a phase space
(virtually) containing $\NN_{tot}=(2^{ 32})^{4N}$ points which cover a
phase space region of desired size $V$ in velocity and $L$ in position with
a lattice of mesh $2^{-32}V$ or $2^{-32}L$ respectively.

Therefore the ``{\it fiction}'' of a discrete phase space, used first by
Boltzmann in his foundational works, \citep[\#42,p.167]{Bo877b}, has
been taken extremely seriously in modern times with the peculiarity that it
is seldom even mentioned in the numerical simulations.

A simulation for a system of $\NN_{tot}$ atoms is a code that operates on
discrete phase space points transforming them into other points. In other
words it is a map $\lis S$ which associates with any point on phase space a
new one with a precise deterministic rule that can be called a {\it
  program} or {\it code}.

All programs for simulating solutions of ordinary differential equations
have some serious drawbacks: for instance, as mentioned above, it is very
likely that the map $\lis S$ defined by a program is not invertible, unlike
the true solution to a differential equation of motion, which obeys a
uniqueness theorem: different initial data might be mapped by $\lis S$ into
the same point.

Since the number $\NN_{tot}$ is finite, all points will undergo a motion,
as prescribed by the program $\lis S$, which will become {\it recurrent},
\ie will become eventually a permutation of a subset of the phase space
points, hence {\it periodic}.  

The ergodic hypothesis was born out of the natural idea that the
permutation would be a {\it one cycle} permutation: every microscopic state
would recur and continue in a cycle, \citep{Ga995a}. In simulations, even if
dealing with discretized time reversible systems, it would not be
reasonable to assume that all the phase space points are part of a
permutation, because of the mentioned non invertibility of essentially any
program. It is nevertheless possible that, once the {\it transient} states
(\ie the ones that never recur, being out of the permutation cycles) are
discarded and motion reduces to a permutation, then the permutation is just
a single cycle one.

So in simulations of motions of isolated systems an ergodic
hypothesis can be defined and it really consists in two parts: first,
the non recurrent phase space points should be ``negligible'' and, second,
the evolution of the recurrent points consists in a single cycle
permutation. Two comments:

\0(a) Periodicity is not in contrast with chaotic behavior: this is a point
that Boltzmann and others (\eg Thomson (Lord Kelvin))
clarified in several papers (\citep[\#39]{Bo877a},\citep{Th874},
for the benefit of the few that at the time listened.
\\
(b) The recurrence times are beyond
any observable span of time (as soon as the particles number $N$ is larger
than a few units), \citep[Sec.88]{Bo896a}.

In presence of dissipation, motions in models based on a continuum
space-time, develop approaching a subset of phase space, {\it the
  attracting set $A$},%
\footnote{\alertr{The notion of attracting set $\AA$ and of attractor
    for a system whose evolution is governed by an ordinary differemtial
    equation are sometimes identified: it is however useful to recall the
    definitions.  An attracting set $\AA$ is a closed set such that all
    initial data close enough to it (\ie in a domain of attraction $U$)
    evolve in time so that their distance to $\AA$ tends to $0$. If points
    chosen randomly in a domain of attraction of $\AA$, with a probabilty
    with a density with respect to the volume, evolve in time generating on
    $\AA$ a {\it unique} statistical distribution $\m$, then any invariant
    subset $B$ of $\AA$ which has $\m$-probability $1$ ($\m(B)=1$) is
    called an attractor. Of course in a discrete model of evolution the two
    notions coincide.}}
 and on it the attractor $B$, which has therefore zero volume: because
 volume is not invariant and is asymptotically, hence forever, decreasing.

 In general the {\it nonrecurrent points will be ``most'' points}: because
 in presence of dissipation the attractor set will have $0$ volume, (even
 in the cases in which the attracting set $\AA$ is the entire phase space,
 like in the small perturbations of conservative Anosov systems,
 \citep{AA966,Si977}).

Nevertheless in the discrete form the ergodic hypothesis can be
extended and formulated also for general nonconservative motions, see below. 

\def\SEC{Nonequilibrium: the ergodic hypothesis}
\section{\SEC}
\iniz\label{sec:13}

The above considerations suggest the following extension of the ergodic
hypothesis: it can be formulated by requiring that
\*
\0{\bf Ergodic hypothesis} (equilibrium and
nonequilibrium): {\it In a phase space discretized on a regular lattice%
  \footnote{\alertr{ As commented above the discretization on
      a regular lattice is essential and simply reflects our empirical
      evidence that the systems must be equivalently described by the usual
      ordinary differential equations.}}%
  non recurrent points are negligible and the recurrent points form the
  attracting set and are cyclically permuted forming a one cycle
  permutation. 
}  \*
Since empirically most systems. in equilibrium as well as out of
equilibrium, show chaotic motions it is convenient to assume that, aside
from exceptions, systems satisfy the stronger:
\*

\0{\bf Chaotic hypothesis}: {\it The above ergodic hypothesis holds and
  furthermore the motion on the attracting set is supposed chaotic.}
\*

Mathematically ``chaotic'' means that the attracting set can be imagined as
a surface and that the motion on it is hyperbolic, see \citep{Ga013b}
(for history and developments).%
\footnote{\alertr{Hyperbolic means that every point is the intersection of
    the phase space velocity with two surfaces, $W_u,W_s$, transversal to
    it and which expand/contract, as time $t$ elapses, under the evolution
    map $S_t$ at a minimum rate $\l>0$.}}

In the latter situations the statistics of the motions will be {\it uniquely
determined} by assigning a probability $\NN^{-1}$ to each of the $\NN$
configurations on the (discrete version of the) attractor: and this will be
the {\it unique} stationary distribution, providing also an answer to the
question raised already by Boltzmann, \citep[\#8,p.255]{Bo871-a}, in the
early days.\*

\0{\it Remarks:} (1) The uniqueness of the stationary distribution is by
no means obvious and, as well, it is not obvious that the motion can
be described by a permutation of the points of a regularly discretized
phase space. Not even in equilibrium.

\0(2) Boltzmann argued, in modern terms, that after all we are interested
in very few observables, in their averages and in their
fluctuations. Therefore we do not have to follow the details of the
microscopic motions and all we have to consider are the time averages of a
few physically important observables $F_1,F_2,\ldots, F_q$, with $q$
small. This means that we have to understand what is now called a {\it
  coarse grained} representation of the motion, collecting together all
points on which the observables $F_1,F_2,\ldots, F_q$ assume the same
values. Such collection of microscopic states is called a {\it macrostate},
\citep{Le993,GGL005}.  \\
The reason why motion appears to reach stationarity and to stay in that
situation is that for the overwhelming majority, \citep{Bo877a,Ul968}, of
the microscopic states, \ie points of a discretized phase space, the
interesting observables have the same values. The deviations from the
averages are observable over time scales that are most often of humanely
appreciable size and have nothing to do with the recurrence
times. Boltzmann gave a very clear and inspiring view of this mechanism by
developing ``Boltzmann's equation'', \citep[\#22]{Bo872}: perhaps
realizing its full implications only a few years later when he had to face
and overcome the conceptual objections of Loschmidt and others,
\citep[\#39]{Bo877a}.

A development of the above chaotic hypothesis together with some of its
consequences, including few implications on the theory of fluctuations, can
be found in the review \citep{Ga013b}.  The first among the consequences is
the identification of the statistics of the stationary states of the
systems: which can considered valid for nonequilibrium as well as for
equilibrium (where it coincides with the microcanonical Gibbs state). And
it can be viewed as the discretized version of Ruelle's theory of chaotic
dynamical systems, \citep{Ru989,Ru995,Ru003a} and originated in
\citep{RT971}.

The discrete conception of phase space also allows, assuming the chaotic
hypothesis, to count the number $\NN$ of configurations in the attracting
sets and to ask the question whether entropy can be defined proportionally
to $\log\NN$ as in equilibrium: in \citep{Ga013b} a (controversial) negative
conclusion is reached together with the (positive) argument that,
neverthess, the evolution towards stationarity might have a Lyapunov
function (actually many) related to $\log \NN$, see \citep{Ga013b}.

The alternative view that entropy should make sense as a function of the
state, possibly defined up to a constant, {\it whether in equilibrium or in a
stationary non equilibrium} or even for non stationary states is held by
many physicists since a long time, see for instance the view of Planck,
\citep[p.239]{Du959}.

\pagina
\appendix
\def\alertr{\alertb}
\renewcommand{\thesection}{\Alph{section}}
\renewcommand{\theequation}{\Alph{section}.\arabic{equation}}
\centerline{\Large \bf Appendices%
}
\setcounter{section}{0}

\def\SEC{Boltzmann' priority claim  (vs. Clausius )}
\section{\SEC}
\label{1871}\iniz
\lhead{\small\thesection: \SEC}
\0{Translation and comments on: 
{\it Zur priorit\"at der auffindung der beziehung zwischen dem zweiten
 hauptsatze der mechanischen w\"armetheo\-rie und dem prinzip
 der keinsten wirkung}, 1871, \citep[\#17,p.228-236]{Bo871-0}.}

\*
\0\alertb{\sl Incipit:}

Mr. Clausius presented, at the meeting of 7 Nov. 1870 of the
``Niederrheinischen Gesellschaft f\"ur Natur und Heilkunde vorgetragenen''
and in Pogg. Ann. {\bf 142}, S. 433, \citep{Cl872}, a work where it is
proved that the second fundamental theorem of the mechanical theory of heat
follows from the principle of least action and that the corresponding
arguments are identical to the ones implying the principle of least
action. I have already treated the same question in a publication of the
Wien Academy of Sciences of 8 Feb. 1866, printed in the volume 53 with the
title {\it On the mechanical meaning of the second fundamental theorem of
  the theory of heat}, [\citep[\#2]{Bo866}];%
\footnote{\alertb{\sl A clear exposition of this work can be found in Dugas'
    book, \citep[p.153-157]{Du959}; see also \citep{Ga013b} where the
    translation error on p.132, l.21 {\it ``...site of the region occupied
      by the body...''}  should be replaced by {\it ``..., whatever the
      state of the body...''}.}}
and I believe I can assert
that the fourth Section of my paper published four years earlier is, in
large part, identical to the quoted publication of
Mr. Clausius. Apparently, therefore, my work is entirely ignored, as well
as the relevant part of a previous work by Loschmidt. It is possible to
translate the notations of Mr. Clausius into mine, and via some very simple
transformation make the formulae identical.  I claim, to make a short
statement, that given the identity of the subject nothing else is possible
but some agreement.  To prove the claim I shall follow here, conveniently,
the fourth section of my work of 8 Feb. 1866, of which only the four
formulae Eq.(23a),(24a),(25a) and (25b), must be kept in mind.%
\footnote{\alertr{\sl The answer in \citep{Cl872}, commented in
    appendix\ref{C1872} below, was to
    apologize for having been unaware of Boltzmann's work and pointed out
    that Boltzmann's formulae became equal to his own after a suitable
    interpretation, requiring assumptions not needed in his work;
    furthermore his version was more general than his: certainly, for
    instance, his analysis takes into account possible variations of
    external forces.}}  \*

To compare this work with Mr. Clausius' we must first translate into each
other the notations. The signs $d$ and $\d$ are used by Mr. Clausius in my
same sense; except that for me the signs $\d$ indicate a general variation
while Mr. Clausius considers variations of a very special kind, by suitably
arranging the times of the varied state which will be compared with those
of the initial state. Once such special kind of variations is employed it
results that, for Mr. Clausius, the variation of the average of a quantity
equals the average of the variation. Likewise we indicate all the
quantities that refer to Thermodynamics $(Q,T,L,Z,S,...)$ in the same
sense. The time that an atom spends, on its closed trajectory, I denote
$t_2-t_1$, and Mr. Clausius denotes it by $i$, the average kinetic energy
of an atom I indicate by
$$\frac1{t_2-t_1}\int_{t_1}^{t_2} \frac{m c^2}2 dt,\quad{\rm
  Mr.\ Clausius\ by}\quad
\frac{m}2\lis{v^2}$$
To make notations of my formulae conform to those of Mr. Clausius
it can be set
$$ (A)\quad\left\{\eqalign{
i&\qquad\otto\qquad t_2-t_1,\cr
i\frac{m}2\lis{v^2}\ {\rm or} \ ih
&\qquad\otto\qquad \int_{t_1}^{t_2}
\frac{m c^2}2 dt,\cr
i\overline{X\d x+Y\d y+ Z \d z}&\qquad\otto\qquad 
\int_{t_1}^{t_2} (X\d x+Y\d y+ Z \d z)\cr}\right.$$
The first notable formulae, to which Mr. Clausius arrives, are his formulae
(17),(18) and (19), which are identical to the first of the equalities (22)
of my work: they are mapped into each other by the sustitutions (A);
in fact my formula (22) so becomes:
$$ i\frac{m}4\lis{\d v^2}=\frac{i\e}2 +
\frac{i}2\overline{X\d x+Y\d y+ Z \d z}.$$
Multiply by $\frac2i$ and consider that, by the method of variation of
Mr. Clausius, $\lis{\d v^2}=\d\lis{v^2}$, so that
$$ \frac{m}2\d\lis{v^2}=\e+\overline{X\d x+Y\d y+ Z \d z}
\eqno{(1)}$$ 
The quantity $\e$ coincides with the one in my formula (23a) which
by the translation table (A) becomes
$$\e=\frac{2\d(i\frac{m}2\lis{v^2})}{i}=
\frac{\d i}i m\lis{v^2}+m\d \lis{v^2}.$$
Substitute these values in equation (1) above and shift all on the left of
the equalities, so it immediately follows:
$$\overline{X\d x+Y\d y+ Z \d z}+\frac{m}2\d\lis{v^2}+\frac{\d i}i
m\lis{v^2} 
=0$$
which is identical to the equations (17),(18) and (19) of the work of
Mr. Clausius. By the successive introduction of the quantities that are
customary in the mechanical theory of heat, we isolate first what I call
the heat produced $\d Q$ and subsequently the work performed is subtracted,
while Mr. Clausius follows the inverse order.  However the total heat
amount transferred equals the sum of the increase of kinetic energy and of
the amount of work received, so that it is not difficult to follow the
context. Namely I define the quantity $\e$ as the amount of heat received
by the body (\ie my two statements ``let now an atom receive an infinitely
small amount of kinetic energy $\e$ and in this way the work done and
the increase of kinetic energy are accounted'' and ``therefore the sum of all
the $\e$ equals the whole quantities of heat received, measured in units of
work''). The quantity $\e$ is then, by my equation (22):
$$\e=\frac{1}{t_2-t_1}\int_{t_1}^{t_2} dt \d\frac{m
  v^2}2-\int_{t_1}^{t_2}\frac{1}{t_2-t_1}dt (X\d x+Y\d y+z\d z)$$
Thus I intend that the amount of heat received by the atom is the sum of
the average increment of its kinetic energy and the average of $-(X\d x+Y\d
y+z\d z)$. Hence it immediately follows that the work output corresponding
to the amount of heat $\d L$, average of $-(X\d x+Y\d y+z\d z)$, is
therefore in Mr. Clausius notations $\d \lis U$, and this is so also in
Mr. Clausius [see his equation (20)]; and again I say: ``if the balance
was not even, we could wait long enough, until thermal equilibrium or
stationary flow are reached, and then extract the average increase in
excess on what was denoted $\e$, per atom''. I also state, as ahead
Mr. Clausius, that the amount of heat is not realized in just one phase
\alertb{[phase in the sense of Clausius]} and define $\e$ as the total heat
intaken.  Also in the theoretical interpretation of the mechanical
propositions we do not differ, although Mr. Clausius puts his equation in a
somewhat different and more detailed way. I can also bring these to
comparison. I consider the quantity $\d L$ only for many material points,
and I develop it in formula (25a). To obtain the formula for one atom, we
take out the summation signs. Then we take again the change of notations
(A), so my formula (25a) is changed in
$$\d L=\frac{\d(ih)}i+h\frac{\d i}i$$
or in the construction of the variation:
$$\d L=\d h+2 h\d\log i$$
wherein immediately are recognized the equations (21),(22) and (23) of
Mr. Clausius. Hence we obtain the disgregation of a material point, as we
bring in my formula (25) the substitution (A) and drop the summation signs.
The mentioned formula becomes then
$$Z=\log(h i)+\log i=\log (h i^2)$$
which again immediately matches with that of Mr. Clausius for the
development of the equation (24) (24) for the value found for the
Disgregation.

I will now remark, that the fact that the forces $X,Y,Z$ in my calculation
likewise are seen subject to variations (which, according to Mr. Clausius'
expression, vary the ergal), indeed about this I said:

``At the same time suppose an infinitely small variation of the volume and
pressure of the body'', of which it was already discussed, as indeed the
forces acting on the body in general should change, because generally, as
long as these forces stay unchanged, there is only a single independent
variable, so that we cannot speak of complete or incomplete differential.%
\footnote{\alertr{\sl Here Boltzmann says that, if no external force is allowed
    to vary, the amount $\frac{d Q}T$ depends on only one parameter, say on
    the temperature, and is necessarily an exact
    differential.\phantomsection\label{2variables} Yet Clausius complaint
    that Boltzmann mentions variation of external forces ``en passant'' and
    then forgets about it is appropriate. The justification that it was
    enough to mention that external forces variations could be present is
    not accepted by Clausius, nevertheless it seems quite convincing after
    some thought: but Boltzmann had been too fast. Actually external forces
    are mentioned again in the 1866 paper towards the end of the paper in
    discussing the action principle, see below.}}
Hence it is to mention that I dedicate, again as Mr. Clausius, my
proposition to treat a single material point which, while it invariantly
describes a closed path, will undergo an infinitely small increase of
kinetic energy and at the same time the action of the acting forces changes.
Since a change of the atomic interactions in nature does not happen I
confine myself in performing the calculations to never give up their
constancy, and to add that the pressure can change.%
\footnote{\alertr{\sl In other words Boltzmann admits that he has not
    considered varying external forces, but claims 
    to have done so for simplicity.\phantomsection\label{simplicity}}}
Here Mr.Clausius obtains the advantage that there is the possibility of
having stressed explicitly such change.

Now Mr. Clausius comes to speak of the principle of least action. He says
setting up some remarks:%
\\
``This equation has the form that for a single moving point the theorem of
least action prescribes.  However in the meaning there is a difference, as
by the solution of our equations we have assumed that the initial and
varied motions develop in closed paths, that in no point are fixed
\alertb{[in particular do not have extermes fixed]} while by the least
action proposition it is supposed that both paths have a common starting
point and a common endpoint.  But this distinction is irrelevant, and the
solution of the equation (24) under both assumptions can be carried in
equal way, if the period $i$ which the moving point needs to come from the
initial location to the end location stays the same for the initial motion
and for the varied one''.
  
This analogy with the principle of least action naturally could not escape
me. My words are the following:\phantomsection{\label{actionpr}}

``It is easily seen that our conclusion on the meaning of the quantities
that intervene here is totally independent from the theory of heat, and
therefore the second fundamental theorem is related to a theorem of pure
mechanics to which it corresponds just as the kinetic energy principle
\alertb{[energy conservation]} corresponds to the first principle; and, as
it immediately follows from our considerations, it is related to the least
action principle, in a form somewhat generalized about as follows.  If a
system of point masses under the influence of forces, for which the kinetic
energy principle holds, performs some motion, and if then all points
undergo an infinitesimal variation of the kinetic energy and are
constrained to move on a path infinitely close to the precedent, then
$$\d\sum \fra{m}2\,\int c\, ds$$ 
equals the variation of the total kinetic energy multiplied by half the
time interval during which the motion develops, when the sum of the product
of the displacements of the points times their speeds and the cosine of the
angles on each of the path elements are equal, for instance the points of
the new path elements are on the normal of the old
paths.\phantomsection{\label{B-variation}} This proposition gives, for the
  kinetic energy transferred and if the variation of the limits of
  integration vanishes, the least action principle in the usual form.''%
\footnote{\alertr{\sl Nevertheless it seems improper to say that we are
    dealing with an extension of the least action principle: the latter
    determines the equations of motion while here a relation is established
    between two motions which are close and which satisfy the equations of
    motion.}}

It is seen that the type and kind that I call proposition, has nature
identical with that of Mr. Clausius, but my proposition is somewhat more
general.  It is in fact clear that if, as Mr. Clausius assumes, the old and
new paths are closed every one of my conditions holds.

Now, in the work of Mr. Clausius, the change from a single material point
to a system of points, follows. Also here the following difference from my
work does not touch the nature of the matter.  So I set the average kinetic
energy of an atom as directly equal to its temperature, Mr. Clausius
instead sets it equal to the temperature multiplied by a constant factor $m
c$, where the factor $m$ has the purpose of fixing a temperature unit and
the factor $c$ to conform to the observations (namely those of Kopps) of the
exceptions to the Dulong-Petit's law. Let us leave aside the question about
the cause of the exceptions: it is anyway clear that the introduction of
such factors would modify our analysis of the second main proposition in an
entirely irrelevant way. Only the transition from closed paths to not
closed paths is treated by Mr. Clausius in a way different from mine,
indeed the applicability for not closed paths of the equalities
$$\int_{t_1}^{t_2} \frac{mc^2}2\cdot dt=
\int_{\t_1}^{\t_2} \frac{mc^2}2\cdot dt$$
to the heat problem still deserves an exact proof. Similar to the
derivation of Mr. Clausius, there is a passage in a work presented on 25
February 1869 at the Wiener Akademie der Wissenshaften by Loschmidt, where
he discusses a second representation of my entire final method
(Sitzungberichte der Wiener Akad., Bd. 59).

Here I will add a list of my final formulae compared to those of
Mr. Clausius.  Mr. Clausius denotes the average kinetic energy of an atom
with $m c T$, while I denote it by
$$\frac1{t_2-t_1}\int_{t_1}^{t_2}\frac{m c^2}2 dt$$
Also, to make the notations match, if must be written in my formulae
$$(B)\qquad \left\{ 
\eqalign{i \quad&\tto\quad t_2-t_1\cr
m c T i  \quad&\tto\quad \int_{t_1}^{t_2}
\frac{m c^2}2 dt\r}\right.$$
therefore the first of the formulae (24) becomes:
$$\d Q=2\sum \frac{\d(m c T i)}i$$
or by multiplication by $T$ and division of the same quantity under the
signs of summation (as I use that $T$ has the same value for all points)
$$\d Q=2 T \d \sum m c \log Ti,$$
which is identical with formula (35) of Mr. Clausius.
My formula (24a) gives us the entropy. Denoting it by $S$ and using the
notations of Mr. Clausius, we obtain therefore

$$S=2\sum m c \log (T i)+C$$
which formula coincides with the equality (36) of Mr. Clausius. The
caloric equivalent of work is missing, as I measure heat in mechanical
units.  Using the change of notations (B) in my formula (25a), we obtain
$$\d L=\sum\frac{\d(m c T i)}i+\sum m c T
\frac{\d i}i= T\d \sum m c \log(T i^2),$$
as the formula (34) of Mr. Clausius, whose division by $T$ gives us
likewise the disgregation \alertb{[free energy]} in agreement with
Mr. Clausius.  I think to have here proven my priority on the discovery of
the mechanical meaning of the second law and I can finally express my
pleasure over that, if an authority of the level of Mr. Clausius
contributes to the diffusion of the knowledge of my works on the mechanical
theory of heat.

Graz, 16 May 1871.


\def\SEC{Clausius' reply (to Boltzmann's claim)}
\section{\SEC}
\label{C1872}\iniz
\lhead{\small\thesection: \SEC}

\0{Translation and comments on: R. Clausius
{\it Bemerkungen zu der priorit\"atreclama\-tion des Hrn. Boltzmann},
Pogg. Ann. {\bf 144}, 265--274, 1871.}
\*

In the sixth issue of this Ann., p.211, Mr. Boltzmann claims to have
already in his 1866 paper reduced the second main theorem of the mechanical
theory of heat to the general principles of mechanics, as I have discussed
in a short publication. This shows very correctly that I completely missed
to remark his paper, therefore I can now make clear that in 1866 I changed
twice home and way of life, therefore naturally my attention and my action,
totally involuntarily, have been slowed and made impossible for me to
follow regularly the literature.  I regret overlooking this all the more
because I have subsequently missed the central point of the relevant paper.

It is plain that, in all point in which his work overlaps mine, the
priority is implicit and it remains only to check the points that agree.

In this respect I immediately admit that his expressions of disgregation
[\alertb{\sl free energy}] and of entropy overlap with mine on two points,
about which we shall definitely account in the following; but his
mechanical equations, on which such expressions are derived are not
identical to mine, of which they rather are a special case.

We can preliminarily limit the discussion to the simplest form of the
equations, which govern the motion of a single point  moving periodically
on a closed path.

Let $m$ be the mass of the point and let $i$ its period, also let its
coordinates at time $t$ be $x,y,z$, and the acting force components be
$X,Y,Z$ and $v$ its velocity. The latter quantities as well as other
quantities derived from them, vary with the motion, and we want to denote
their average value by over-lining them.  Furthermore we think that near
the initially considered motion there is another one periodic and
infinitesimally different, which follows a different path under a
different force.  Then the difference between a quantity relative to the
first motion and the one relative to the varied motion will be called
``variation of the quantity'', and it will be denoted via the symbol
$\d$. And my equation is written as:

$$-\lis{X\,\d x+Y \d\, y+Z\, \d Z}=\fra{m}2\d\lis{v^2}+m\lis
{v^2}\d\log i\eqno{(I)}$$
or, if the force acting on the point admits an ergale [\alertb{\sl
    potential}], that we denote $U$, for the initial motion,%
\footnote{\alertr{\sl For Clausius' notation used here see Sec.\ref{sec:3}
    above, or \citep{Cl871}.  The (I) implies that in the following (Ia)
    there should be $\lis{\d U}$: but the accurate definition of variation
    by Clausius is such that it is $\lis{\d U}=\d {\lis U}$. An important
    technical point of Clausius's paper is that it establishes a notion of
    variation implying that the averages of the variations, in general of
    little interest because quite arbitrary (the arbitrariness being due,
    for instance, to possible differences $\d i$ in the periods which could
    lead to differences of order $\d i$ depending on the correspondence
    established between the times of the two motions), coincide with the
    variations of the averages.}}

$$\d\lis U=\fra{m}2\,\d\lis{v^2}+m\,\lis {v^2}\,\d\log i\eqno{(Ia)}$$
Boltzmann now asserts that these equations are identical to the equation
that in his work is Eq.(22), if elaborated with the help of the equation
denoted (23a). Still thinking to a point mass moving on a closed path and
supposing it modified into another for which the point has a kinetic
energy infinitely little different by the quantity $\e$, 
then Boltzmann's equation, after its translation into my notations, is

$$\fra{m}2\lis{\d v^2}=\e+\lis{X\,\d x+Y \d\, y+Z\, \d Z}\eqno(1)$$
and thanks to the mentioned equation becomes:
\footnote{\alertr{\sl Adding $\frac{m}2 \lis {v^2}$ to both sides.}}

$$\e=\fra{\d i}{i} m \lis {v^2}+ m\d\lis{v^2}\eqno(2)$$
The first of these Boltzmann's equations will be identical to my Eq.(I), if
the value assigned to $\e$ can be that of my equation.
\footnote{\alertr{\sl%
    A problem is that in Boltzmann $\e$ appears, at least at first,
    not clearly defined.}}

I cannot agree on this for two reasons.

The first is related to a fact that already Boltzmann casually mentions
but, as it seems to me, leaves it aside afterwards,
\citep[l.11,p.267]{Cl872}. In his equations both quantities $\lis{\d v^2}$
and $\d \lis{v^2}$ (\ie the average value of the variation $\d v^2$ and the
variation of the average value of $v^2$) are fundamentally different from
each other, and therefore it happens that his and my equations cannot be
confronted.
\\
I have dedicated, in my research, extreme care to avoid leaving
variations vaguely defined. And I use a special treatment of the variations
by means of the notion of {\it phase}.  This method has the consequence
that for every varied quantity the average of the variation is the
variation of the average, so that the equations are significantly simple
and useful. Therefore I believe that the introduction of such special
variations is essential for the subsequent researches, and do not concern a
point of minor importance.%
 \footnote{\alertr{\sl The definition of variation in Boltzmann is not
     really specified until, see p.\pageref{B-variation}, towards the end
     of the paper. Still Clausius will agree that the variation used by
     Boltzmann can be properly understood.}}

If now my variations are inserted in Boltzmann's Eq.(1) the following is
deduced:

$$\fra{m}2 \d \lis{v^2}=\e+ \lis{X\,\d x+Y \d\, y+Z\, \d Z}\eqno(1a)$$
and if next we suppose that the force acting on the point has an ergale
[\alertb{\sl potential}], which we denote $U$, the equation becomes
$\fra{m}2\d\lis {v^2}=\e -\d\lis U$, alternatively written as

$$\e=\fra{m}2\d \lis{v^2}+\d\lis{ U}.\eqno(1b)$$
If the value of  $\e$ is inserted in Eq.(2) my Eq.(I),(Ia) follow. 
In spite of the changes in Eq.(1a) and (1b) 
Boltzmann's equations so obtained are not identical to mine for a second
and very relevant reason.

{\it I.e.} it is easy to recognize that both Boltzmannian equations and
Eq.(1) and (2) hold under certain limiting conditions, which are not
necessary for the validity of mine. To make this really evident, we shall
instead present the Boltzmannian equations  as the most general equations,
not subject to any condition. However we shall suppose
more conveniently that they take the form taken when the force acting on
the point has an ergale [\alertb{\sl potential}].

Select, in some way, on the initial trajectory a point as initial point of
the motion, which starts at time $t_1$ as in Boltzmann, and denote the
corresponding values of $v$ and $U$ with $v_1$ and $U_1$. Then during the
entire motion the equation

$$\fra{m}2 v^2+U=\fra{m}2 v_1^2+U_1\eqno(3)$$
will hold; thus, likewise, we can set for the average values:

$$\fra{m}2 \lis {v^2}+\lis U=\fra{m}2 v_1^2+ U_1\eqno(4)$$
About the varied motion suppose that it starts from another point, with
another initial velocity and takes place under the action of other
forces. Hence we shall suppose that the latter have an ergale $U+\m V$,
where $V$ is some function of the coordinates and $\m$ an infinitesimal
constant factor \alertb{[so $W=0,\m V\equiv \d_eW$ in the notation of
    Sec.\ref{sec:3} above]}. Consider now again the two specified points, on
the initial trajectory and on the varied one, so instead of $v^2$ we shall
have in the varied motion the value $v^2+\d v^2$ and instead of $U$ the
value $U+\d U+\m(V+\d V)$; therefore, since $\m \,\d V$ is a second order
infinitesimal, this can be written $U+\d U+\m V$. Hence for the varied
motion Eq.(3) becomes:

$$\fra{m}2v^2+\fra{m}2\d v^2+U+\d U+\m V=
\fra{m}2v_1^2+\fra{m}2\d v_1^2+U_1+\d U_1+\m V_1\eqno(5)$$
so that my calculation of the variation leads to the equation:
$$\fra{m}2\lis {v^2}+\fra{m}2\d \lis {v^2}+\lis U+\d \lis U+\m \lis V=
\fra{m}2v_1^2+\fra{m}2\d v_1^2+U_1+\d U_1+\m V_1\eqno(5)$$
Combining the last equation with the Eq.(4) it finally follows

$$ \fra{m}2\d v_1^2+\d U_1+\m (V_1-\lis V)=\fra{m}2\d \lis {v^2}+\d \lis
U.\eqno(7)$$ 
This is the equation that in a more general treatment should be in place of
the different Boltzmannian Eq.(1b). Thus instead of the Boltzmannian Eq.(2)
the following is obtained:
\footnote{\alertr{\sl Noting that the varied motion will in general have no
    points in common with the initial motion this is not very clear and
    might be interpreted as follows.
%
    Computing (as done in Sec.\ref{sec:3}) the variation of the averages
    {\it to first order}, in the following Eq.(8) a first order correction
    appears instead of $\m (V_1-\lis V)$, due to the change of the external
    potential by $\m V$ (in Sec.\ref{sec:3} $\m V$ is denoted $\d_e W$
    and, here, it is supposed $W=0$, which is no loss of generality as noted
    in Sec.\ref{sec:3}), equal to $\m \lis V$ which is the external work
    performed, because of the external forces variation. The Eq.(8) is
    changed, see Sec.\ref{sec:3}, into $\d\lis {K+ U}+\m\lis V=\fra{\d
      i}i m \lis {v^2}+m \d{\lis v^2}$ which is what is needed, because the
    \lhs can be interpreted as the heat, $\d Q$, that is sent out of the
    system.  This might also clarify the final comment about the vanishing
    contribution of $\m(V_1-\lis V)$ which, also not very clear, shows that
    Clausius concludes that Boltzmann's final equations are correct even in
    presence of external forces; it nevertheless remains that Boltzmann has
    not considered the external work term: a neglection that he justifies,
    see previous appendix, convincingly as a matter of
    simplicity.\phantomsection\label{deltacomment}}}
$$\fra{m}2 \d v_1^2+\d U_1+\m(V_1-\lis V)=\fra{\d i}i m \lis {v^2}+m
\d{\lis v^2}.\eqno(8)$$
As we see, since such new equations are different from the Boltzmannian
ones, we must treat more closely the incorrect quantity $\e$. 
As indicated by the found infinitesimal variation of the kinetic energy due
to the variation of the motion, it is clear that in the variation $\e$ of
the kinetic energy at the initial time one must understand, and hence set:

$$\e=\fra{m}2\d v_1^2.$$
Hence of the three terms, that are to the left in Eq.(7) and (8), the
Boltzmannian equations should only contain the first.

Mr. Boltzmann, whose equations incompleteness I have, in my view, briefly
illustrated, pretends a wider meaning for $\e$ in his claim to contain at
the same time the kinetic energy of the motion and the work, and consequently
one could set

$$\e= \fra{m}2\d v^2_1+\d U_1.$$
But I cannot find that this is said anywhere, because in the mentioned
places where the work can be read it seems to me that there is a gain that
exchanges the kinetic energy with another property of the motion that can
transform it into work, which is not in any way understandable, and from
this it does not follow that the varied trajectory could be so transformed
that is has no point in common with the original while, also, in the
transformation the points moved from one trajectory to the other could be
moved without spending work.

Hence if one wishes to keep the pretension on the mentioned meaning of
$\e$, then always two of the three terms appearing in Eq.(7) and (8) are
obtained, {\it the third of them, \ie $\m(V_1-\lis V)$ no doubt is missing
  in his equations}.
\footnote{\alertr{\sl It seems, however, that the pretension can be reasonably
    assumed if the (admittedly somewhat obscure) comment in
    \citep[\#5,l.-11]{Bo866}, which was repeated in
    \citep[\#17,l.12]{Bo871-0} to prevent Clausius comment, is
    interpreted properly, while the next comment on $\m(\lis V -V_1)$ is
    about a ``gap'' already admitted by Boltzmann who essentially said that
    he did not discuss
    it for simplicity, see footnote at p.\pageref{simplicity} above.}}

On this point he writes: ``The term $\m(\lis V -V_1)$ is really missing in my
equations, because I have not explicitly mentioned the possibility of the
variation of the ergale. Certainly all my equations are so written that
they remain correct also in this case. The advantage, about the possibility
of considering some small variation of the ergale and therefore to have at
hand the second independent variable in the infinitesimal $\d U$ exists
and from now on it will not be neglected...''.
\citep[p.271]{Cl872}.

I must strongly disagree with the remark in the preceding reply, that all
his equations are written so that also in the case in which the ergale 
varies still remain valid. The above introduced Eq.(1) and (2), even if the
quantity  $\e$ that appears there receives the extended meaning $\fra{m}2\d
v_1^2+\d U_1$, are again false in the case in which by the variation of the
motion of a point the ergale so changes that the term $\m(\lis V_1-V)$.  has
an intrinsic value.\footnote{\alertr{\sl This is a typo for $\m(V_1-
    \lis V)$.}}

It cannot be said that my Eq.(I) is {\it implicitly}
contained in the Boltzmannian work, but the relevant equations
of his work represent, also for what concerns my method of realizing the
variations, only a special case of my equations.

Because I must remark that the development of the treatment of the case in
which the ergale so changes is not almost unessential, but for researches
of this type it is even necessary.

It is in fact possible to consider a body as an aggregate of very many
point masses that are under the influence of external and internal
forces. The internal forces have an ergale, depending only on the points
positions, but in general it stays unchanged in all states of the body; on
the contrary this does not hold for the external forces. If for instance
the body is subject to a normal pressure $p$ and later its volume $v$
changes by $dv$, then the external work $p \,dv$ will be performed. This
term, when $p$ is varied independently of $v$, is not an exact differential
and the work of the external force cannot, consequently, be representable
as the differential of an ergale. The behavior of this force can be so
represented: for each given state of the body in which its components are
in a state of stationary type it is possible to assign an ergale also to
the external forces which, however, does not stay unchanged, unlike that of
the internal forces, but it can undergo variations while the body evolves
into another state, independent of the change of position of the
points.

Keep now in mind the equations posed in the thermology of the changes of
state to build their mechanical treatment, which have to be reconsidered to
adapt them to the case in which the ergale changes.

I can say that I looked with particular care such generalizations. Hence it
would not be appropriate to treat fully the problem, but I obtained in my
mechanical equations the above mentioned term $\m(V_1-\lis V)$, for which
the corresponding term cannot be found in the mechanical equations. I must
now discuss the grounds for this difference and under which conditions such
term could vanish. I find that this will be obtained if the ergale
variation is not happening instantaneously at a given moment, but gradual
and uniform while an entire cycle takes place.%
\footnote{\alertr{\sl In other words these are the quasi static
    transformations, as considered by Boltzmann at the very beginning of
    his paper.}}
 And at the same time I claim that the same result is obtained if it is
 supposed that we do not deal with a single moving point but with {\it very
   large numbers of equal points}, all moving in the same way but with
 different phases, so that at every moment the phases are uniformly
 distributed and this suffices for each quantity to be evaluated at a point
 where it assumes a different value thus generating the average value. The
 latter case arises in the theory of heat, in which the motions, that we
 call heat, are of a type in which the quantities that are accessible to
 our senses are generated by many equal points in the same way. Hence the
 preceding difficulty is solved, but I want to stress that such solution
 appears well simpler when it is found than when it is
 searched.\footnote{\alertr{\sl See the footnote before Eq.(8) above.}}

The circumstance that for the motions that we call heat those terms
disappear from the averages had as a result that Boltzmann could obtain for
the digregation [\alertb{\sl free energy}] and the entropy, from his more
restricted analysis, results similar to those that I obtained with a more
general analysis; but it will be admitted that the real and complete
foundation of this solution can only come from the more general treatment.

The validity condition of the result, which remains hidden in the more
restricted analyses, will also be evident.

In every case Boltzmann restricts attention to motions that take place along
closed trajectories. Here we shall consider motions on non closed curves,
hence it now becomes necessary a special argument.%
\footnote{\alertr{\sl The case of motions taking place on non closed
    trajectories is, {\it however}, treated by Boltzmann.}}

Here too I took a different approach with respect to Boltzmann, and this is the
first of the two points mentioned above, in which Boltzmann's result on
disgregation and entropy differ. In his method taking into account of
time is of the type that I called {\it characteristic time of the period of
a motion}, essentially different. The second point of difference is found
in the way in which we defined temperature. The special role of these
differences should be followed here in detail, but I stop here hoping to
come back to it elsewhere.%
\footnote{{\small While this article was in print I found in a parallel
  research that the doubtful expression, to be correct in general, requires
  a change that would make it even more different from the Boltzmannian
  one.}}

Finally it will not be superfluous to remark that in another of my
published works the theorem whereby in every stationary motion {\it the
  average kinetic energy equals the virial} remains entirely outside of the
priority question treated here. This theorem, as far as I know, has not
been formulated by anyone before me.

\def\SEC{Collision analysis and equipartition}
\section{\SEC}
\label{1868}\iniz
\lhead{\small\thesection: \SEC}

\0{Quotes and comments on: {\it Studien {\"u}ber das
    Gleichgewicht der lebendigen Kraft zwischen bewegten materiellen
    Punkten}, 1868, \citep[\#5,p.49-96]{Bo868}.}  \*

All principles of analytic mechanics developed so far are limited to the
transformation of a system of point masses from a state to another,
according to the evolution laws of position and velocity when they are left
in a motion unperturbed for a long time and are concerned, with rare
exceptions, with theorems of the ideal, or almost ideal, gas.  This might
be the main reason why the theorems of the mechanical theory of heat which
deal with the motions so far considered are so uncorrelated and
defective. In the following I shall treat several similar examples and
finally I shall establish a general theorem on the probability that the
considered point masses occupy distinct locations and velocities.
\footnote{\alertr{\sl It will be what today is called microcanonical
    distribution.}}

\*
\0{\bf I. The case of an infinite number of point masses}
\*

Suppose we have an infinite number of elastic spheres of equal mass and
size and constrained to keep the center on a plane. A similar more general
problem has been solved by Maxwell (Phil.Mag., March 1868);%
\footnote{\alertr{\sl This could be \citep[1866]{Ma867-b}.}}
however partly because of the non complete exposition partly also because
the exposition of Maxwell in its broad lines is difficult to understand,
and because of a typo (in formulae (2) and (21) on the quantities called
$dV^2$ and $dV$) will make it even more difficult, I shall treat here the
problem again from the beginning.

It is by itself clear that in this case every point of the plane is a
possibly occupied location of the center of one of the elastic spheres and
every direction has equal probability, and it remains to determine the
speeds.  Let $\f(c)dc$ be the sum of the time intervals during which the
speed of one of the spheres happens to have a value between $c$ and $c+dc$
divided by a very large total time: which is also the probability that $c$
is between $c$ and $c+dc$ and let there be $N$ spheres per unit area, then
$$N\,\f(c)\,dc$$
denotes the number of spheres whose center is within the unit of surface
where the velocities are between $c$ and $c+dc$.\footnote{\alertr{\sl So the
  probability of speed in $dc$ is the fraction of time spentby the
  particles with speed in $dc$.}}

\ifnum\pdf=1
.\kern3cm\hbox{\includegraphics[width=150pt]{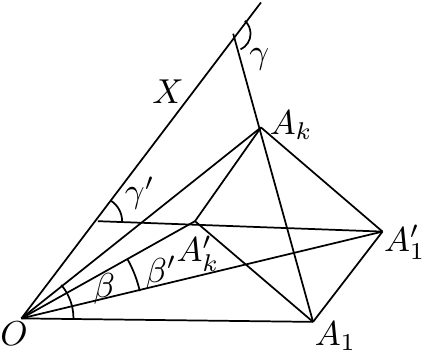}}%
\kern2cm\raise45pt\hbox{\small fig.2}
\fi
\ifnum\pdf=0
\eqfig{190}{109}
{
}
{Fig2}{Fig1}
\fi

\0Consider now a sphere, that I call $I$, with speed $c_1$ towards $OA_1$,
Fig.1, represented in size and direction, and let $OX$ the line joining the
centers at the impact moment and let $\b$ be the angle between the
velocities $c_1$ and $c_k$, so that the velocities components of the two
spheres orthogonal to $OX$ stay unchanged, while the ones parallel to $OX$
will simply be interchanged, as the masses are equal; hence let us
determine the velocities before the collision and let $A_1A'_1$ be parallel
to $OX$ and let us build the rectangle $A_1 A'_1 A_k A'_k$; $OA'_1$ and
$OA'_k$ be the new velocities and $\b'$ their new angle.  Consider now the
two diagonals of the rectangle $A_1A_k$ and $A'_1A'_k$ which give the
relative velocities of the two spheres, $g$ before and $g'$ after the
collision,  and call $\g$ the angle between the lines $A_1A_k$ and $OX$,
and $\g'$ that between the lines $A'_1A'_k$ and $OX$; so it is easily
found:
.......
\*
\0[\alertb{\sl A {\bf long} analysis follows about the relation between the
    area elements $d^2\V c_1 d^2\V c_k$ and the corresponding $d^2\V c'_1
    d^2\V c'_k$.  Collisions occur with the collision direction and the
    line connecting the centers forming an angle, between $\b$ and $\b+d\b$
    with probability proportional to $\s(\b)d\b$ where $\s(\b)$ is the
    cross section (equal to $\s(\b)=\fra12r \sin\b$ in the case, studied
    here, of disks of radius $r$).  Then the density $f(\V c) d^2\V c$ must
    have the property $\f(\V c_1) f(\V c_k)\s(\b)d\b= \f(\V c'_1) f(\V
    c'_k)$ $\s(\p-\b)d\b\,\cdot J$ where $J$ is the ratio $\fra{d^2\V c'_1
      d^2\V c'_k}{d^2\V c_1 d^2\V c_k}$, if the momentum and kinetic energy
    conservation relations hold: $\V c_1+\V c_k=\V c'_1+\V c'_k$ and $\V
    c_1^2+\V c_k^2=\V c^{'2}_1+\V c^{'2}_k$ and if the angle between $\V
    c_1-\V c_k$ and $\V c'_k-\V c'_1$ is $\b$.
    \\ The analysis leads to
    the conclusion, well known, that $J=1$; therefore it must be $f(\V c_1)
    f(\V c_k)= f(\V c'_1) f(\V c'_k)$ for all four velocities that satisfy
    the conservation laws of momentum and energy: this implies that $f(\V
    c)=const \,e^{-h c^2}$.  Boltzmann uses always the directional
    uniformity supposing $f(\V c)=\f(c)$ and therefore expresses the
    probability that the modulus of the velocity is between $c$ and $c+dc$
    as $\f(c) c dc$, hence (keeping in mind that the $2$-dimensional case
    is considered so that $\int_0^\infty f(c)c dc=1$) the result is
    expressed by $\f(c)=b\,e^{-h c^2}$, with $b=2h$ a normalization
    constant.
    \\ In reality if systematic account was taken of the volume
    preserving property of canonical transformations (\ie to have Jacobian
    $1$) the strictly algebraic part of evaluating the Jacobian, would save
    several pages of the paper. It is interesting that Boltzmann had to
    proceed to rediscover this very special case of a general property of
    Hamiltonian mechanics.
    \\ Having established this result for planar systems of elastic disks
    (the analysis has been {\it very} verbose and complicated and Boltzmann
    admits that ``Maxwell's argument was simpler'', but he says to have
    chosen on purpose a direct approach based on ``very simple examples'',
    \citep[\#5,p.58]{Bo868}), Boltzmann considers a $3$--dimensional
    case with an interaction which is more general than elastic collision
    between rigid spheres, and admits that it is described by a potential
    $\ch(r)$, {\it with very short range}. However he says that, since
    Maxwell has treated completely the problems analogous to the ones just
    treated in the planar case, he will study a new problem in subsection
    2. Namely the one dimensional problem of a ball $M$ moving on a line on
    which move also hard point particles $m$: $M$ is subject to a central
    force of potential $\chi(x)$ towards a fixed center $O$ and the
    particles $m$ can collide with $M$ but, a long time having elapsed,
    their system has acquired a stationary velocity distribution $\f(c)
    dc$. The aim is to determine the distribution of position $x$ and speed
    $C$ of the particle $M$ and to show that is generalizes the hard balls
    Maxwellian into $2B e^{-h(\frac{C^2}{2M}-\chi(x))} dC dx$ and obtain
    that the other balls have a Maxwellian velocity distribution, $\f(c)$
    proportional to $e^{-h\frac{c^2}{2m}}$, with $h$ the inverse
    temperature.}] \*

\centerline{\bf I.2. (p.61)}
\*
\0Along a line $OX$ an elastic ball of mass $M$ is moving attracted
  by $O$ by a force depending only on the distance. Against $O$ are moving
  other elastic balls of mass $m$ and their diverse velocities during
  disordered time intervals dart along the same line, so that if one
  considers that all flying balls have run towards $O$ long enough on the
  line $OX$ without interfering with each other, the number of balls with
  velocity between $c$ and $c+dc$, which in average are found in the unit
  length, is a given function of $c$, \ie $N \f(c) dc$.

The potential of the force, which attracts $M$ towards $O$ be $\chi(x)$,
hence as long as the motion [of $M$] does not encounter collisions it will
be

$$\frac{M C^2}2=\chi(x)+A\eqno{(9)}$$
where $C$ is the speed of the ball $M$ and $x$ the distance between its
center and $O$. Through the three quantities $x,A$ and $c$ the kind of
collision is fixed....
\*

\0[\alertb{\sl The discussion continues and the conclusion is that the joint
    distribution of $x,C,c$ for finding the particle $M$ within $dx$ with
    speed in $dC$ and a particle $m$ with speed in $dc$ is the canonical
    distribution $B e^{-h(\frac{M}2 C^2-\ch(x)) -h \frac{m}2c^2} dx dC dc$.
It has to be kept in mind that only events on the line $OX$ are considered
so that the problem is $1$--dimensional. The argument builds a kind of
Boltzmann's equation (like the analysis in subsection 1) very much
resembling Maxwell's treatment in \citep{Ma867-b} and is a
precursor of the later development of the Boltzmann's equation,
\citep[\#22]{Bo872}, and follows the same path. The analysis, here and in
the similar but increasingly complex problems in the rest of Section II,
will reveal itself very useful to Boltzmann in the other two papers of the
1871 ``trilogy'', and eventually in the 1872 construction of the
Boltzmann's equation, because it contains all technical details to be put
together to derive it.  \\
The $3$-dimensional corresponding problem is treated in Sec.I, subsection
3 (p.64).  \\
In Sec.I subsection 4 (p.70) all points are again on a line and the
interaction between the particle $M$ (now called $I$) with a fixed center
at $O$ is as in subsection 2, but the interaction between the $m$ point
particles (now called $II$) and $M$ is no longer hard core but is governed
by a potential with finite range $\ell$. It is supposed that the fraction of
time the point II has speed between $c$ and $c+dc$ (the problem is again
$1$-dimensional) is $N\f(c)dc$ and that events in which two or more
particles II come within $\ell$ of $I$ can be neglected.  The analysis leads
to the ``same'' results of subsections 2 and 3 respectively for the $1$ and
$3$ dimensional cases.
\\ In Sec.I subsection 5 (p.73) a similar problem is studied: the particle
$I$ and the particles $II$ move in space rather than on a line. It is more
involved as the angular momentum also enters into the equations. Again the
results are as expected.
\\ The problems of increasing complexity are shown to admit solutions that can
be viewed to fit in a general context that unifies them. The purpose is to
consider simple cases in which, however, not all interactions are hard core
interactions and softer potentials are treated.\\
The series of examples in Sec.$I$ prepares for the main part of the paper
where Boltzmann solves the general problem of the equilibrium distribution
of a system of interacting particles, discovering the microcanonical
distribution. The technical preparation is presented in Sec.II (see below)
where, still assuming that the interaction between the particles is very
short range, the Maxwell distribution for the velocities is derived for the
first time {\it via a combinatorial argument} in a discretized
representation of phase space, implicitly assuming that all microscopic
configurations are visited during an equal fraction of time. }] \*

\0{\bf II. On the equipartition of the kinetic energy for a finite number of
  point masses} (p.80) \* 
\0[\alertb{\sl In this section finitely many particles confined in a
    bounded container are studied in various (stationary) cases, evaluating
    the finite volume corrections and showing that equipartition of kinetic
    energies always holds exactly. The first case, in subsection II.6,
    deals with a planar case. Here the distribution of kinetic energy in
    shells $k_i,k_i+d k_i$ (with $k_i=i\frac1p$ integer multiple of a small
    quantity $\frac1p$) is determined to be uniform (under the constraint
    of fixed total kinetic energy, of the $n$ particles, at $n\k$),
    p.83. This is used to determine the fraction of time the energy of a
    particle is in the shell between $k$ and $k+dk$; determined by the
    combinatorial count of the number of ways the total kinetic energy can
    be split in the various shells and assuming such number proportional to
    the fraction of time spent in the shell; \ie the fraction of time in
    which the energy is in the shell $dk$ equals the average number of
    particles with energy in $dk$: a forerunner of the ergodic
    hypotesis. The arguments are as follows:}] \*

In a very large area, bounded in every direction, let there be $n$ point
masses, of masses $m_1,m_2,\ldots,m_n$ and velocities $c_1,c_2,\ldots,c_n$,
and between them act arbitrary forces, which {\it just begin to act at a
  distance which vanishes compared to their mean distance} [\alertb{\sl
    italics added}].%
\footnote{\alertr{\sl This assumption is meant to allow still neglecting the
  interaction time and consider collisions instantaneous so that the total
  kinetic energy is conserved.}}
Naturally all directions in the plane are equally probable for such
velocities. But the probability that the velocity of a point be within
assigned limits and likewise that the velocity of the others be within
other limits, will certainly not be the product of the single
probabilities; the others will mainly depend on the value chosen for the
velocity of the first point. The velocity of the last point depends from
that of the other $n-1$, because the entire system must have a constant
amount of kinetic energy.

I shall identify the fraction of time during which the velocities are so
partitioned that $c_2$ is between $c_2$ and $c_2+dc_2$,
likewise $c_3$ is between $c_3$ and $c_3+dc_3$ and so on until $c_n$, 
with the probability $\f_1(c_2,c_3,\ldots,c_n) dc_2\,dc_3\ldots
dc_n$ for this velocity configuration.

The probability that $c_1$ is between $c_1$ and $c_1+dc_1$ and the
corresponding velocities different from $c_2$ are between analogous limits
be $\f_2(c_1,c_3,\ldots,c_n)\cdot$ $dc_1\,dc_3\ldots dc_n$, {\it etc.}.

Furthermore let
$$\frac{m_1 c_1^2}2=k_1,\ \fra{m_2 c_2^2}2=k_2,\ \ldots\ \fra{m_n
c_n^2}2=k_n$$
be the kinetic energies and let the probability that $k_2$ is between $k_2$
and $k_2+dk_2$, $k_3$ is between $k_3$ and $k_3+dk_3\,\ldots$ until $k_n$
be $\ps_1(k_2,k_3,\ldots,k_n)\,dk_2$ $dk_3\ldots dk_n$. And analogously
define $\ps_2(k_1,k_3,\ldots,k_n)\,dk_1\,dk_3\ldots dk_n$
\etc., so that
$$\eqalign{
&m_2 c_2\cdot m_3 c_3\ldots m_nc_n \,
\psi_1(\fra{m_2 c_2^2}2,\fra{m_3 c_1^3}2,\ldots,
\frac{m_n c_n^2}2)=\f_1(c_2,c_3,\ldots,c_n)\quad{\rm or}\cr
&\f_1(c_2,c_3,\ldots,c_n)=2^{\frac{n-1}2}\,\sqrt{m_2m_3\ldots m_n}\,
\sqrt{k_2 k_3\ldots k_n}\,\ps_1(k_2,k_3,\ldots,k_n)\cr}$$
and similarly for the remaining $\f$ and $\ps$.

Consider a collision involving a pair of points, for instance $m_r$ and
$m_s$, which is such that $c_r$ is between $c_r$ and $c_r+d c_r$, and $c_s$
is between $c_s$ and $c_s+dc_s$. Let the limit values of these quantities
after the collision be between $c'_r$ and $c'_r+d c'_r$ and $c'_s$ be
between $c'_s$ and $c'_s+dc'_s$.

It is now clear that the balance of the kinetic energy will remain valid
always in the same way when many point, alternatively, come into collision
and are moved within different limits, as well as the other quantities
whose limits then can be remixed, among which there are the velocities of
the remaining points.

The number of points that are between assigned limits of the velocities,
which therefore have velocities between $c_2$ and $c_2+dc_2\ldots$, are
different from those of the preceding problems because instead of the
product $\f(c_r)dc_r\f(c_s)dc_s$ appears the function
$\f_1(c_2,c_3,\ldots,c_n) dc_2\,dc_3\ldots$ $dc_n$. This implies that
instead of the condition previously found
$$\fra{\f(c_r)\cdot\f(c_s)}{c_r\cdot
c_s}=\fra{\f(c'_r)\cdot\f(c'_s)}{c'_r\cdot c'_s}$$
the new condition is found:
$$\fra{\f_1(c_2,c_3,\ldots ,c_n)}{c_r\cdot
c_s}=\fra{\f_1(c_2,\ldots,c'_r,\ldots ,c'_s,\ldots, c_n)}{c'_r
\cdot c'_s}$$
The same holds, of course, for $\f_2,\f_3,\ldots$. If the function $\f$ is
replaced by $\ps$ it is found, equally,
$$\ps_1(k_2,k_3,\ldots,
k_n)=\ps_1(k_2,k_3,\ldots,k'_r,\ldots,k'_s,\ldots, k_n),\qquad {\rm
if}\ k_r+k_s=k'_r+k'_s.
$$
Subtract the differential of the first of the above relations 
$\fra{d\ps_1}{dk_r}dk_r  +\fra{d\ps_1}{dk_s} dk_s=
 \fra{d\ps_1}{dk'_r}dk'_r+\fra{d\ps_1}{dk'_s}dk'_s$
that of the second  [$dk_r+dk_s=dk'_r+dk'_s$] multiplied by $\l$
and set equal to zero the coefficient of each differential,
so that it is found:
$$\l=\fra{d \ps_1}{dk_r}=\fra{d \ps_1}{dk_s}=\fra{d \ps_1}{dk'_r}=
\fra{d \ps_1}{dk'_s}.$$
{\it I.e.}, in general,
$\fra{d\ps_1}{dk_2}=\fra{d\ps_1}{dk_3}=\fra{d\ps_1}{dk_4}=
\ldots \fra{d\ps_1}{dk_n}$, hence $\ps_1$ is function of
$k_2+\ldots+k_n$. Therefore we shall write $\ps_1(k_2,\ldots,k_n)$ 
in the form $\ps_1(k_2+k_3+\ldots+k_n)$.  We must now find the meaning of the
equilibrium between $m_1$ and the other points. And we shall determine the
full $\ps_1$.

It is obtained simply with the help of the preceding $\ps$ of which of
course the $\ps_1$ must be a sum. But these are all in reciprocal
relations. If in fact the total kinetic energy of the system is $n\k$, it is
$$k_1+k_2+\ldots+k_n=n\k$$
It follows that $\ps_1(k_2+k_3+\ldots+k_n) dk_2dk_3\ldots dk_n$
can be expressed in the new variables%
\footnote{\alertr{\sl In the formula $k_2$ and
$k_1$ are interchanged.}}
$$k_3,k_4,\ldots, n\k-k_1-k_3-\ldots-k_n=k_2$$
and so it must be for $\ps_2(k_1+k_3+\ldots+k_n) dk_1dk_3\ldots dk_n$. Hence
$\ps_1(k_2+k_3+\ldots+k_n)$ can be converted in $\ps_1(n\k-k_1)$ and
$dk_2dk_3\ldots dk_n$ in $dk_1dk_3\ldots dk_n$. Hence also
$$\ps_1(n\k-k_1)=\ps_2(k_1+k_3+\ldots+k_n)=\ps_2(n\k-k_2)$$
for all $k_1$ and $k_2$, therefore all the $\ps$ are equal to the same
constant $h$. This is also the probability that in equal time intervals it
is $k_1$ between $k_1$ and $k_1+dk_1$,  $k_2$ is between $k_2$ and $k_2+dk_2$
\etc, thus for a suitable $h$, it is $h\, dk_1\,dk_2\,\ldots\,dk_n$
were again the selected differential element must be absent. Of course this
probability that at a given instant $k_1+k_2+k_3+\ldots$ differs from $n\k$
is immediately zero.

The probability that $c_2$ is between $c_2$ and $c_2+dc_2$, $c_3$ between
$c_3$ and $c_3+dc_3\ldots$ is given by
$$\f_1(c_2,c_3,\ldots,c_n)\, dc_2\,dc_3\,\ldots\,dc_n=
m_2m_3\ldots m_n \cdot h \cdot c_2c_3\ldots c_n dc_2\,dc_3\,\ldots\,dc_n.$$
Therefore the point $c_2$ is in an annulus of area
$2\p c_2 dc_2$, the point $c_3$ in one of area $2\p c_3 dc_3$ \etc, 
that of  $c_1$ on the boundary of length $2\p c_1$ of a disk
and all points have equal probability of being in such annuli.

Thus we can say: the probability that the point $c_2$ is inside the area 
$d\s_2$, the point $c_3$ in $d\s_3$ \etc, while  $c_1$ is on a line element
$d\o_1$, is proportional to the product
$$\fra1{c_1} \,d\o_1\,d\s_2\,d\s_3\, \ldots\, d\s_n,$$
if the mentioned locations and velocities, while obeying the principle of
conservation of the kinetic energy, are not impossible.

We must now determine the fraction of time during which the kinetic energy of
a point is between given limits $k_1$ and $k_1+dk_1$, without considering
the kinetic energy of the other points. For this purpose subdivide the entire
kinetic energy in infinitely small equal parts $(p)$, so that if now we have
two point masses, for $n=2$ the probability that $k_1$ is in one of the $p$
intervals $[0,\fra{2\k}p]$, $[\fra{2\k}p,\fra{4\k}p]$,
$[\fra{4\k}p,\fra{6\k}p]$ \etc is equal and the problem is solved.

For $n=3$ if $k_1$ is in $[(p-1)\fra{3\k}p,p\fra{3\k}p]$, then
$k_2$ and $k_3$ must be in the interior of the  $p$ intervals. If $k_1$
is in the next to the last interval, \ie if
$$(p-2)\fra{3\k}p\le k_1\le (p-1)\fra{3\k}p$$
two cases are possible ....
\*

\0[\alertb{\sl Follows the combinatorial calculation of the number of
ways to obtain the sum of $n$ multiples $p_1,\ldots,p_n$ of a unit $\e$ and
$p_1\e=k_1$ such that $\sum_{i=2}^{n-1} p_i\e=n\k-p_1\e$, and Boltzmann chooses
$\e=\fra{\k}p$ with $p$ ``infinitely large'': \ie
$$\sum_{p_2=0}^{n\k/\e- p_1}\ \sum_{p_3=0}^{n\k/\e-p_1-p_2}\ldots\ldots
\sum_{p_{n-1}=0}^{n\k/\e-p_1-\ldots-p_{n-2}} 1$$
the result is obtained by explicitly treating the cases $n=2$ and $n=3$
and inferring the general result in the limit in which $\e\to0$.
\\
The ratio between this number and the same sum performed also on $p_1$
is, fixing $p_1\in [k_1/\e,(k_1+dk_1)/\e]$, 
$$\eqalign{&\fra{dk_1\int_0^{n\k-k_1}dk_2\int_0^{n\k-k_1-k_2} dk_3\,\ldots\,
\int_0^{n\k-k_1-k_2-\ldots-k_{n-2}} dk_{n-1}}
{\int_0^{n\k}dk_1\int_0^{n\k-1}dk_2\int_0^{n\k-k_1-k_2} dk_3\,\ldots\,
\int_0^{n\k-k_1-k_2-\ldots-k_{n-2}} dk_{n-1}}\cr
=&\fra{(n-1) (n\k-k_1)^{n-2} dk_1}{(n\k)^{n-1}},\cr}$$
This is, however, remarked {\it after} the explicit combinatorial analysis
of the cases $n=2$ and $n=3$ from which the last equality is inferred in
general (for $\e\to0$).
\\
Hence the ``remark'' is in reality a proof simpler than the combinatorial
analysis of the number of ways to decompose the total energy as sum of
infinitesimal energies. The choice of Boltzmann is a sign of his
preference for arguments based on a discrete view of physical quantities.
And as remarked in \citep{Ba990} this remains, whatever interpretation is
given to it, an important analysis.
\\
In the successive limit, $n\to\infty$, the Maxwell's distribution (in
dimension $2$) is obtained.
$$\fra1\k e^{-k_1/\k}dk_1$$
concluding the argument.
\*
In the next subsection II.7 (p.86) Boltzmann repeats the analysis in the
$3$-dimensional case obtaining again the Maxwellian distribution for the
speed of a single particle in a system of $n$ point masses in a finite
container with perfectly elastic walls, not necessarily flat. Particles
interact only if closer that their average distance and so close that the
total kinetic energy can be considered constant.
\\
Finally in Sec. II.8 the case is considered in which $N$ particles
interact as in Sec. II.7 and with $n$ other particles which do not
interact with each other but do interact with the first $N$; again the
interaction range is so short that collisions can be considered
instantaneous, as assumed at the beginning of Sec. II.
\\
Then the final (surprising and original) solution to the general problem is
in Sec.III (see below) where 'no assumption', see the comment by Maxwell at
p.\pageref{Max1879} below, is made on the density of particles and on the
interaction potential to compute the probability of microscopic
configurations. }]
\*
\0{\bf III. General solution of the kinetic energy equipartition problem}
\*
\0[\alertb{\sl The section III, p.92, is remarkable for the generality, for
    the derivation of the microcanonical distribution and 
    for the implicit emergence of the Ergodic Hypothesis, as commented
    in \citep{Ma879-c}:} \phantomsection{\label{Max1879}}
  \*

\0{\it ``The only assumption which is necessary for the direct proof is
  that the system, if left to itself in its actual state of motion, will,
  sooner or later, pass through every phase which is consistent with the
  equation of energy.  Now it is manifest that there are cases in which
  this does not take place.  The motion of a system not acted on by
  external forces satisfies six equations besides the equation of energy,
  so that the system cannot pass through those phases, which, though they
  satisfy the equation of energy, do not also satisfy these six equations.
  Again, there may be particular laws of force, as for instance that
  according to which the stress between two particles is proportional to
  the distance between them, for which the whole motion repeats itself
  after a finite time. In such cases a particular value of one variable
  corresponds to a particular value of each of the other variables, so that
  phases formed by sets of values of the variables which do not correspond
  cannot occur, though they may satisfy the seven general equations.  But
  if we suppose that the material particles, or some of them, occasionally
  encounter a fixed obstacle such as the sides of a vessel containing the
  particles, then, except for special forms of the surface of this
  obstacle, each encounter will introduce a disturbance into the motion of
  the system, so that it will pass from one undisturbed path into
  another. The two paths must both satisfy the equation of energy, and they
  must intersect each other in the phase for which the conditions of
  encounter with the fixed obstacle are satisfied, but they are not subject
  to the equations of momentum. It is difficult in a case of such extreme
  complexity to arrive at a thoroughly satisfactory conclusion, but we may
  with considerable confidence assert that except for particular forms of
  the surface of the fixed obstacle, the system will sooner or later, after
  a sufficient number of encounters, pass through every phase consistent
  with the equation of energy.}
\* \0\alertb{Remark: \phantomsection{\label{degeneracy}} Here the particles
  interact with a rather general force, no restriction to its
  range. Furthermore the microscopic configurations are counted via the
  occupation number of shells in kinetic energy and in cubic boxes in
  position space. However the shells in kinetic energy space are measured
  in terms of the speed and are appropriately given the weight $c^2dc$ or
  $\sqrt{k}dk$ which would amounts to imagine the shell covered by cubic
  boxes in the velocity space. Then the distribution of both velocity and
  positions is counted simply using the combinatorial study of the previous
  section: the result is a uniform distribution under the condition that
  the amount of kinetic energy of a microscopic configuration located in a
  cubic ($3n$-dimensional) box centered at a point $x\in R^{3n}$ is
  $n\k-\ch(x)$ (p.95,l.25): the result, p.95,l.31 and following, would be
  called today a microcanonical distribution, and Boltzmann integrates it
  over all momenta but one, to obtain the equipartition theorem. The
  following arguments are given:}] \*

Consider $n$ points-mass given in space with masses $m_1,\ldots,m_n$.  Let
their interaction be between pairs and arbitrarily depending on the
reciprocal distance. Also arbitrary forces with arbitrary fixed centers
(points with infinite mass) can act. The forces by which every point will
be affected are subject to the condition of depending on the position and
of admitting a potential. In all systems the total energy, still denoted
$n\k$ (as in the preceding sections was instead denoted the kinetic
energy), is assigned; we ask the probability that each point can be found
in a given place with its velocity having a given value. We denote the
positions of the point $m_1$ by three rectilinear coordinates
$x_1,y_1,z_1$, of the point $m_2$ by $x_2,y_2,z_2$, \etc, and the three
components of the velocity $c_1$ of the point $m_1$ with the three
coordinates $u_1,v_1,w_1$ and those of the point $m_2$ with $u_1,v_2,w_2$,
\etc. Denote the time spent with $x_1$ in the interval $x_1,x_1+dx_1$,
$y_1$ in $y_1,y_1+dy_1$ and $z_1$ in $z_1+dz_1$ and also $m_1$ in the
volume element $dx_1 dy_1 dz_1$, and likewise for the point $m_2$ ......and
the velocities ...  $u_1$ in $u_1+d_1$, $v_1$ in $v_1,v_1+dv_1$, $w_1$ in
$w_1,w_1+w_1$, .... $c_1$ in $ d\s_1=du_1dv_1dw_1$, .... , $c_{n-1}$ in
$d\s_{n-1}= d\s_{n-1}=du_{n-1}dv_{n-1}dw_{n-1}$, and finally the $c_n$ in
the surface element $d\o_n$.
\footnote{\alertr{\sl the last velocity is determined by its
  direction, because the total energy is fixed to $n\k$.}}
Divide the length of these  time intervals by the total time elapsed
obtaining:
$$f(x_1,y_1,z_1,\ldots,u_1,v_1,w_1,\ldots)
dx_1dy_1dz_1\cdots du_1dv_1dw_1\cdots du_ndv_nd\o_n $$
\0[\alertb{\sl The analysis continues and yields a long derivation of the
    formula that in modern notations would be that the probability of a
    phase space volume element $d^{3n}p d^{3n}q$ for a system with total
    energy $n\k$ is proportional to
$$\d(\frac1{2m} p^2+\ch(q)- n\k) d^{3n}p d^{3n}q
$$
and the number of particles with position in $d^{3n}q$, or $ds_1\ldots ds_n$
in Boltzmann's notation, is (integrating over the momenta) obtained from
the probability that the $n$ particles are in the volume elements
$ds_1,\ldots,ds_n)$. It is proportional to:
$$(n\k-\ch(q))^{\frac{3n-2}2} ds_1\ldots ds_n$$
\ie it is the microcanonical ensemble distribution of the velocities
(Maxwellian in the velocities for large $n\k$). The analysis of the earlier
sections was carried, taking $\chi\sim0$ to keep conserved the kinetic
energy (aside from a neglegible fraction of time) and under a form of the
ergodic hypothesis (saying that the system spends equal time in equal phase
space volumes): it implied equipartition of kinetic energy. The last
section analysis is a strong extension of all the previous sections results
and yields equipartition in presence of a rather arbitrary interaction
potential and even the microcanonical distribution. He also remarks that
positions and velocities that would seem permitted, because $n\k>\ch(q)$,
may not be such if the domain where $n\k>\ch(q)$ is not connected.  \\
As noted above (see the remark, p.\pageref{degeneracy}) on the implicit
split of the momentum shell $c,c+dc$ equal cells in number proportional to
$c^2 dc$) the discretisation used can be thought as done by dividing phase
space in small parallelepipedal cells (in the $6n$-dimensional phase
space): this is an essential point. Boltzmann does not comment on that: but
he wants to think, as he said, later but repeatedly, since he started
realizing that his ideas were misunderstood, that derivatives and integrals
are just ``approximations'' of certain ratios or of certain sums. See
more detailed comments at p.\pageref{regularcells} above.
The highlight of the work, and its conclusion, is:}]
\*

\0{\bf p.96.} As special case from the first theorem it follows, as already
remarked in my paper on the mechanical interpretation of the second
theorem, that the kinetic energy of an atom in a gas is equal to that of
the progressive motion of the molecule.%
\footnote{\alertr{\sl to which the atom belongs}}
The latter demonstration also contains the solution of others that were
left incomplete: it tells us that for such velocity distributions the
balance of the kinetic energies is realized in a way that could not take
place otherwise.

An exception to this arises when the variables $x_1,y_1,z_1,x_2,\ldots,v_n$
are not independent of each other.  This can be the case of all systems of
points in which the variables have special values, which are constrained by
assigned relations that remain unaltered in the motions, but which under
perturbations can be destroyed (labile equipartition of kinetic energy),
for instance when all points and the fixed centers are located on a
mathematically exact line or plane.  A stable balance is not possible in
this case, when the points of the system are so assigned that the variables
for all initial data within a certain period come back to the initial
values, without having consequently taken all values compatible with the
energy conservation.%
\footnote{\alertr{\sl This last paragraph refers to possible lack of
  equipartition in cases in which the system admits constants of motion due
  to symmetries that are not generic and therefore are destroyed by ``any''
  perturbation (like the harmonic chains).}
}

Therefore such way of achieving balance is always so infinitely more
possible that immediately the system ends up in the set of domains
discussed above when, for instance, a perturbation acts on a system of
points which evolves within a container with elastic walls or, also, on an
atom in free motion which collides against elastic walls.%
\footnote{ \alertb{[Editorial comment added to the Wissenshaftliche version
      of the paper]} This last paragraph is questionable. It is not proved
  that despite reflexions against elastic walls or despite the action of a
  single atom modifying the kinetic energy equations, or other integrals, the
  system may not go through all compatible values.}

\* \0[\alertr{\sl Boltzmann identifies the probability of visiting a cell
    in the full phase space with the fraction of time spent in it by the
    phase space point discretized on a regular lattice. Being interested in
    the equipartition integrates over the momenta: not using the delta
    function, as it would be done today, his argument becomes somewhat
    involved.  \\
In \citep{Ba990} the question is raised on whether Boltzmann would have
discovered the Bose-Einstein distribution before Planck referring to the
way he employs the discrete approach to compute the number of ways to
distribute kinetic energy among the various particles, after fixing the
value of that of one particle, in \citep[\#5,p.84,85]{Bo868}. This is
an interesting and argumented view, however Boltzmann considered here the
discrete view a ``fiction'', see also \citep[\#42,p.167]{Bo877b}, and
the way the computation is done would not distinguish whether particles
were considered distinguishable or not: the limiting case of interest would
be the same in both cases (while it would be quite different if the
continuum limit was not taken, leading to a Bose-Einstein like
distribution, see \citep{Ga000}). This may have prevented him to be led to
the extreme consequence of considering the difference physically
significant and to compare the predictions with those that follow in the
continuum limit with the distribution found with distinguishable particles,
discussed later in \citep[\#42]{Bo877b}, see also
\citep[Sec.(2.2),(2.6)]{Ga000}.}]

\def\SEC{Density in phase space: example}
\section{\SEC}
\label{1868b}\iniz
\lhead{\small\thesection: \SEC}

\0{Comments on: {\it L\"osung eines mechanisches
    Problems}, 1868, \citep[\#6,p.97-105]{Bo868b}.}
\*
  
\0[\alertb{\sl An example about the general theory of Sec.III of
    \citep[\#5]{Bo868} is attempted here.  The aim of the
    example it to exhibit a simple case in which the difficult problem of
    finding the region of phase space visited by a trajectory is the set of
    all configurations with a given energy so that, on the basis of the
    previous work, the probability of finding a point mass occupying a
    given position with given velocity can be solved and shown to be given
    by equidistribution (\ie microcanonical).}]  \\
    \alertb{\sl Boltzmann considers the simple system consisting in a point
      mass [\alertb{\sl mass $=1$}] subject to a central gravitational
      force with potential $-\frac\a{2R}$ and to a centrifugal barrier
      augmented by a potential $+\fra\b{2R^2}$: furthermore the point mass
      is reflected by an obstacle consisting in a straight line, \eg
      $y=\g>0$.}
    ]
\\ 
\ifnum\pdf=1
.\kern3.5cm\hbox{\includegraphics[width=100pt]{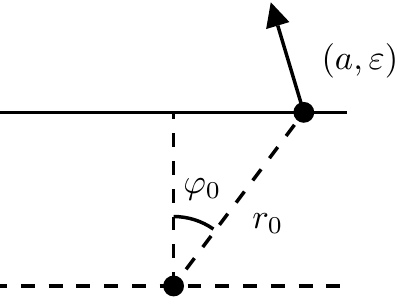}}\kern4cm
\raise45pt\hbox{\small fig.3}
\fi
\ifnum\pdf=0
\eqfig{105}{85}
{
\ins{77}{13}{$\f_0$}
\ins{63}{37}{$r_0$}
\ins{93}{70}{$(a,\e)$}
}
{Fig3}{fig3}
\fi
\*
\0 \alertb{\sl The discussion is an interesting example of a problem in
  ergodic theory for a two variables map, ``{\it not really easy to find}''
  as a soluble case among the great variety of cases. Angular momentum $-a$
  is conserved between collisions and the motion is explicitly reducible to
  an elementary quadrature which  (here $r\equiv R$ and
  $A$ is a constant of motion equal to twice the total energy, constant
  because collisions with the line are supposed elastic)
  yields a function:
$$F(r,a,A)\defi \frac{a}{\sqrt{a^2+\b}}{\rm
  arccos}(\frac{{2(a^2+\b)}/r-\a}
{\sqrt{\a^2+4 A(a^2+\b)}})$$ 
such that the polar angle at time $t$ is $\f(t)-\f(0)=
F(r(t),a_0,A)-F(r(0),a_0,A)$.
\\Let $\e_0\defi \f(0)-F(r(0),a_0,A)$, then if $\f_0,a_0$ are the initial
polar angle and the angular momentum of a motion that comes out of a
collision at time $0$ then $r(0)\cos\f_0=\g$ and $\f(t)-\e_0=F(r(t),a_0,A)$
until the next collision. Which will take place when
$\f_1-\e_0=F(\frac\g{\cos\f_1},a_0,A)$ and if $a_1$ is the outgoing angular
momentum from then on $\f(t)-\e_1=F(r(t),a_1,A)$ with $\e_1\defi
\f_1-F(\frac\g{\cos\f_1},a_1,A)$.  \\
    Everything being explicit Boltzmann computes the Jacobian of the just defined map
    $S: (a_0,\e_0)\to (a_1,\e_1)$ and shows that it is $1$ (which is carefully
    checked without reference to the canonicity of the map). The map is
    supposed to exist, \ie that the Poincar\'e's section defined by the timing
    event ``hit of the fixed line'' is transverse to the solution flow (which
    may be not true: if $A\ge0$ or,  even if $A<0$, unless $\g$ is small
    enough).
\\ Hence the observations timed at the collisions, \ie the evolution of the
values $a,\e$ at the successive collisions, admit an invariant measure $d\e
d a$. If the allowed values of $a,\e$ vary in a bounded set (which
certainly happens if $A<0$) the measure $\frac{d\e da}{\int d\e da}$ is an
invariant probability measure, \ie the microcanonical distribution, which
can be used to compute averages and frequency of visits to the points of
the plane $\e,a$.
\\ The interest of Boltzmann in the example was to show that, unless the
interaction was very special (\eg $\b=0$), the motion would invade the
whole energy surface so that the stationary probability distribution would
be positive on the entire phase space, in essential agreement with the idea
of ergodicity (of the earlier work, see Appendix \ref{1868}).
\\ However the stationary distribution that is studied is just one
invariant distribution: Boltzmann did not yet suspect that there might be
other invariant distributions. The doubt will arise in his mind soon: as he
will point out that other invariant distributions might exist (even if
described by a density in phase space), \citep[\#18,p.255]{Bo871-a} and
Appendix\ref{1871-a} below, and he  will be troubled by the problem.
\\
Boltzmann proves that the map $(a,\e)\to (a',\e')$, of the data of a
collision to those of the next, has Jacobian $1$.%
\footnote{\alertr{\sl The
  analysis is long and detailed: in modern language it could be carried
  by remarking that the action angle coordinates for the unconstrained
  (integrable) motion are $(E,t),(a,\e)$ so that the motion leaves
  invariant the volume $d Ed t da d\e$ and this implies that $dad\e$ is
  invariant under the map.}} %
From this he infers that the frequency of visit to a rectangle $da\,d\e$ is
a constant function of $(a,\e)$ and actually he claims that a stationary
distribution of the successive values of $(a,\e)$ is, therefore, uniform in
the set in which $a,\e$ vary. And he then uses the distribution to derive its
form in other variables and to evaluate various averages.
A related, very simplified, version is in Sec.\ref{sec:4} above. See
also \citep[\#39]{Bo877a}.  \\
The conclusion, \citep[\#6,p.102]{Bo868}, is however somewhat fast: because
he is assuming that the invariant distribution has a density in $a,\e$ and
that it is the only distribution with a density (and furthermore, strictly
speaking, that the distribution is ergodic in the modern sense of the
notion).  See Fig.4. for a simple {\it heuristic} test of density in the
above case and in a second simple case (Boltzmann concludes the paper by
briefly discussing even more general cases).}
\footnote{\alertr{\sl Actually the problem of the ergodicity of the map
    $(a,e)\to (a',e')$ in systems integrable by quadratures (hence quasi
    periodic) in presence of reflections like the ones considered by
    Boltzmann might even be, as far as I know, an open problem, even in the
    case $\b=0$, \ie in the Kepler problem case or in the similar harmonic
    potential case.  And it might be that the confined motions of this
    system (integrable in absence of the obstacle) are even periodic for
    open regions of initial data at least if, in absence of obstacles, the
    system has only periodic motions: the stress on $\b\ne0$ suggests that
    Boltzmann might have thought so.}}

\def\CAP4{\alertb{\small Fig.4: Evolution of $(x,a)$, with $x$ abscissa of
    the collision point, different from $(e,a)$ chosen by Boltzmann in his
    example but equivalent for the purpose of the density problem and,
    perhaps, simpler. Two orbits (drawn by points, although they look
    continuous) are enclosed in each domain of given energy (delimited by
    the external dotted loop). Uniqueness is not violated by the
    intersecting orbits as, at constant energy, the map to $(x,a)$
    variables is not $1\otto1$.}}
\ifnum\pdf=1
{
\includegraphics[width=150pt,angle=-90,scale=.656]{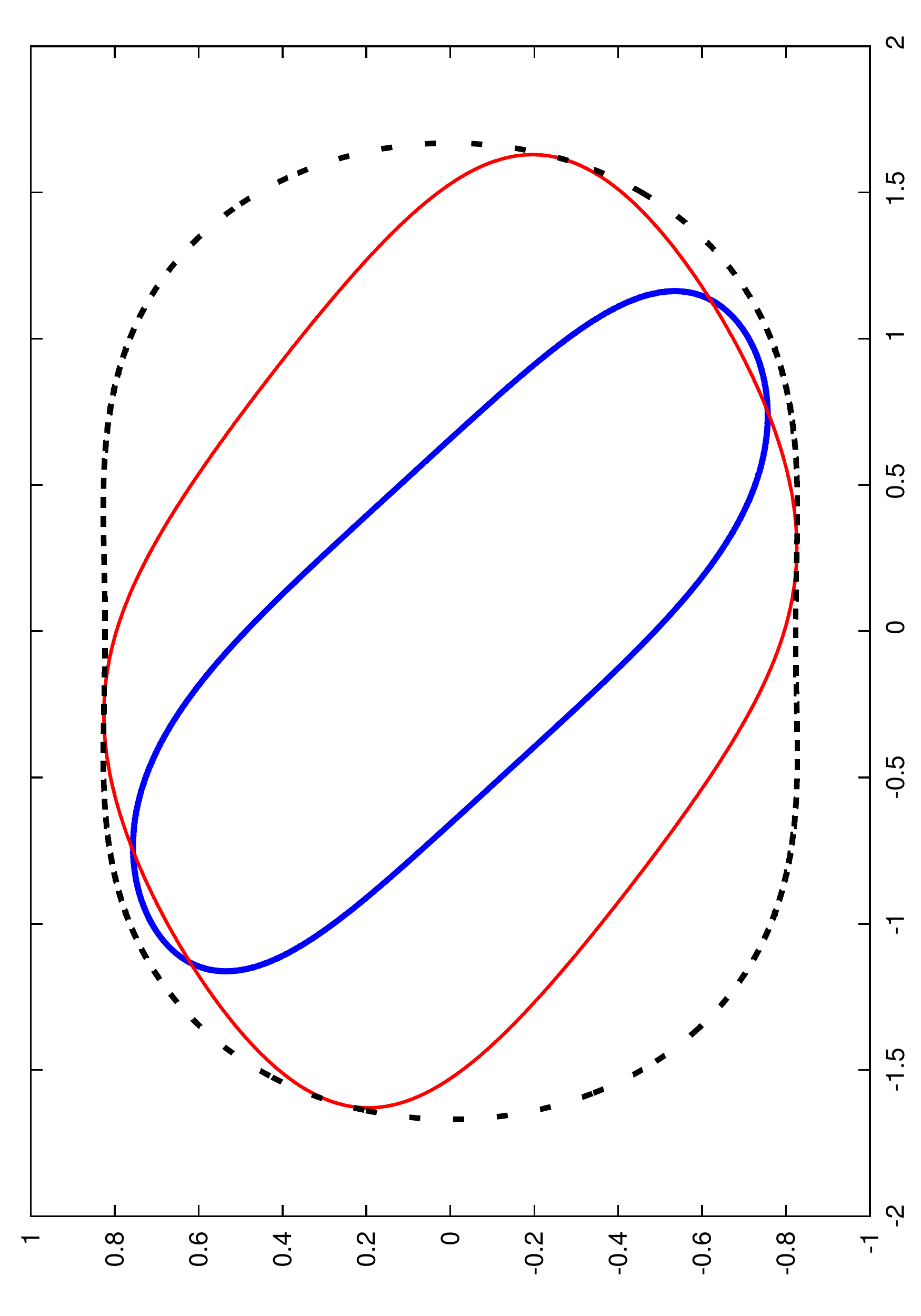}%
\includegraphics[width=150pt,angle=-90,scale=.656]{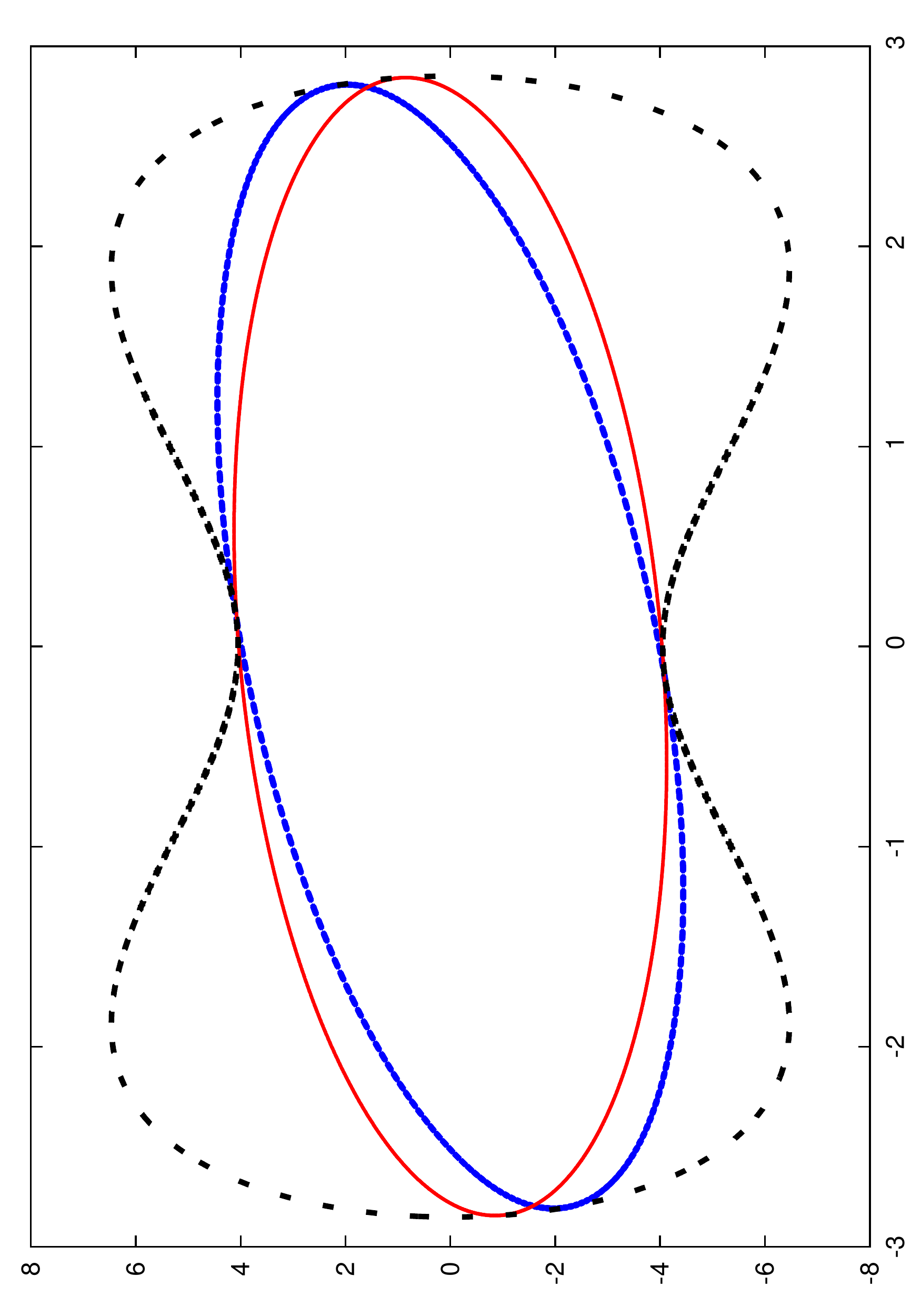}%
\kern.4cm\lower50pt\hbox{fig.4}
}

\0\CAP4
\fi
\ifnum\pdf=0
\eqfig{75}{200}{}{Fa}{}

\kern-5mm
\0\CAP4
\fi

\*\*

\0\alertb{\sl In each of Fig.4 two initial data with same energy move
  subject to a central force sum of a Newtonian attraction (left figure) or
  of a harmonic potential (right figure) and, furthermore, are elastically
  reflected upon collision with a straight line: the simulation appears to
  indicate (after $6.10^3$ iterations) that the motion is not dense on the
  energy surface.  \\
  Simulations for the interaction considered by Boltzmann as well as the
  cases of purely harmonic or purely Newtonian potential (still with the
  straight line obstacle), lead to similar results at least for wide energy
  ranges. Density of the trajectories on energy surfaces can be expected in
  the above cases when the obstacle is ``rougher'' than a straight line,
  possibly already with an obstacle of equation $1+\e\sin x$ with $\e$
  large. For small $\e$ there could be chaotic islands trapped
  between quasi periodic orbits and for large $\e$ motion might be fully
  chaotic.}
\footnote{\alertb{\sl Data for figure on the right are:
    $H=\frac12(p_x^2+p_y^2)+\frac{b}2(x^2+y^2) -\frac{g}{\sqrt(x^2+y^2)}$,
    simulated with error of $5$-th order, at step $h=10^{-3}$, with
    $b=2.,g=0.1$, $x=2.8,y=1$ and $u=.3, v=.7$ or  $u\otto v$,
    $H=E=9.096366$. In figure on the left
    $H=\frac12(p_x^2+p_y^2)+\frac{b}{2(x^2+y^2)}
    -\frac{g}{\sqrt(x^2+y^2)}$, with $b=.4,g=1.$,$x=1.0824, u=.15189,
    v=.477131, y=1.$ or $u\otto v$, $H=E=-.5302125$.  It suggests
    that, in the considered models, the energy surface is not in general
    covered densely by a single trajectory.}}%
\\ \alertb{{\it Heuristic analysis} \sl of the simulations results: from
  the first of the above figures it appears that there are data which seem
  to run on a quasi periodic orbit that remains well within the energy
  surface: and (via simulations) this remains true for data in a close
  enough region. Analysis of various cases seems to suggest that a purely
  gravitational potential ($\b=0$) or a purely harmonic potential remain
  integrable and quasi periodic even in spite of collisions with the
  barrier. It has been pointed out\footnote{I am indebted to I. Jauslin for
    informing me of the results of simulations that he performs to
    test existence of chaotic motions for large $\b$, hence non-integrability
    in general for $\b\ne0$.}  that if $\b>0$ is
  large the motions are chaotic on the energy surface.

  Then for $\b$ small, if the potential is augmented by a centrifugal
  barrier ($\b>0$) and, at least within a region in which motions do not
  touch the boundary of the constant energy domain, the map
  $(x,a)\to(x',a')$ remains smooth and generates quasi periodic motions on
  smooth curves: this would be consistent because introduction of the extra
  centrifugal barrier could be considered a perturbation and the KAM
  theorem, \cite{Mo962}, might be applicable and guarantee that large
  portions of the energy surface are covered by quasi periodic motions,
  developing on smooth curves with possible chaotic motions between them.
  Deformation of the obstacle into $y=1+\e \sin(x)$ is also a perturbation
  to which KAM theorem should be applicable (if all motions are periodic in
  absence of any added centrifugal barrier ($\b=0$)). Hence, for $\e$
  small, we could expect {\it coexistence of chaotic motions trapped
    between near pairs of motions on closed orbits run quasi periodically}
  (KAM) and delocalization or, at least, large chaotic regions can be
  expected at large $\b$ (and observed) or large $\e$; could full
  delocalization (\ie density of motion of almost all initial data) in the
  energy surface be expected?

  It should also be remarked that the cases in which the obstacle line
  passes through the attraction point (\ie $\g=0$) are maifestly integrable,
  even in presence of the centrifugal force: and a relation with KAM theory
  could also be attempted by considering $\g$ as a perturbation parameter.
  }

\def\SEC{Equilibrium inside molecules (Trilogy: \#1)}
\section{\SEC}
\label{1871-a}\iniz
\lhead{\small\thesection: \SEC}

\0{Quotes and comments on: 
{\it {\"U}ber das {W\"a}rme\-gleichge\-wicht zwischen mehratomigen
{G}asmolek{\"u}len}, 1871, \citep[\#18,p.237-258]{Bo871-a}.}

\* \0[\alertb{\sl This work complements the last section of
    \citep[\#5]{Bo868} in the sense that it can be seen as a different
    viewpoint towards an application of the general result contained in it.
    It is remarkable particularly for its Sec.II where Maxwell's
    distribution is derived as a consequence of the of the assumption that,
    {\it because of the collisions with other molecules}, the atoms of the
    molecule visit all points of the phase space of the molecule.  It is
    concluded that the distribution of the atoms inside a molecule is, in
    the center of mass frame, a function of the center of mass average
    kinetic energy, \ie of the temperature $\frac1h$, and of the total
    energy in the center of mass, sum of  kinetic energy $K$, in the
    center of mass, and of the internal potential energy
    $\ch$, proportionally to $e^{-h (K+\ch)}$.\\
    So the distribution of the coordinates of the body will depend on the
    total average energy (just kinetic as the distance between the
    particles is assumed to be very large compared with the interaction
    range). The form of the distribution is not obtained by supposing the
    particles energies discretized regularly, as done in
    \citep[\#5]{Bo868}, but exploiting invariance properties. The
    question of uniqueness of the canonical distribution is explicitly
    raised, \citep[\#18,p.255]{Bo871-a}:
 the determined distribution (\ie the canonical distribution) might be not
 the unique invariant distribution: this shows that a {\it doubt now
   entered Boltzmann's mind} which had not arisen at the time of the works
 \citep[\#5]{Bo868} and \citep[\#6]{Bo868b} (see above), where an
 invariant distribution was found but its possible uniqueness was not
 brought up, see Appendix \ref{1868}.
The
 invariance arguments used leave open the doubt about uniqueness. The
 relation of the ``Trilogy'' papers with Einstein's statistical mechanics
 is discussed in \citep{Re997}.}]

\*
According to the mechanical theory of heat every molecule of gas while in
motion does not experience, by far for most of the time, any collision; and
its baricenter proceeds with uniform rectilinear motion through space.
When two molecules get very close, they interact via certain forces, so
that the motion of each feels the influence of the other.

The different molecules of the gas take over all possible configurations %
\footnote{\alertr{\sl In this paper Boltzmann imagines that a molecule of gas, in due
    time, goes through all possible states: this is not yet the ergodic
    hypothesis because this is attributed to the occasional interaction of
    the molecule with the others.  The hypothesis is used to extend the
    hypothesis formulated by Maxwell for the monoatomic systems to the case
    of polyatomic molecules. For these he finds the role of the internal
    potential energy of the molecule, which must appear together with the
    kinetic energy of its atoms in the stationary distribution. The results
    agree with the earlier work \citep[\#5]{Bo868} where, without even
    assuming very low density, the microcanonical distribution for the
    entire system was found by supposing that the equilibrium state
    distribution had a unique positive density on the available phase space
    and was determined by a combinatorial count of the possible
    configurations. Here no combinatorial count is done and only uniqueness
    of a distribution with positive density is used together with
    negligibility of multiple collisions: the latter assumption however, as
    it is stressed, casts doubts on the uniqueness of the distribution.}}
 and it is clear that it is of the utmost importance to know the
 probability of the different states of motion.

We want to compute the average kinetic energy, the average potential
energy, the mean free path of a molecule \&tc. and, furthermore, also the
probability of each of their values. Since the latter value is not known we
can then at most conjecture the most probable value of each quantity, as we
cannot even think of the exact value.

If every molecule is a point mass, Maxwell provides the value of the
probability of the different states (Phil. Mag.,
March\footnote{\alertr{\sl Maybe February?}}  1868), \citep{Ma867-b}. In
this case the state of a molecule is entirely determined as soon as the
size and direction of its velocity are known.  And certainly every
direction in space of the velocity is equally probable, so that it only
remains to determine the probability of the different components of the
velocity.

If we denote $N$ the number of molecules per unit volume, Maxwell finds
that the number of molecules per unit volume and speed between $c$ and
$c+dc$, equals, \citep[Eq.(26),p.187]{Ma867-b}:

$$4\sqrt{\fra{h^3}\p} N e^{-h c^2}c^2\, dc,$$
where $h$ is a constant depending on the temperature.
We want to make use of this expression: through it the velocity
distribution is defined, \ie it is given how many molecules have a speed
between $0$ and $dc$, how many between $dc$ and $2dc$, $2dc$ and $3dc$,
$3dc$ and $4dc$, etc. up to infinity.

Natural molecules, however, are by no means point masses. We shall get
closer to reality if we shall think of them as systems of more point masses
(the so called atoms), kept together by some force. Hence the state of a
molecule at a given instant can no longer be described by a single variable
but it will require several variables. To define the state of a molecule at
a given instant, think of having fixed in space, once and for all, three
orthogonal axes. Trace then through the point occupied by the baricenter
three orthogonal axes parallel to the three fixed directions and denote the
coordinates of the point masses of our molecule, on every axis and at time
$t$, with
$\x_1,\h_1,\z_1,\x_2,\h_2,\z_2,\ldots,\x_{r-1},\h_{r-1},\z_{r-1}$.  The
number of point masses of the molecule, that we shall always call atoms, be
$r$. The coordinate of the $r$-th atom be determined besides those of the
first $r-1$ atoms from the coordinates of the baricenter.  Furthermore let
$c_1$ be the velocity of the atom 1, $u_1,v_1,w_1$ be its components along
the axes; the same quantities be defined for the atom 2, $c_2,u_2,v_2,w_2$;
for the atom 3 let them be $c_3,u_3,v_3,w_3$ \&tc. Then the state of our
molecule at time $t$ is given when the values of the $6r-3$ quantities
$\x_1,\h_1,\z_1,\x_2,\ldots,\z_{r-1},u_1,v_1,w_1,u_2,\ldots,w_r$ are known
at this time. The coordinates of the baricenter of our molecule with respect
to the fixed axes do not determine its state but only its position.

We shall say right away, briefly, that a molecule is at a given place when
its baricenter is there, and we suppose that in the whole gas there is an
average number $N$ of molecules per unit volume. Of such $N$ molecules at a
given instant $t$ a much smaller number $dN$ will be so distributed that,
at the same time, the coordinates of the atom 1 are between $\x_1$ and
$\x_1+d\x_1$, $\h_1$ and $\h_1+d\h_1$, $\z_1$ and $\z_1+d\z_1$, those of
the atom 2 are between $\x_2$ and $\x_2+d\x_2$, $\h_2$ and $\h_2+d\h_2$,
$\z_2$ and $\z_2+d\z_2$, and those of the $r-1$-{th} between $\x_{r-1}$ and
$\x_{r-1}+d\x_{r-1}$, $\h_{r-1}$ and $\h_{r-1}+d\h_{r-1}$, $\z_{r-1}$ and
$\z_{r-1}+d\z_{r-1}$, while the velocity components of the atom 1 are
between $u_1$ and $u_1+du_1$, $v_1$ and $v_1+dv_1$, $w_1$ and $w_1+dw_1$,
those of the atom 2 are between $u_2$ and $u_2+du_2$, $v_2$ and $v_2+dv_2$,
$w_2$ and $w_2+dw_2$, and those of the $r-1$-th are between $u_{r-1}$ and
$u_{r-1}+du_{r-1}$, $v_{r-1}$ and $v_{r-1}+dv_{r-1}$, $w_{r-1}$ and
$w_{r-1}+dw_{r-1}$.

I shall briefly say that the so specified molecules  are in the domain
(A). Then it immediately follows%
\footnote{\alertr{\sl Here it is assumed the the equilibrium
    distribution has a density on phase space.
}}
that
$$dN=f(\x_1,\h_1,\z_1,\ldots,\z_{r-1},u_1,v_1,\ldots,
w_r)d\x_1d\h_1d\z_1\ldots d\z_{r-1}du_1dv_1\ldots dw_r.$$
I shall say that the function $f$ determines a distribution of the states
of motion of the molecules at time $t$. The probability of the different
states of the molecules would be known if we knew which values has this
function for each considered gas when it is left unperturbed for a long
enough time, at constant density and temperature. For monoatomic molecules 
gases Maxwell finds that the function $f$ has the value
$$4\sqrt{\fra{h^3}\p} N e^{-h c^2}c^2\, dc.$$
The determination of this function for polyatomic molecules gases seems
very difficult, because already for a three atoms complex it is not
possible to integrate the equations of motion. Nevertheless we shall see
that just from the equations of motion, without their integration, a value
for the function $f$ is found which, in spite of the motion of the
molecule, will not change in the course of a long time and therefore,
represents, at least, a possible distribution of the states of the
molecules.%
\footnote{\alertr{\sl  Remark the care with which the possibility is not
    excluded of the existence invariant distributions different from the
    one that will be determined here.}}

That the value pertaining to the function $f$ could be determined without
solving the equations of motion is not so surprising as at first sight
seems. Because the great regularity shown by the thermal phenomena induces
to suppose that $f$ be almost general and that it should be independent
from the properties of the special nature of every gas; and also that the
general properties depend only weakly from the general form of the
equations of motion, except when their complete integration presents
difficulties not unsurmountable.%
\footnote{\alertr{\sl  Boltzmann is aware that special behavior could
    show up in integrable cases: he was very likely aware of the theory of
    Lagrange's solution of the harmonic chain, \citep[Vol.I]{La867}.}}

Suppose that at the initial instant the state of motion of the molecules is
entirely arbitrary, \ie think that the function $f$ has a given
value.\footnote{\alertr{\sl  This is the function called ``empirical
    distribution'', \citep{GL003,GGL005}.}}  As time elapses the state of
each molecule, because of the motion of its atoms while it follows its
rectilinear motion and also because of its collisions with other molecules,
becomes steady; hence the form of the function $f$ will in general change,
until it assumes a value that in spite of the motion of the atoms and of
the collisions between the molecules will no longer change.

When this will have happened we shall say that the states of the molecules
are distributed in {\it thermal equilibrium}. 
From this immediately the problem is posed to find for the function $f$ a
value that will not any more change no matter which collisions take
place. For this purpose we shall suppose, to treat the most general case,
that we deal with a mixture of gases. Let one of the kinds of gas (the kind
G) have $N$ molecules per unit volume. Suppose that at a given instant $t$
there are $dN$ molecules whose state is in the domain (a). Then as before
$$dN=f(\x_1,\h_1,\z_1,\ldots,\z_{r-1},u_1,v_1,\ldots,
w_r)d\x_1d\h_1d\z_1\ldots d\z_{r-1}du_1dv_1\ldots dw_r.\eqno(1)$$
The function $f$ gives us the complete distribution
of the states of the molecules of the gas of kind G at the instant $t$. 
Imagine that a certain time $\d t$ elapses. At time $t+\d t$ the
distribution of the states will in general have become another, hence the
function $f$  becomes different, which I denote $f_1$, so that at time
$t+\d t$  the number of molecules per unit volume whose state in the domain
(A) equals:

$$f_1(\x_1,\h_1,\ldots ,w_r)\, d\x_1\,d
h_1\,\ldots\, dw_r.\eqno(2)$$
\*
\centerline{{\bf \S{}I.} Motion of the atoms of a molecule}
\*

\0[(p.241):\alertb{\sl Follows the analysis of the form of $f$ if the
    collisions can be regarded as instantaneous compared to the average
    free flight time: Liouville's theorem, for pair collisions, is derived
    from scratch and it is shown that if $f$ is invariant then it has to be
    a function of the coordinates of the molecules through the integrals of
    motion. This is a wide extension of the argument for
    monoatomic gases in \citep{Ma867-b}.}]

\*
\centerline{{\bf \S{}II.} Collisions between molecules}
\*

\0[(p.245):\alertb{\sl It is shown that to have a stationary distribution
    also in presence of binary collisions it suffices (p.254,l.6)
    that the function $f$ has the form $A e^{-h \f}$ where $\f$ is the
    total energy, sum of the kinetic energy and of the potential energy of
    the atoms of the molecule, in the center of mass. Furthermore if the
    gas consists of two species then $h$ must be the same constant for the
    distribution of either kinds of molecules and it is identified with the
    inverse temperature.\\
    Since a gas, monoatomic or not, can be considered as a giant molecule
    it is seen that this is also a derivation of the microcanonical
    distribution if no other constants of motion besides the total energy
    (implicit in the last section of the \citep[\#5]{Bo868} work, see
    comments at the end of Appendix\ref{1868}) is assumed for the entire
    gas.\\
    The kinetic energies equipartition and the ratios of the specific heats
    is deduced. It becomes necessary to check that this distribution ``of
    thermal equilibrium'' generates average values for observables
    compatible with the heat theorem: but this will be done in the
    successive paper of the trilogy, \citep[\#20]{Bo871-c}. There it
    will also be checked that the hypothesis that each
    group of atoms that is part of a molecule passes through all states
    compatible with the value of the energy (possibly with the help of the
    collisions with other molecules) leads to the same result if the number
    of molecules is infinite (with finite density) or very large (in a
    container).\\
    It is instructive that Maxwell comments (somewhat unfairly?) the
    assumptions of short duration of the collisions and the lack of
    multiple collisions as ``\alertn{\rm It is true that in following the
      steps of the investigation, as given either by Boltzmann or by
      Watson, it is difficult, if not impossible, to see where the
      stipulation about the shortness and the isolation of the encounters
      is made use of. We may almost say that it is introduced rather for
      the sake of enabling the reader to form a more definite mental image
      of the material system than as a condition of the demonstration.}'',
    \citep[p.714,l.2]{Ma879-c}.\\
    The uniqueness of the equilibrium distribution of the atoms in a
    molecule rests on the nonexistence, for isolated molecules, of other
    constants of motion besides the energy and the question is left open as
    explicitly stated at p.255, see below: the problem arises because the
    argument is based on the conservation law in pair collisions and here
    Boltzmann acknowledges that there might be problems in presence of
    multiple collisions (actually there is a problem even in absence of
    other constants of motion for isolated molecules); nevertheless he
    points out that the distribution found is a possible stationary
    distribution:}] \*

\0{\bf p.255, \rm(l.21)} Against me is the fact that, until now, 
the proof that these distributions are the only ones that do not change in
presence of collisions is not complete. Nevertheless remains the fact that
[the distribution shows] that the same gas with equal temperature and
density can be in many states, depending on the given initial conditions,
{\it a priori} improbable and which will even never be observed in
experiments.

\* \0[\alertb{\sl The uniqueness problem suggests the idea, see the next
    Appendix\ref{1871-b}, of considering $n$ copies of the entire system
    (of molecules or atoms in a gas) not interacting at all and apply again
    the argument of this paper (that the distribution of the states has to
    be a (positive, being a frequency of visit) function in the region
    compatible with the assigned values of the constants of motion): in
    absence of other constants of motion, beyond the total energy, it will
    follow that the equilibrium distribution must be the microcanonical one
    which will also imply a canonical distribution for the single
    molecules: this will be the subject of the immediately following work,
    \citep[\#19]{Bo871-b}, which can be considered the first in which an
    ``ensemble'' in the modern sense appears, \citep{Gi902,In015}.  It is
    worth reporting a comment in \citep[p.715]{Ma879-c}, on this
    matter: ``\alertn{But whether we are able or not to prove that the constancy
    of this function is a necessary condition of a steady distribution, it
    is manifest that if the function is initially constant for all phases
    consistent with the equation of energy, it will remain so during the
    motion. This therefore is one solution, if not the only solution, of
    the problem of a steady distribution}''. \\
    This paper\label{Bo-Ma} is also important as it shows that
    Boltzmann was well aware of the paper \citep{Ma867-b}: in which 
    key steps towards the Boltzmann's equation is discussed in great
    detail. One can say that Maxwell's analysis yields a form of ``weak
    Boltzmann's equation'', namely several equations which can be seen as
    equivalent to the time evolution of averages of one particle observable
    with what we call now the one particle distribution of the
    particles. Boltzmann will realize that the one
    particle distribution itself obeys an equation (Boltzmann's equation)
    and will obtain in this way a major conceptual simplification of Maxwell's
    approach and derive the $H$-theorem, \citep[\#22]{Bo872}.}]

\def\SEC{Microcanonical ensemble and ergodic hypothesis (Trilogy: \#2)}
\section{\SEC}
\label{1871-b}\iniz
\lhead{\small\thesection: \SEC}

\0{Quotes and comments on: 
{\it Einige allgemeine S{\"a}tze {\"u}ber {W\"a}rme\-gleichgewicht},
(1871), \citep[\#19]{Bo871-b}.}
\*

\0{\bf \S{}I.} {\it Correspondence between the theorems on the polyatomic
molecules behavior and Jacobi's principle of the last multiplier.}
\footnote{\alertr{\sl  This title is quoted by Gibbs in the introduction
    to his {\it Elementary principles in statistical mechanics},
    \citep{Gi902}, thus generating some confusion because this title is not
    found} \alertr{ in the list of publications by Boltzmann.}}  \*

The first theorem that I found in my preceding paper {\it\"Uber das
  {W\"a}rme\-gleich\-gewicht zwischen mehratomigen {G}asmolek{\"u}len},
1871, \citep[\#18]{Bo871-a} (see also Appendix\ref{1871-a} above),%
\footnote{\alertr{\sl It might be useful to summarize the logic behind this
    reference.\\ In the quoted paper, \citep[\#5]{Bo868}, a gas of $n$
    molecules was supposed to be at so low density that the duration of the
    collisions could be negelected: so kinetic energy would be conserved
    and molecules, as systems of atoms, would evolve independently for all
    but a neglegible fraction of time. If the position of the baricenter of
    each molecule was not considered, the state of the system could be seen
    as a collection of molecules of $r$ atoms each, evolving in time
    independently.\\
    Each molecule would be in a state represented by a point in a $6r-3$
    dimensional phase space and a state of the gas would be a point in a
    $(6r-3)n$ dimensional phase space. At collisions, assuming that
    multiple collisions could be neglected, the states of the colliding
    pairs would vary: but, in a stationary state of the gas, the changes
    due to collisions would not cause a change in the average number of
    molecules with a given state.\\
The latter has to be identified with the fraction of time each molecule
spends in the state. Hence one has only to look at the whole region in the
$(6r-3)n$-dimensional phase space, in which can be found the point
representing the state, in order to find the (hopefully unique) probability
density invariant under the collisions dynamics (between pairs of
molecules).\\
    A first problem then is to find the phase space region 
      that will be visited by the system: it will be determined ``under an
      hypothesis'', namely that this will be the $(6r-3)n-1$ dimensional
      region in which the total kinetic energy (denoted $n\k$) has a given
      value, and a second problem is to find the density (which Boltzmann
      implicitly assumes to be smooth) representing the count of molecules
      near a phase space point. The second problem was solved easily if
      multiple collisions and pairs collisions duration were
      negligible: there is only one possibility (as it is reluctantly
      admitted, see \citep[\#18,p.251]{Bo871-a} and comments to the
      Appendix \ref{1871-a} above)%
 for the density, namely it has to be a function of the total kinetic
 energy, as he shows (with an argument going back to Maxwell).
%
%
 It follows that the density is constant on the surface and all it remained
 to do was to evaluate the probability of events of interest.}}
 is strictly related to a theorem, that at first sight has nothing to do
with the theory of gases, \ie with Jacobi's principle of the last
multiplier.

To expose the relation, we shall leave aside the special form that the
mentioned equations of the theory of heat have, whose relevant developments
will be generalized here later.

Consider a large number of systems of point masses (as in a gas containing a
large number of molecules of which each is a system of point masses). The
state of a given system of such points at a given time is assigned by $n$
variables $s_1,s_2,\ldots,s_n$ for which we can pose the following
differential equations:

$$\fra{d s_1}{dt}=S_1,\fra{d s_2}{dt}=S_2,\ldots,\fra{d s_n}{dt}=S_n.$$
Let $S_1,S_2,\ldots,S_n$ be functions of $s_1,s_2,\ldots,s_n$ and
possibly of time. Through these equations and the initial value of the $n$
variables $s_1,s_2,\ldots,s_n$ are obtained the values of such quantities at
any given time. To arrive to the principle of the last multipliers, we can
use many of the conclusions reached in the already quoted paper; hence we
must suppose that between the point masses of the different systems of
points never any interaction occurs. As in the theory of gases the
collisions between molecules are neglected, also in the present research
the interactions will be excluded.
\*
\0[\alertb{\sl In the present paper the purpose it to replace the
    assumption on the absence of multiple collisions with an hypothesis
    that, in an equilibrium state, all configurations compatible with the
    given value of the total energy must be visited by the phase space
    point representing the state of the system. The  new hypothesis is
    the ergodic hypothesis.\\
 As an application the microcanonical distribution for the total rarefied
 gas and, as its consequence, the canonical distribution of the states of the
 single molecules are derived.}

 \alertb{\sl This paper is not only important because it goes beyond the low
   density assumption in the general case and the no multiple collisions
   assumptions in the case of rarefied gas of molecules. Boltzmann imagines
   the general system as a giant molecule and, following the idea applied
   to the molecules of a gas, the statistics is given by considering a
   large number of copies of the system without reciprocal interaction (as
   the molecules in a rarefied gas of \citep[\#18]{Bo871-a}) and supposing
   that the entire phase space is the set of given total energy $n\k$
   (kinetic)}
 \alertb{\sl (plus potential) is visited by their independent motions. This
   is the hypothesis that no other constants of motion beyond the
   conservation laws exist%
\footnote{\alertr{\sl It has to be kept in mind that Boltzmann usually only
    considers smooth functions ({\it e.g.}, if functions of time, twice
    differentiable, see his axiom 2 of Mechanics,
    \citep[p.10]{Bo897}).}}
    and that the phase point representing the state visits all
    neighborhoods of the phase space. At the same time it is the first
    appearance, as noted in \citep{Gi902}, of what is often called an
    ``ensemble'': for a thorough discussion of the notion and use of the
    ensembles see the recent analysis in \citep{In015}.\\
Conceiving the state
    of a system as described by the density of frequency of visit to
    regions of phase space or by the number of identical non interacting
    copies of the system that fall in the regions is equivalent if the
    density of frequency or if, respectively, the number of copies falling
    in the regions, are invariant under the time evolution: the first view
    is mostly used by Boltzmann, except in this work, and the second is
    preferred by Maxwell in his last work on statistical mechanics,
    \citep{Ma879-c}.}

\alertb{\sl The case in which there is only one constant of motion
  was already solved in the earlier work. Furthermore in this paper the
  case of several constants of motion is also considered. Actually most of
  the paper is dedicated to the computation of the Jacobian determinant
  necessary to compute the surface element of the surface defined by the
  values of the constants of motion. The calculation is based on the
  Lecture 12, about the ``principle of the last multiplier'', in
  Jacobi's Lectures on Mechanics, \citep{Ja884}, see also
  \citep[p.277]{Wh917},}%
\footnote{\alertr{\sl Lagrange's multipliers method of enforcing holonomic
    constraints consists in adding extra forces suitably determined
    to impose the constraints. In Jacobi's lectures, assuming that in a
    $n$-dimensional ODE $n-1$ constants of motion are known, a last
    one, the last constant of motion, can be found (and its construction is
    reducible to a quadrature of an exact differential) and can be
    associated with the ``missing'' or ``last'' Lagrange multiplier. The
    method involves a long analysis of various Jacobian determinants which
    are used by Boltzmann to evaluate the Jacobian determinants needed in
    his work in cases in which the number of constants of motion is $1$
    (total energy) or at most $7$ (energy, momentum and angular momentum).}}
 \alertb{\sl which Boltzmann rederives here, presenting it from scratch (in 24
   pages!).
Extensions of Liouville's theorem are
    discussed and used to express the form of the distribution at fixed
    values of of $n-k$ constants of motion; concluding that the
    distribution is deduced by dividing by the appropriate Jacobian
    determinant of the transformation expressing $k$ free coordinates in
    terms of the $n-k$ constants of motion: the last multiplier of Jacobi
    is just the Jacobian determinant of the change of coordinates. } \\

\* \0{\bf \S{}II.} {\it Thermal equilibrium for a finite number of point
  masses.}  \*

\0[(p.269){\alertb{\sl In this section the method of \S I is applied to
      compute the average kinetic energy and the average potential energy
      in a system with $n$ point particles and $n-k$ constants of motion
      $\f_n,\ldots,\f_{k+1}$. The arguments of Sec. I,II in modern language
      would be the following.  It is assumed, with no discussion, that the
      motion visits the entire region where $\f_j$ have given values and
      which is connected to the initial values of interest (usually this
      region is connected), (p.271,l.15).
    It is also assumed that there are no further constants of motion
    (keeping in mind that Boltzmann never really considers non smooth
    functions, except occasionally when dealing with hard core
    collisions). Then, if a distribution defined on the whole phase space
    is stationary, it must have the form $f(\f_n,\f_{n-1},\ldots,\f_{k+1})
    ds_1 ds_2\ldots ds_n$: but we are interested on the distribution with
    given values $a_n,\ldots,a_{k+1}$ of the constants of motion. Hence we
    would say that it has the form
$$f(\f_n,\f_{n-1},\ldots,\f_{k+1})\,(\prod_{j=k+1}^n\d(\f_j-a_k))
    ds_1 ds_2\ldots ds_n$$
    Therefore the probability distribution density for $s_1,\ldots,s_k$ is
$$ \int f(\f_n,\f_{n-1},\ldots,\f_{k+1})\,(\prod_{j=k+1}^n\d(\f_j-a_k))
    ds_{k+1} ds_{k+2}\ldots ds_n$$
where the integral is over the variables
$s_{n},\ldots,s_{k+1}$; hence developing the delta functions it is
$$  f(\f_n,\f_{n-1},\ldots,\f_{k+1})\,
\frac1
    {\frac{\dpr(\f_n,\ldots,\f_{k+1})}{\dpr(s_n,\ldots,s_{k+1})}}
    \Big|_{\f_n=a_n,\ldots \f_{k+1}=a_{k+1}}$$
This formula is derived, in 24 pages, via use of the principle of the last
multiplier and illustrated by a few examples on quasi periodic motions in
simple $2$ dimensional force fields, in which there are up to two constants
of motion (energy and, possibly, angular momentum).\\
It is then applied, at the end of Sec.II, to the derivation of the
microcanonical ensemble in the general case with $k=n-1$ (\ie the energy
being the only conserved quantity), already obtained in general by a
combinatorial argument in \citep[\#5]{Bo868} and via the conservation laws
in binary collisions and assuming instantaneous duration of collisions, see
Appendix\ref{1871-a}, in \citep[\#18]{Bo871-a}.\\
Here the two assumptions are replaced by positivity of the distribution
density on the surface defined by all the constants of motion: therefore it
is an assumption implicit in \citep[\#5]{Bo868}, as recognized in
\citep{Ma879-c} and quoted above, but the method is very different from both
the earlier derivations.  \\
Assuming non existence of other constants of motion is a form of the
ergodic hypothesis if non smooth constants of motion are not allowed; a
natural assumption, see \citep[p.10,V.1]{Bo897}, whose importance has not
been always realized: \eg it led Fermi to think to have proved the generic
ergodicity in mechanical systems, \citep{Fe923,Fe923a} (later he brilliantly
corrected himself in \citep[Vol.2,p.977]{Fe965}).}}]

\*
\0{\bf \S{}III.({\rm p.284})} {\it Solution for the thermal equilibrium for
the molecules of a gas with a finite number of point masses under an
hypothesis.}
\*
Finally from the equations derived we can, under an assumption which it
does not seem to me of unlikely application to a warm body, directly access
to the thermal equilibrium of a polyatomic molecule, and more generally of
a given molecule interacting with a mass of gas. The great chaoticity of
the thermal motion and the variability of the force that the body feels
from the outside makes it probable that the atoms get in the motion, that
we call heat, all possible positions and velocities compatible with the
equation of the kinetic energy, and even that the atoms of a warm body
can take all positions and velocities compatible with the last equation
considered.%
\* \0[\alertb{\sl The hypothesis, used already in the context of general
    evolution equations at p.271,l.15, and in similar form earlier (see
    above), that is formulated here explicitly for systems of atoms is that
    the atoms of a ``warm body'' containing a molecule take all positions
    and velocities. This is the first time the ergodic hypothesis is
    formally stated almost in the form in which it is still intended today
    (the exceptions occurring on a set of zero volume in phase space are
    not mentioned: not because they are overlooked but, as discussed in
    Sec.\ref{sec:7},\ref{sec:8},\ref{sec:9} above, because of the discrete
    view of phase space.): \\
The logic of the paper seems to be the followimg. To compute the time
averages consider the accessible phase space points, \ie compatible with
the constraints. Consider all points {\it as independent systems} placed in
phase space with a density which has to be invariant under time evolution
because the points will move and others occupy their places.\\
  Boltzmann considers only distributions with a density and appears to
  think that the density has to be positive (being proportional to a
  frequency of visit). Of course if the system admits constants of motion
  $\f_n,\ldots,\f_{k+1}$ the phase space is reduced to the surface, that I
  shall call $\Si$, defined by the constants of motion: therefore the
  density, which would be constant on the energy surface if the energy was
  the only constant of motion, on $\Si$ is obtained in modern language by
  the distribution $\prod_{i=n}^{k+1} \d(\f_i-a_i)$ times a function
  invariant on $\Si$, hence constant (by the hypothesis); the evaluation of
  the delta functions requires the computation of a Jacobian (the last
  multiplier).\\
 The Jacobi's principle, in this paper, is the theorem that expresses the
 volume element in a system of coordinates in terms of that in another
 through a ``multiplier'' (the ``Jacobian determinant'' of the change
 of coordinates). Boltzmann derived, already in the preceding paper, what
 we call today ``Liouville's theorem'' for the conservation of the volume
 element of phase space and here he gives a version that takes into account
 the existence of constants of motion, such as the energy.  \\
 At the same time this work {\it lays the foundations of the theory of the
   statistical ensembles}, as recognized by Gibbs in the introduction of
 his treatise on statistical mechanics, \citep{Gi902}: curiously Gibbs
 quotes this paper of Boltzmann attributing to it a title which, instead,
 is the title of its first Section, \cite[see]{Bo871-d}.  From the uniform
 distribution (``microcanonical'') on the surface of constant total energy
 (suggested by the above interpretation of the ergodic hypothesis) the
 canonical distribution of subsystems (like molecules or atoms) follows by
 integration and use of the formula $(1-\fra{c}{\l})^\l=e^{-c}$ if $\l$
 (total number of molecules) is large.%
\\ It appears that Boltzmann by making the ``Hypothesis'', {\it that the
  atoms of a warm body can take all positions and velocities} compatible
with the equation of the kinetic energy, assumes that the only such
distribution is the one which, as he shows, is constant on the energy
surface, because of the Liouville's theorem that he derives in several
others among his papers, and in particular in the first of the trilogy and
of the idea that the distribution is positive in the regions visited by the
phase space point. This has been mathematized as the ``metric
transitivity'' form of ergodicity. The ``doubt'' that ``perturbed'' him
in the first paper of the trilogy, see above, does not show up in this work
(the reason is that he was certainly not thinking to the possible existence
of non smooth constants of motion, see \citep[p.10,V.1]{Bo897}).
\\ The above form of the ergodic hypothesis assumed for the whole gas, is
used in this last Section to derive the canonical distribution for the
velocity and position distributions both of a single molecule and of an
arbitrary number of them ($\ll n$). It goes beyond the preceding paper
deducing the {\it microcanonical} distribution, on the assumption of the
ergodic hypothesis, and finding as a consequence the {\it canonical}
stationary distribution of the atoms of each molecule or of an arbitrary
number of them by integration on the positions and velocities of the other
molecules. %
\\
 Hence, imagining the gas large, the canonical distribution follows for
 every finite part of it, be it constituted by $1$ or by $10^{19}$
 molecules: a finite part of a gas is like a giant molecule.}]  \*

Let us accept this hypothesis, and thus let us make use of the formulae to
compute the equilibrium distribution between a gas in interaction with a
body supposing that only $r$ of the mentioned $\l$ atoms of the body
interact with the mass of gas.

Then $\ch$ [\alertb{\sl potential energy}] has the form $\ch_1+\ch_2$ where
$\ch_1$ is a function of the coordinates of the $r$ atoms, $\ch_2$ is a
function of the coordinates of the remaining $\l-r$. %
\footnote{\alertr{\sl Additivity of the potential reflects the
    assumption that density is low and intermolecular
  interaction,} \alertr{%
    important to reach a stationary state, can be neglected in
    studying the molecules distributions in equilibrium. Thus the potetnial
    energy is just the interaction energy of the atoms inside the molecules
    to which they belong.}}
\\
\ldots
\*

\0[\alertb{\sl The derivation of the canonical distribution for the
    molecules from the microcanonical one for the whole gas, at small
    density to reproduce the results of \citep[\#18]{Bo871-a}: the
    derivation is the one still taught today.}]  \*
  \ldots%
These equations must, under our hypothesis, hold for an arbitrary body in a
mass of gas, and therefore also for a molecule of gas. In the considered
case it is easy to see that this agrees with the formulae of my work {\it
  {\"U}ber das {W\"a}rme\-gleichgewicht zwischen mehratomigen
  {G}asmolek{\"u}len}, 1871, \citep[\#18]{Bo871-a}. We also arrive there
in a much easier way to what found there. However the proof, in the present
section, makes use of the hypothesis about the warm body, certainly
acceptable but which had not yet been proved: thus I avoided it in the
quoted paper achieving the proof in a way independent from that
hypothesis.%
\footnote{\alertr{\sl He means that he had proved the invariance of the
    canonical distribution (which implies the equidistribution) without the
    present hypothesis.  However even that was not completely satisfactory
    for him (and for us), as he had also stated in the quoted paper, since
    he had not been able to prove the uniqueness of the solution found
    there (that we know today to be not unique in general).}}

\def\SEC{Heat theorem without dynamics (Trilogy: \#3)}
\section{\SEC}
\label{1871-c}\iniz
\lhead{\small\thesection: \SEC}

\0{Quotes and Comment on: 
{\it {A}nalytischer {B}eweis des zweiten {H}auptsatzes der
mechanischen {W\"a}rmetheorie aus den {S\"a}tzen {\"u}ber das
{G}leichgewicht des leben\-digen {K}raft}, 1871,
\citep[\#20,p.288-308]{Bo871-c}.}
\*

\0\alertb{\sl %
Here dynamics enters only through the conservation laws and the hypothesis
(see the first trilogy paper \citep[\#18,Sec.III]{Bo871-a}) that never a
multiple collision takes place and collisions duration is negligible
because of low density. Alternatively it could be based on the ergodic
hypothesis of \citep[\#19]{Bo871-b}. It is used at the beginning to
start from the results of the earlier works leading to the  microcanonical
distribution for the entire warm body and to the canonical distribution for
the smaller subsystems, see also Appendices\ref{1871-a},\ref{1871-b} above.\\
  Referring to \citep[\#18]{Bo871-a} for motivation and going back to
  the quasi static processes of his early work, \citep[\#2]{Bo866}, and
  assuming right away the canonical distribution, it is shown that it also
  follows that defining the heat $dQ$ received by the body, in an
  infinitesimal step of the process, as the variation of the total average
  energy $dE$ plus the work $dW$ done by the system on the outside
  (variation of the time average of the potential energy due to a change of
  the values of the external parameters) it follows that $\frac{d Q}T$ is
  an exact differential if $T$ is proportional to the average kinetic
  energy. \\
This is the first introduction of the notion of ``thermodynamic
analogy'' or ``model of thermodynamics'': an example is exhibited by the
canonical distribution; the novelty with respect to the 1866 theory is that
the probability distribution of the microscopic configurations is also
determined explicitly and reference to periodicity is formally avoided.
The periodic motion of the system is replaced by a stationary
probability distribution in the phase space: the two notions coincide if
the ergodic hypothesis of \citep[\#19]{Bo871-b} is accepted.
\footnote{\alertr{\sl In the 1866 work Boltzmann determined mechanical
    quantities analogus to temperature, specific heat, free energy directly
    from the definitions of the latter in rarefied gases kinetic theory,
    while entropy was obtained through the new idea that atoms follow
    periodic trajetories: distribution of the phase space points is even
    mentioned.}}%
\\
It
%
will be revisited in the paper of 1884, \citep[\#73]{Bo884}, with the
general theory of statistical ensembles and of the states of thermodynamic
equilibrium.\\
This work also preludes to the {\it molecular chaos}
analysis in the paper, following this a little later, developing
``Boltzmann's equation'', \citep[\#22]{Bo872}. \\
The first seven pages, from p.288 to p.294 are translated and commented as
they are useful to understand the main ideas and the difference with the
earlier works. The rest of the work deals with the concept of thermodynamic
analogy which will be taken up and further developed in later works,
\citep[1877-84]{Bo877a,Bo877b,Bo884}, see the Appendices below.}%

\* Let (K) be an arbitrary body, consistig of $r$ mass points (Atoms). We
shall denote the coordinates of its atoms $x_1,y_1,\ldots,z_r$ and the
velocities $c_1,c_2,\ldots,c_r$ and their corresponding three components
$u_1,v_1,w_1,u_2,\ldots,w_r$. Finally let $\ch$ be the potential energy,
also a function of of $x_1,y_1,\ldots,z_r$, whose negative gradient with
respect to the coordinates of an atom give the force acting in the
direction of the coordinates. Let us leave unchanged for a very long time
the other circumstances (temperature and the external forces) to which the
body is subject, and the fraction of $T$ in which the variables
$x_1,y_1,\ldots,z_r,u_1,v_1,\ldots,w_r$ are in the region
$$\leqno{(A)} \qquad
x_1\ {\rm and}\ x_1+dx_1,\ y_1\ {\rm and}\ y_1+dy_1,\ldots,\ w_r \ {\rm and}\ 
w_r+dw_r$$
will be denoted $\t$. I denote the value $\t/T$ as the time, during which
{\it in average} the state of the body is enclosed in the region $(A)$.  I
found in the quoted work, %
\footnote{See note p.237 of Wissenshaeftliche Abhandlungen
  \alertr{[\sl This footnote is added to
  the WA: however it points at \citep[\#18]{Bo871-a}: but it seems that it
  should point instead at \citep[\#19]{Bo871-b}]. The remark here is that the
  main object in this paper, \ie the canonical distribution of the
  molecules, can be obtained via the work \citep[\#19]{Bo871-b} with its ergodic
  hypothesis or via \citep[\#18]{Bo871-a} with the low density, no multiple
  collisions assumptions.}}%
aware of an underlying hypothesis and if the body is in contact with
infinitely many molecules of a gas, that this ratio has the value

$$dt=\frac{e^{-h\f}dx_1dy_1dz_r,du_1dv_1\ldots dw_r}{
  \int \int \ldots e^{-h\f}
  dx_1dy_1dz_r,du_1dv_1\ldots dw_r}\eqno{(1)}$$
where
$$\f=\ch+\sum \frac{m c^2}2$$
This formula appears as the last formula of the quoted
work%
\footnote{\alertr{\sl \citep[\#19,p.287]{Bo871-b}}}
and there also is expressed the same probability of all the velocity
components and is determined the suitable constant $\l''$.%
\footnote{\alertr{\sl Should be $C''$.}} 

We know this expression, even without the deliberate key hypothesis, from
my work \citep[\#18]{Bo871-a}.  Let one molecule, among the ones of which the
body is composed, contain $\r$ atoms. Let the coordinates of the baricenter
of this molecule be $x,y,z$, let the coordinates if its atoms with respect
to coordinate axes through the baricenter be $\x_1,\h_1,\ldots,\z_\r$ and
the velocities of the atoms parallel to the same axes be
$\a_1,\b_1,\ldots,\g_\r$.  The total kinetic energy contained in the
molecule will be denoted $\f^*$.%
\footnote{\alertr{\sl  For consistency with the last formula $\f^*$
    should be the total energy in the center of mass, not just the kinetic
    energy. Also here it is somewhat confusing that from now on the body (K)
    will not contain the selected molecule although up to this point the
    molecule was part of (K).}} 
Finally let the gas contain $N$ molecules per unit volume. I find in the
work \citep[\#18]{Bo871-a} that the number of those molecules of this
gas per unit volume, for which at the same time
$\x_1,\h_1,\ldots,\z_\r,\a_1,\b_1,\ldots,\g_\r$ are in the region
$$\leqno{(B)}\qquad
\x_1\ {\rm and}\ \x_1+d\x_1,\ \h_1\
  {\rm and}\ \h_1+d\h_1,\ldots,\ \g_\r \ {\rm and}\ 
\g_\r+d\g_\r$$
equals
$$ dN=a\, e^{-h\f^*}\,d\x_1d\h_1\ldots d\g_\r $$
where $a$ and $h$ are constants. The beginning of a collision of one of
these molecules with the body (K), or of a collision between themselves,
should again be characterized by a suitable function, of the relative
position of the atoms of the body and of those of the colliding molecule,
$F(x_1,y_1,\ldots,z_r,x,y,$ $z,\x_1,\h_1,\ldots,\z_{\r-1})$, indicating the
beginning of a collision when it assumes a 
certain value $b$.%
\footnote{\alertr{\sl Remark that $F$ does not depend on velocities.}}%
 Re-acquisition of this value should mark the end of the collision. We
 shall denote temporarily the not yet known time $\t$ within which, during
 a very long time $T$, the state of the body (K) is within the region (A), by
 $T\cdot g \,dx_1dy_1\ldots dw_r$, so that $g \,dx_1dy_1\ldots dw_r$ is
 also the \alertr{[fraction of]} time during which the state of the body is
 within the region (A).%
 \footnote{\alertr{\sl  This anticipates the assumption, that will
     be made shortly, that the collisions duration can be neglected together
     with the time interval between collisions of some molecule of the body
     (K) and the extra molecule, as well as collisions between the molecule
     and more than just one molecule of  (K).}}
  Hence $g$ is a function of the state of the body at the moment of the
 collision via $x_1,y_1,\ldots,z_r,u_1,v_1,\ldots,w_r$.  The total state at
 the beginning of the collision is completely determined if we know the
 values of the $6r+6\r-1$ quantities  $x_1,y_1,\ldots,z_r,\x_1,\h_1,
 \ldots,\z_{\r-1},\a_1,\b_1,\ldots,\g_\r,x,y$ at this instant;
 $z$ is determined at the collision beginning by the
 equality $F=b$.%
\footnote{\alertr{\sl  These are coordinates for the state of (K) and of
    the molecule, not just of the state of  (K).}}
By an analysis closely similar to the one I have employed in the work
\citep[\#18]{Bo871-a}, it follows that the number of collisions that
take place during the time $T$ and in which, at the moment of their
beginning, the atoms are in the region
$$(C)\kern8mm\left\{ \eqalign{&{\rm variables\ of\ the\ body\ (K)} \ \in\ (A)\cr
  &{\rm variables\ of\ the\ molecule} \ \in\ (B)\cr
  &{\rm at\ the\ same\ time}\ x \in(x,x+dx),\ y \in(y,y+dy),\cr}\right.
$$
is equal to
$$ dm=\t \,dN\,\o dx dy= T g a e^{-h\f^*}\o dx_1dy_1\ldots dw_rd\x_1d\h_1\ldots
d\g_\r dx dy\eqno{(2)}$$
where\footnote{\alertr{\sl The sign of $\o$ is not taken into account. This is
    similar to arguments familiar in the context of the Boltzmann equation:
    $dN\,\dot F\,\d(F-b)$ is the number of collisions per unit time and
    $\int dN\,\dot F\,\d(F-b) dt$. The $dN$ is $e^{-h\f^*} dx_1\ldots dw_r
    d\x_1\ldots\z_{\r-1}dxdydz$ and the formula follows by eliminating $z$
    via the delta function and integrating over $t$ (obtaining $\t$).}}
$$\o=\frac1{\frac{\dpr F}{\dpr z}}\Big(\frac{\dpr F}{\dpr x_1}\frac{d
  x_1}{d t}+ \frac{\dpr F}{\dpr y_1}\frac{d y_1}{d t}+\ldots+\frac{\dpr
  F}{\dpr z}\frac{d z}{d t}\Big)$$
The variables determining the state of the interacting body, when at the
beginning of the collision are in the region (C),  shall be in the region
$$\ (D)\kern8mm\left\{ \eqalign{&X_1\ {\rm and}\ X_1+dX_1,\
  Y_1\ {\rm and}\ Y_1+dY_1,\ldots W_r{\rm and}\ W_r+dW_r\cr
  &\X_1\ {\rm and}\ \X_1+d\X_1,\
   H_1\ {\rm and}\  H_1+d H_1,\ldots Y{\rm and}\ Y+dY
\cr}\right.
$$
Then, as in my quoted work, also  follows the equation
$$\o\cdot D=\O\eqno{(3)}$$
where $D$ is the determinant
$$\sum \pm \frac{\dpr x_1}{\dpr X_1}.\frac{\dpr y_1}{\dpr Y_1}\ldots
\frac{\dpr w_r}{\dpr W_r}.\frac{\dpr \x_1}{\dpr \X_1}
.\frac{\dpr \h_1}{\dpr  H_1}\ldots\frac{\dpr y}{\dpr Y}$$
With $\O,G,\F^*$ we shall denote the quantities into which $\o,g,\f$
are changed, if the $X_1,Y_1,\ldots$ replace $x_1,y_1,\ldots$.
The number of collisions, taking place in the time $T$, in which the
variables lie at the beginning in the region (D) is
\footnote{\alertr{\sl In Eq. (4) it seems that $a$ should be instead $A$.}}
$$ dM=T G a e^{-h\F^*} dX_1 dY_1\ldots dW_r d\X_1\ldots dY\eqno{(4)}$$
We can thus express $X_1,Y_1,\ldots$ as functions of $x_1,y_1,\ldots$ or
viceversa $x_1,y_1,\ldots$  as functions of $X_1,Y_1,\ldots$. So does the
last terms in the formula (2) become 
$$dm= T g a e^{-h\f^*}\o D  dX_1 dY_1\ldots dW_r d\X_1\ldots dY\eqno{(5)}$$
This is the number of collisions which, during the time $T$ take place and
at their end  have the variables in the region (D). It is possible to think
also $g,\o$ and $\f^*$ as expressed as functions of $X_1,Y_1,\ldots$.
Let us integrate $dm$ over all possible variables of the
state of the molekule $\X_1,H_1,\ldots,Z_{\r-1},X,Y$, thus we obtain the
number of all collisions which during the time $T$ take place so that at
their end the state of the body its variables are in the region
$$(E)\qquad X_1\ {\rm and}\ X_1+dX_1,\ Y_1\ {\rm and}\ Y_1+dY_1,\ldots
W_r\ {\rm and}\ W_r+d W_r$$
while the state of the molecule are not subject to any restriction.  This
number will be denoted $\int dm$ (where the integration is extended to the
state variables for the molecule). We now wish to make a special
assumption, although not realized in nature, but dealing with a conceivable
case which will facilitate the calculation in the general cases. Namely we
wish to assume that the collisions happen so frequently that immediately if
a collision terminates again the next begins, while the body never
experiences collisions involving two molecules, so that if the variables
determining the state of the body are in the region (E) at the end of a
collision again in a new collision in no way (or neglegibly) the movement
of the atoms without collisions escapes this region. Hence, while the body
is involved in a collision with a molecule no second collision occurs, so
that we want to exclude from the time $T$ these times and understand them
as not included in $T$ the totality of all instants in which no molecule
endures a collision.  Under this assumption the number of collisions, which
so take place, in which at their end the variables are in the region (E)
have to be equal to the number which at the \alertr{[collision]} beginning
are in (E).  Then first is the frequency in which the variables are
entering these regions and secondly also the frequency of exiting, as,
according to our assumption, the number of entrances and exits as a result
of the movement of the atoms without collisions with the molecule can be
neglected.  The first number is $\int dm$, the second gives the expression
of the integral of equation (4) over all values determining the state of
the molecule. It will be denoted $\int dM$.  It is also clear that this is
the only condition on the determination of the function $g$. But the
condition $\int dm=\int dM$ is satisfied, if
$$g=A e^{-h \f}$$
is set, and it can be justified as follows. Replace  in the equality $\int
dm=\int dM$ this value for $g$ and also divide the values (4) and (5)
for $dM$ and $dm$  by $T a A dX_1dY_1\ldots dW_r$ to obtain
$$\int\int\ldots e^{-h(\f+\f^*)}\o \cdot D\cdot d\X_1dH_1\ldots dY
=\int\int  e^{-h(\F+\F^*)}\O \cdot d\X_1dH_1\ldots dY$$
or because of the equation (3)
$$\int\int\ldots e^{-h(\f+\f^*)}\O \cdot d\X_1dH_1\ldots dY
=\int\int  e^{-h(\F+\F^*)}\O \cdot d\X_1dH_1\ldots dY$$
Here all the work \alertr{[potential energy]} and kinetic energy involved
are in $\f$ and $\f^*$ before the collision and in $\F$ and $\F^*$ after
the collision (it is possible to express the variables $x_1,y_1,\ldots$ as
functions of $X_1,Y_1,\ldots$, and on the other hand $\F$ and $\F^*$ and
again $x_1,y_1,\ldots$ interchanged with $X_1,Y_1,\ldots$).

Here $\f$ and $\f^*$ include the kinetic energy and the work
\alertr{[potential energy]} that do not change in the collision so that in
general $\f+\f^*=\F+\F^*$ hence also both integrals in formula (6) are
equal, because the function inside the integral signs are equal for all
values of the integration variables. The value adopted for $g$ fulfills in
fact the condition $\int dm=\int dM$ and consequently also the time, while
some state of the body is in the region (A) of formula (1). Because in this
formula $g$ has in fact those values. (There just the constant $A$ has been
suitably determined).
\\
But now I have already in the work \citep[\#18]{Bo871-a} shown that a
distribution of states given through formula (1) will not be changed in the
motion of the atoms without collisions; therefore the probability of the
different states of the body will also still be given by formula (1) even
if it does not hold the assumption of collisions so frequent that the
motion of the atoms without collisions is not influent. The formula (1)
remains even valid is the collisions should so develop that the
simultaneous collisions of several molecules can be neglected. By the way
this teaches a difficult result that such simultaneous collisions, even if
take place somewhat often, do not influence the distribution formula
(1).
\\However in bodies which in nature are in contact with gases, there
certainly are always such simoultaneous collisions; but in the parts of the
bodies, in which the collisions of several molecules happen, there are no
appreciable direct interaction: consequently, in its absence, it is as if
over the body a collision took place with only one molecule. We want now to
suppose that the probability distribuiton of the states in a warm body,
also when it is not in contact with a gas, remains the same; under this
assumption formula (1) holds for any warm body and I will now show how with
the same ease will follow an analytic proof of the main theorem of the
mechanical theory of heat.%
\footnote{\alertr{\sl This is the important claim that assuming the
    canonical distribution for any system the main heat theorem $\oint
    \frac{dq}T=0$ follows as well as all the main formulae of
    thermodynamics. For this reason it has been sometimes claimed that
    Boltzmann abandoned the dynamical derivation of the heat theorem:
    however it should be kept in mind that the basis of the discussion is
    the indentity between the the probability of a state (\ie microscopic
    configuration) and the frequency of visit to the parallelepipedal cell
    defining it: for an isolated system this is of course the ergodic
    hypothesis which in this form is already present in the earlier works
    in which it is shown that this assumption implies a canonical
    distribution for subsystems, \citep[1868-71]{Bo868,Bo871-a,Bo871-b,Bo871-c}.}}
\*

\0[\alertb{\sl(p.294-308): If the canonical distribution is accepted for a
    system in equilibrium with the surroundings it is possible to define
    and compute several averages which admit physical interpretations, like
    for instance temperature, pressure, volume, energy, specific heat ...
    ($T,p,V,U,C_v\ldots$). Boltzmann dedicates the remaining part of the
    work to check that such average values obey the relations that are
    expected from the theory of thermodynamics, \ie they define a {\it
      thermodynamic analogy} or {\it model of thermodynamics}: a notion
    that will be formalized later. He first studies a process in which the
    parameters of the canonical ensemble vary, and defines the amount of
    heat $\d Q$ that enters in the system and the amount of work $\d W$
    that the system performs on the external bodies (\eg through the volume
    variation) and checks that $\frac{\d Q}T$ is a two paramenters exact
    differential: he fulfills, at the same time, the promise made to
    Clausius in the priority paper that he will take into acount the
    variation of the external potential. Furthermore he shows that even the
    microcanonical distribution has the same property that the natural
    definition of $\frac{\d Q}T$ in a process is an exact
    differential.  \\
    However it is hard to understand the need of the long analysis
    preceding p.294 and translated above. The canonical distribution was
    derived already in \citep[\#5]{Bo868} for the first time in general, as
    stressed in \citep{Ma879-c} quoted above, and again in
    \citep[\#18]{Bo871-a} with special assumptions on the density and free
    flight time, and again in \citep[\#19]{Bo871-b}. Once the canonical
    distribution is accepted the analysis from (p.294-308) can be
    performed. \\
    The derivation will be reproduced in the later paper
    \citep[\#73]{Bo884} and in Appendix\ref{1884},
    p.\pageref{olodecanonico} below.  Therefore in this paper the theory of
    the equilibrium ensembles is essentially fully developed. }]

\def\SEC{Irreversibility: Loschmidt and ``Boltzmann's sea''}
\section{\SEC}
\label{1877-a}\iniz
\lhead{\small\thesection: \SEC}\index{Loschmidt}\index{Boltzmann's sea}

\0{Quotes and comments on: 
{\it Bemerkungen {\"u}ber einige Probleme der mechanischen
{W}{\"a}rmetheo\-rie}, 1877, \citep[\#39,p.112-148]{Bo877a}.}

\* \0[\alertb{\sl Sec.I deals with problems about specific heat of liquid
  droplets in saturated vapors. Again there is some overlapping of part of
  Sec.I with Clausius' work as pointed out at footnote${}^1$, p.116.  \\In
  Sec.II the discussion is really about a deep and clear interpretation of
  the second law and of the conflict between irreversibility and
  microscopic reversibility.  \\
  And Sec.III deals with the example of one
  dimensional motions to illustrate the heat theorem (a simple case is
  described in Sec.\ref{sec:4} above): the example was discussed also in
  earlier papers of Boltzmann but this is particularly important as it
  introduces\sl formally the notion of model of thermodynamics, which will
  be the basis of the theory of the ensembles in the 1884 work, see
  Appendix\ref{1884}, on monocyclic systems. }]  \*

\0{\bf \S{}II.} {\bf On the relation between a general mechanics theorem
  and the second main theorem of the theory of heat} (p.116)
\*

In his work on the states of thermal equilibrium of a system of bodies,
with attention to the force of gravity, Loschmidt formulated an opinion,
according to which he doubts about the possibility of an entirely
mechanical proof of the second theorem of the theory of heat.  With the
same extreme sagacity he suspects that for the correct understanding of the
second theorem an analysis of its significance is necessary, deeper than
what appears indicated in my philosophical interpretation, in which perhaps
various physical properties are found which are still difficult to
understand, hence I shall immediately here undertake their explanation with
other words.

We want to explain in a purely mechanical way the law according to which
all natural processes proceed so that
$$\int \fra{dQ}T\le 0$$
and so, therefore, behave bodies consistent of aggregates of point masses.
The forces acting between these point masses are imagined as functions of
the relative positions of the points. If they are known as functions of
these relative positions we say that the interaction forces are
known. Therefore the real motion of the point masses and also the
transformations of the state of the body will be known once given the
initial positions and velocities of the generic point mass. We say that the
initial conditions must be given.

We want to prove the second theorem in mechanical terms, founding it on the
nature of the interaction laws and without imposing any restriction on the
initial conditions, knowledge of which is not supposed.  We look also for
the proof that, always whatever may be the initial conditions, the
transformations of the body always take place so that
$$\int \fra{dQ}T\le 0$$
Suppose now that the body is constituted by a collection of point-like, or
virtually such, masses.  The initial condition be so given that the
successive transformation of the body proceed so that

$$\int \fra{dQ}T\le 0$$
We want to claim immediately that, provided the forces stay unchanged, it
is possible to exhibit another initial condition for which it is

$$\int \fra{dQ}T\ge 0.$$
We can consider the values of the velocities of all point masses reached at
a given time $t_1$ and we now want to consider, instead of the preceding
initial conditions, the following: at the beginning all point masses have
the same positions reached, starting form the preceding initial conditions,
in time $t_1$ but with all velocities inverted. In such case we want to
remark that the evolution of the state towards the future retraces exactly
the preceding evolution towards the time $t_1$.

It is clear that the point masses retrace the same states followed by the
preceding initial conditions, but in the opposite direction.  We shall see
the initial state that before we had at time $0$ to be realized at time
$t_1$ [{\sl with opposite velocities}]. Hence if before it was

$$\int \fra{dQ}T\le 0$$
we shall have now $\ge0$. 

On the sign of this integral the interaction cannot have influence, but it
only depends on the initial conditions. In all processes in the world in
which we live, experience teaches us this integral to be $\le 0$, and
this is not implicit in the interaction law, but rather depends on the
initial conditions. If at time $0$ the state [{\sl of the velocities}] of
all the points of the Universe was opposite to the one reached after a very
long time $t_1$ the evolution would proceed backwards and this would imply
$$\int \fra{dQ}T\le 0$$
Every experimentation on the nature of the body and on the mutual
interaction law, without considering the initial conditions, to check that
$$\int \fra{dQ}T\le 0$$
would be vain. We see that this difficulty is very attractive and we must
consider it as an interesting sophism. To get close to the fallacy that is
in this sophism we shall immediately consider a system of a finite number
of point masses, which is isolated from the rest of the Universe.

We think to a very large, although finite, number of elastic spheres, which
are moving inside a container closed on every side, whose walls are
absolutely still and perfectly elastic. No external forces be supposed
acting on our spheres. At time $0$ the distribution of the spheres in the
container be assigned as non uniform; for instance the spheres on the right
be denser than the ones on the left and be faster if higher than if lower
and of the same order of magnitude.  For the initial conditions that we have
mentioned  the spheres be at time $t_1$ almost uniformly mixed. 
We can then consider instead of the preceding initial conditions, the ones
that generate the inverse motion, determined by the initial conditions
reached at time $t_1$.%
\\ Then, as time evolves, the spheres come back; and at time $t_1$ will
have reached a non uniform distribution although the initial condition was
almost uniform. We then must argue as follows: a proof that, after the time
$t_1$ the mixing of the spheres must be with absolute certainty uniform,
whatever the initial distribution, cannot be maintained.  This is taught by
the probability itself; every non uniform distribution, although highly
improbable, is not absolutely impossible.  It is then clear that every
particular uniform distribution, that follows an initial datum and is
reached in a given time is as improbable as any other even if not uniform;
just as in the lotto game every five numbers are equally probable as the
five $1,2,3,4,5$. And then the greater or lesser uniformity of the
distribution depends on the greater size of the probability that the
distribution becomes uniform, as time goes.

It is not possible, therefore, to prove that whatever are the initial
positions and velocities of the spheres, after a long enough time, a
uniform distribution is reached, nevertheless it will be possible to prove
that the initial states which after a long enough time evolve towards a
uniform state will be infinitely more than those evolving towards a
nonuniform state, and even in the latter case, after an even longer time,
they will evolve towards a uniform state.%
\footnote{\alertr{\sl  Today this important discussion is referred as the
  argument of {\it Boltzmann's sea}, \citep{Ul968}.}}

Loschmidt's proposition teaches also to recognize the initial states that
really at the end of a time $t_1$ evolve towards a very non uniform
distribution; but it does not imply the proof that the initial data that
after a time $t_1$ evolve into uniform distributions are not infinitely
many more. Contrary to such statement is even the proposition itself which
enumerates as infinitely many more uniform distributions than non uniform,
and the number of the states which, after a given time $t_1$ arrive to
uniform distribution must also be much larger than those which arrive to
nonuniform distributions, and these are just the configurations that arise
in the initial states of Loschmidt becoming non uniform at time $t_1$.

It is actually possible to calculate the ratio of the numbers determining
the probabilities of the different initial states, perhaps leading to
``an interesting method to calculate the thermal equilibria''%
\footnote{\alertr{\sl The idea will be implemented in
    \citep[\#42]{Bo877b}, see Appendix\ref{1877-b}.}}
exactly analogous to the one that leads to the second theorem.  It is at
least in some special cases successfully checked when a system undergoes a
transformation from a nonuniform state to a uniform one, then $\int
\fra{dQ}T$ will be intrinsically negative, while it will be positive in the
inverse case. Since there are infinitely many more uniform than nonuniform
distributions of the states, therefore the last case will be extremely
improbable: and in practice it could be considered impossible that at the
beginning a mixture of oxygen and nitrogen are given so that after one
month the chemically pure oxygen is found in the upper part while the
nitrogen is in the lower, an event that probability theory states as
improbable but not as absolutely impossible.

Nevertheless it seems to me that the Loschmidtian theorem has a great
importance, since it tells us how intimately related are the second
principle and the calculus of probabilities. For all cases in which $\int
\fra{dQ}T$ can be negative it is also possible to find an initial condition
very improbable in which it is positive.  It is clear to me that for closed
atomic trajectories $\int\fra{dQ}T$ must always vanish. For non closed
trajectories it can also be negative. Now a peculiar consequence of the
Loschmidtian theorem which I want to mention here, \ie that the state of
the Universe at an infinitely remote time, with fundamentally equal
confidence, can be considered with large probability both as a state in
which all temperature differences have disappeared, and as the state in
which the Universe will evolve in the remote future.%
\footnote{\alertr{\sl 
  Reference to the view of Clausius who claims that in the remote future
  the Universe will be in an absolutely uniform state. Here Boltzmann says that
  the same must have happened, with equal likelihood in the remote past.}}

This is analogous to the following case: if we want that, in a given gas at
a given time, a non uniform distribution is realized and that the gas
remains for a very long time without external influences, then we must
think that as the distribution of the states was uniform before so it will
become again entirely uniform.

In other words: as any nonuniform distribution evolves at the end of a time
$t_1$ towards a uniform one the latter, if inverted, as the same time $t_1$
elapses again comes back to the initial nonuniform distribution (precisely
for the said inversion). The [{\sl new}] but inverted initial condition,
chosen as initial condition, after a time $t_1$ similarly will evolve to a
uniform distribution.%
\footnote{\alertr{ {\it I.e.}\sl  if once having come back we continue the
  evolution for as much time again a uniform distribution is reached.}}

But perhaps such interpretation relegates in the domain of probability
theory the second principle, whose universal use appears very questionable,
and nevertheless just because of the theory of probability it will be
realized in every laboratory experimentation.
\*

\0[\alertb{\sl \S{}III, p.127 and following: a check of the heat theorem
    is presented in the case of a central motion, which will be revisited
    in the papers of 1884 by v. Helmholtz and Boltzmann. The cases of
    central motion with a repulsive potential $\frac{b}{r^2}$
    plus either a gravitational potential or a harmonic potential are treated
    in great detail.\\
    The result is again that $\frac{\d Q}{T}$ is an exact differential
    only if the potential $\d b=b=0$, see also \citep[p.205]{Ga013b} and
    \citep[p.45]{Ga000}. %
    \\
    The main purpose and conclusion of Sec. III seems however to be that
    when there are several constants of motion (which it would be tempting
    to consider parameters of the states of the system) it cannot be
    expected that the average kinetic energy is an integrating factor for
    $dQ=dE-\media{\dpr_{\V c} \f\cdot \d\V c}$. The Newtonian potential is
    a remarkable exception. Other exceptions are the $1$--dimensional
    systems, obtained as special cases of the central potentials cases with
    zero area velocity, $\b\equiv 0$. Several special one-dimensional cases
    associated with $2$-dimensional central motions are considered: the
    matter will be discussed a few years laterin 
    \citep{He884a,He884b}, and immediately afterwards in
    \citep[\#73]{Bo884}, leading to a general theory of the
    ensembles.}]

\def\SEC{Discrete phase space, count of its points and entropy.}
\section{\SEC}
\label{1877-b}\iniz
\lhead{\small\thesection: \SEC}

\0{\it Quotes and comments on: 
{\it \"Uber die Bezie\-hung zwischen dem zweiten Hauptsatze der
mechanischen W\"arme\-theo\-rie und der Wahr\-scheinlich\-keitsrechnung,
respektive den S\"atzen \"uber das W\"armegleichgewicht}, 1877,
\citep[\#42,p.164-233]{Bo877b}.}
\*
\0[\alertb{\sl This work takes up the method already used in \citep[\#5]{Bo868}
    of dividing phase space into cells to count the number of microscopic
    states: implementing the statement, in \citep[\#39,p.121]{Bo877a}, that
    the count is possible and may lead to an interesting method to calculate
    the thermal equilibria. The results will be later, by other authors,
    summarized in the formula $S=k_B\log N$, $N$ being the number of ways
    of realizing the microscopic configurations which assign to few
    thermodynamic observables value equal to their averages, in systems
    with many particles. In the 1868 work phase
    space was simply discretized and the microcanonical distribution
    derived (in Sec.III) by assuming that all cells, in the
    $(6N-1)$-dimensional phase space, which could be visited were indeed
    visited and the frequency of visit was assumed equal to the number of
    ways the cells could be occupied so that the total energy has the fixed
    value. \\
    The novelty is that Boltzmann uses the combinatorial count details to
    consider the probabilities of all configurations and shows that just
    the ones that maximize the frequency of visit are by far sufficient to
    compute the averages of thermodynamic observables. That played an
    important role in the work that he had to do to defend and clarify his
    ideas and results.  \\
    To perform the count the energy is supposed to take values regularly
    spaced and with a degeneracy depending on the space
    dimensionality. Hence it is imagined that the momentum space is divided
    into small cubic cells of sides $dc$; the ones in which the speed is in
    the shell between $e=k\e$ and $e+de$ (with $\e$ the size of the
    discrete kinetic energy jumps and $k\ge0$ integer) or speed between $c$
    and $c+dc$, with $c=\sqrt{\frac{2 k\e}{m}}$, is $\frac{4\p}m c d e=4\p
    c^2dc$, \ie it is proportional to $c^2 dc$, while in dimension $2$ it
    is proportional to $dc$ and in $1$ dimension it is just the constant
    $2$ corresponding to $\pm c$.  \\
    At the end of the work also the positions are considered and the cells
    become parallelepipeds in phase space: however the forces considered
    are only external forces and the system is still a low density gas.
    \\
    It should be remarked that although the general strategy is very clear,
    and the role of discreteness is stressed and used to give a new
    definition and interpretation of entropy, the analysis is closely
    related, less general but much more detailed and with\sl new perspectives,
    like entropy even out of equilibrium, with respect to the early work
    \citep[\#5]{Bo868}. There, in Sec.II. cells were constructed in
    momentum space and in Sec.III in phase space, \ie were parallelepipeds
    in a regular lattice in phase space, see Appendix\ref{1868} above. And
    mainly the system was not a rarefied gas but the internal forces were
    rather general conservative forces, as stressed by Maxwell,
    \citep{Ma879-c}.  \\
It seems that this restriction on the forces, not commented by Boltzmann,
reflects his intention to connect the combinatorial work with the Boltzmann
equation and the related $H$-theorem.}]

  \* \0{\bf p.166} ... We now wish to solve the problem which on my above
  quoted paper {\it Bemerkungen \"uber einige Probleme der mechanischen
    {W}{\"a}rmetheorie}, \citep[\#39]{Bo877a}, I have already
  formulated clearly, \ie the problem of determining the ``ratios of the
  number of different states of which we want to compute the
  probabilities''.

We first want to consider a simple body, \ie a gas enclosed between
absolutely elastic walls and whose molecules are perfect spheres absolutely
elastic (or centers of force which, now, interact only when their
separation is smaller than a given quantity, with a given law and
otherwise not; this last hypothesis, which includes the first as a special
case does not at all change the result).  Nevertheless in this case the use
of probability theory is not easy. The number of molecules is not infinite
in a mathematical sense, although it is extremely large. The number of the
different velocities that every molecule can have, on the contrary, should
be thought as infinite. Since the last fact renders much more difficult the
calculations, thus in the first Section of this work I shall rely on easier
conceptions to attain the aim, as I often did in previous works (for
instance in the {\it Weiteren Studien}, \citep[\#22]{Bo872}).
\*
.....
\*
\0{\bf\S{}I.} {\bf The number of values of the kinetic energy is
  discrete.}(p.167) \*

We want first to suppose that every molecule can assume a finite number of
velocities, for instance the velocities

$$0,\fra1q,\fra2q,\fra3q,\ldots,\fra{p}q,$$
where $p$ and $q$ are certain finite numbers. At a collision of two
molecules will  correspond a change of the two velocities, so that the state
of each will have one of the above mentioned velocities, \ie

$$0, \ {\rm or}\ \fra1q, \ {\rm or}\ \fra2q, \ {\it \&tc\ until\
}\,\fra{p}q,$$
It is plain that this fiction is certainly not realized in any
mechanical problem, but it is only a problem that is much easier to treat
mathematically, and which immediately becomes the problem to solve if $p$
and $q$ are allowed to become infinite.

Although this treatment of the problem appears too abstract, also it very
rapidly leads to the solution of the problem, and if we think that all
infinite quantities in nature mean nothing else than going beyond a bound,
so the infinite variety of the velocities, that each molecule is capable of
taking, can be the limiting case reached when every molecule can take an
always larger number of velocities.\index{continuum limit}

We want therefore, temporarily, to consider how the velocities are related
to their kinetic energy. Every molecule will be able to take a finite number
of values of the kinetic energy. For more simplicity suppose that the values
of the kinetic energy that every molecule can have form an arithmetic
sequence,

$$0,\e,\ 2\,\e,\ 3\,\e,\ldots, p\,\e$$
and we shall denote with $P$ the largest of the possible values of $p$.

At a  collision each of the two molecules involved
will have again a velocity
$$0, \ {\rm or}\ \e, \ {\rm or}\ 2\,\e, \ {\rm etc.}\ldots ,\ p\,\e,$$
and in any case the event will never cause that one of the molecules will
end up with having a value of the kinetic energy which is not in the
preceding sequence.

Let $n$ be the number of molecules in our container. If we know how many
molecules have a kinetic energy zero, how many $\e$, \&tc, then we say that
the distribution of the kinetic energy between the molecules is given.

If at the initial time a distribution of the molecules states is given, it
will in general change because of the collisions. The laws under which such
changes take place have been often object of research.  I immediately
remark that this is not my aim, so that I shall not by any means depend on
how and why a change in the distribution takes place, but rather to the
probability on which we are interested, or expressing myself more
precisely, I will search all combinations that can be obtained by
distributing $p+1$ values of kinetic energy between $n$ molecules, and hence
examine how many of such combinations correspond to a distribution of
states.  This last number gives the probability of the relevant
distribution of states, precisely as I said in the quoted place of my {\it
Bemerkungen \"uber einige Probleme der mechanischen W\"armetheorie}
(p.121), \citep[\#39]{Bo877a}.

Preliminarily we want to give a purely schematic version of the problem to
be treated. Suppose that we have $n$ molecules each susceptible of assuming
a kinetic energy

$$0,\e,\ 2\e,\ 3\e,\ldots, p\e.$$
and indeed these kinetic energy will be distributed in all possible ways
between the $n$ molecules, so that the sum of all the kinetic energy
stays the same; for instance is equal to $\l \e=L$.

Every such way of distributing, according to which the first molecule has a
given kinetic energy, for instance $2\e$, the second also a given one, for
instance $6\e$, \&tc until the last molecule, will be called a
``complexion'', and certainly it is easy to understand that each single
complexion is assigned by the sequence of numbers (obviously after division
by $\e$) to which contribute the kinetic energy of the single molecules.
We now ask which is the number ${\cal B}$ of the complexions in which $w_0$
molecules have kinetic energy $0$, $w_1$ kinetic energy $\e$, $w_2$ kinetic
energy $2\e$, {\it \&tc}, $\ldots\ w_p$ kinetic energy $p\e$.
\footnote{The word ``complexion'', defined here, is no longer often
  employed: here it denotes a partition of the total kinetic energy defined
  by the number of particles, out of a total of $n$, that have the same
  kinetic energy. It can also denote, more generally, the partition of the
  total kinetic energy of the particles that occupy distinct volume
  elements $ds$ around points $q$ in $R^{3n}$, \ie assigning for each
  volume element $ds$ the number of particles that have kinetic energy
  $n\k-\ch(q)$, if $n\k$ is the {\it total energy} and $\ch(q)$ the
  potential energy, as in the derivation of the microcanonical ensemble in
  \citep[\#5,p.95,l.-6]{Bo868} (where, however, it was not given a special
  name). In the latter sense it is now called a ``microstate'' or a
  ``microscopic configuration'' in the $n$-particles phase space
  $R^{6n}$.}%

We have said before that when is given how many molecules have zero kinetic
energy, how many $\e$ {\it \&tc}, then the distribution of the states among
the molecules is given; we can also say: the number ${\cal B}$ gives us how
many complexions express a distribution of states in which $w_0$ molecules
have zero kinetic energy, $w_1$ kinetic energy $\e$, {\it \&tc}, or it gives us
the probability of every distribution of states. Let us divide in fact the
number ${\cal B}$ by the total number of all possible complexions and we
get in this way the probability of such state.

It does not follow from here that the distribution of the states gives
which are the molecules which have a given kinetic energy, but only how many
they are, so we can deduce how many ($w_0$) have zero kinetic energy, and how
many ($w_1$) among them have one unit of kinetic energy $\e$, {\it \&tc}. All
of those values zero, one, {\it \&tc} we shall call elements of the
distribution of the states.

(p.170) ........... (p.175)

We shall first treat the determination of the number denoted above ${\cal
B}$ for each given distribution of states, \ie the permutability of such
distribution of states. Then denote by $J$ the sum of the permutabilities
of all possible distributions of states, and hence the ratio $\fra{{\cal
B}}{J}$ gives immediately the probability of the distribution that from now
on we shall always denote $W$.

We also want right away to calculate the permutability ${\cal B}$ of the
distributions of states characterized by $w_0$ molecules with zero kinetic
energy, $w_1$ with kinetic energy $\e$ {\it \&tc}. Hence evidently

$$w_0+w_1+w_2+\ldots+w_p=n\eqno(1)$$
$$w_1+2w_2+3 w_3+\ldots+p w_p=\l,\eqno(2)$$
and then $n$ will be the total  number of molecules and $\l\e=L$ their
kinetic energy.

We shall write the above defined distribution of states with the method
described, so we consider a complexion with $w_0$ molecules with zero
kinetic energy, $w_1$ with unit kinetic energy {\it \&tc}. We know that the
number of permutations of the elements of this complexion, with in total
$n$ elements distributed so that among them $w_0$ are equal between them,
and so $w_1$ are equal between them ... The number of such complexions is
known to be%
\footnote{\alertr{\sl Here particles are considered distinguishable and the
    total number of complexions is $P^n$.}}
$${\cal B}=\fra{n!}{(w_0)!\,(w_1)!\ldots}\eqno(3)$$
The most probable distribution of states will be realized for those choices
of the values of $w_0,w_1,\ldots$ for which ${\cal B}$ is maximal and
quantities $w_0,w_1,\ldots$ are at the same time constrained by the
conditions (1) and (2). The denominator of ${\cal B}$ is a product and
therefore it will be better to search for the minimum of its logarithm, \ie
the minimum of
$$M=\ell[(w_0)!]+\ell[(w_1)!]+\ldots\eqno(4)$$
where $\ell$ denotes the natural logarithm.

..............
\*

\0[\alertb{\sl Follows the discussion of this simple case in which the energy
levels are not degenerate (\ie this is essentially a $1$-dimensional case)
ending with the Maxwell distribution of the velocities.}]
\*
\0\rm Sec.II, {\it The kinetic energies contuinously permute
      each  other}, (p.186)\\
\0[\alertb{\sl Boltzmann goes on to consider the case of $2$-dimensional and of
    $3$--dimensional cells (of sides $da,db$ or $da,db,dc$) in the space of
    the velocities (to be able to take degeneracy into account), treating
    first the discrete case and then taking the continuum limit: getting
    the canonical distribution. The system considered is a very rarefied
    collection of molecules whose state is simply characterized by the
    value of the velocity coordinates each of which is a multiple of a
    small quantity $\e$. Hence, see formulae (30),(31), at p.191,
    the cells in momentume space as small cubes of side $\e$.}]\\%
\0\rm Sec.III, {\it Consideration of polyatomic molecules and external
  forces }, (p.198)\\
\0[\alertb{\sl Deals with the case of polyatomic molecules and external
    forces. Therefore it becomes necessary also to consider the
    discretization of position space (of the molecules atoms with respect
    to their center of mass), and if there is also an external force
    (gravity is considered) it becomes necessary to consider, and
    discretize, for each molecule also the baricenter coordinates. }]\\
\0\rm Sec.{IV}, {\it On the conditions on the maximal exponent values in
  the free products in the function determining the distribution of
  states}, (p.204)\\
\0[\alertb{\sl Studies the maximum value of the product $\prod_{i=1}^p w_i$
    under the conditions that $\sum_{i=1}^p w_i=n$ and $\e\,\sum_{i=1}^p
    i\,w_i=L$ (at given $n,\e,L$ aiming at $n\to\infty,
    L/n=const,\e\to0$).
    \\
An accurate analysis of the combinatorics developed in this paper, together
with \citep[\#5]{Bo868}, is in \citep{Ba990} where two ways of computing the
distribution when the energy levels are discrete are discussed pointing out
that {\it unless the continuum limit, as considered by Boltzmann in the two
  papers, was taken} would lead to a distribution of Bose-Einstein type or
of Maxwell-Boltzmann type: \citep[Sec.(2.2),(2.6)]{Ga000}.}]\\
\0\rm Sec.V, {\it Relation between entropy and the quantities that I
    have introduced as probability distributions}, (p.215),\\
\0[\alertb{\sl The link between the probability distributions found in the
    previous sections with entropy is discussed, dealing with examples
    derived through the combinatorial analysis of the previous sections,
    and expressing it through the ``permutability'' $\O$ (\ie the count of
    the possible states). Examples are computation of the free gas entropy,
    the entropy change expansion of a free gas in a half empty container,
    the entropy of a free gas in the gravity field (barometric formula).
    \\
On p.218 the following celebrated statement is made ({\it in italics in the
  original}) about ``permutability'' (\ie number of ways in which a given
(positions-velocities) distribution can be achieved) and is illustrated
with the example of the expansion of a gas in a half empty container:}] %

\* {\it Let us think of an arbitrarily given system of bodies, which
  undergo an arbitrary change of state, without the requirement that the
  initial or final state be equilibrium states; then always the measure of
  the permutability of all bodies involved in the transformations
  continually increases and can at most remain constant, until all bodies
  during the transformation are found with infinite approximation in
  thermal equilibrium.}%
\footnote{\alertr{\sl  After the last word appears in parenthesis and still in
  italics {\it (reversible transformations)}, which seems to mean ``{\it or
    performing reversible transformations}''.}}
\*

\0[\alertb{\sl The last comment underlines that the new view of entropy is
    not restricted to equilibrium states: it makes sense as a ``Lyapunov
    function'' also in the evolution towards equilibrium; Klein remarks:}]
``{\it ... $\log P$ was well defined whether or not the system is in
  equilibrium, so that it could serve as a suitable generalization of
  entropy}'', \citep[p.82]{Kl973}.

\def\SEC{Monocyclic and orthodic systems. Ensembles}
\section{\SEC}
\label{1884}\iniz
\lhead{\small\thesection: \SEC}

\0{\sl Quotes and comments on: {\it {\"U}ber die
    {E}igenshaften monozyklischer und anderer damit verwandter {S}ysteme",
    1884, \citep[\#73,p.122-152]{Bo884}.}}%
\* \0[\alertb{\sl Thermodynamic analogies are fully discussed starting with
    the works of Helmoltz to whom the notion, and examples, are attributed
    by Boltzmann although he (as well as Clausius) had already several
    times employed them, even though without a formal definition,
    \citep{Bo866,Cl871},\citep[\#39,p.127-148]{Bo877a}. Boltzmann goes far
    beyond Helmoltz and fully develops the theory of ensembles, which had
    been left at an early stage since his first works
    \citep[1868-71]{Bo868,Bo871-a,Bo871-b}.}] \*

{\it The most complete proof of the second main theorem is manifestly based
  on the remark that, for each given mechanical system, equations that are
  analogous to equations of the theory of heat hold.}%
\footnote{\alertr{\sl italics added; see Sec.\ref{sec:10} for the notions
    of thermodynamic analogies or thermodynamics models.}}
But, on the one hand, it is evident that the proposition, in this
generality, cannot be valid and, on the other hand, because of our scarce
knowledge of the so called atoms, we cannot establish the exact mechanical
properties with which the thermal motion manifests itself: hence the task
arises to search in which cases and up to which point the equations of
mechanics are analogous to the ones of the theory of heat.  We should not
refrain to list the mechanical systems with behavior congruent to the one
of the warm bodies, rather than to look for all systems for which it is
possible to establish stronger or weaker analogies with warm bodies. The
question has been posed in this form by Mr. von Helmoltz %
\footnote{{\small Berl.Ber, 6 and 27 March 1884.}} 
and I have aimed, in what follows and before proceeding to general
propositions, to treat a very special case,of the analogy that he
discovered between the thermodynamic behavior and that of systems, that he
calls monocyclic, and to follow the propositions, to which I will refer, of
the mechanical theory of heat intimately related to monocyclic systems.
\footnote{{\small A very general example of monocyclic system is
    offered by a current without resistance (see Maxwell, ``Treatise on
    electricity'', Sec.579-580, \alertb{[p.224-225]},
    where $x$ and $y$ represent the v. Helmholtzian
    $p_a$ and $p_b$).}}
\*
\centerline{\bf\S1}

Let a point mass move, according to Newton's law of gravitation, around a
fixed central fixed body $O$, on an elliptic trajectory.  Motion is not in
this case monocyclic; but it can be made such with a trick, that I already
introduced in the first Section of my work ``{\it Einige allgemeine
  s{\"a}tze {\"u}ber {W\"a}rme\-gleichgewicht}$\,$''%
\footnote{\small Wiener. Berl. {\bf 63}, 1871, \citep[\#18]{Bo871-a},
  [\alertb{\sl
      see also \citep[Sec.1A.1]{Ga000}.}]} and that also Maxwell
\footnote{{\small Cambridge Phil. Trans.  {\bf 12}, III, 1879 (see also
  Wiedemanns  Beibl\"atter, {\bf 5}, 403, 1881).}} has again followed.

Imagine that the full elliptic trajectory is filled with mass, which at
every point has a density (of mass per unit length) such that, while time
elapses, density in each point of the trajectory remains unchanged. As it
would be a Saturn ring thought as a homogeneous flow or as a homogeneous
swarm of solid bodies so that, fixed one of the rings among the different
possible ones a stationary motion would be obtained. The external force can
accelerate the motion or change its eccentricity; this can be obtained
increasing or diminishing slowly the mass of the central body, so that
external work will be performed on the ring which, by the increase or
diminution of the central body mass, in general is not accelerated nor
decelerated in the same way.  This simple example is treated in my work
{\it Bemerkungen {\"u}ber einige Probleme der mechanischen
{W}{\"a}rmetheo\-rie}%
\footnote{{Wien, Ber. {\bf 75}.  See also Clausius, Pogg. Ann. {\bf 142},
  433; Math. Ann. von Clebsch, {\bf4}, 232, {\bf 6}, 390,
  Nachricht. d. G\"ott. Gesellsch. Jarhrg. 1871 and 1871; Pogg. nn. {\bf
    150}, 106, and Erg\"angzungs, {\bf 7},215.}}
where in Section 3 are derived formulae in which now we must set $b=0$ and
for $m$ we must intend the total mass of the considered ring%
\\
$\ldots$ 
\*

\0[\alertb{\sl Detailed comparison with the work of Helmoltz follows.  A
    monocyclic system is imagined to have its orbits filled with points
    with a density proportional to the time spent in its arcs (here the
    example of the Saturn rings is evoked).
    \\ Varying the parameters of the orbit the state changes (\ie the orbit
    changes). Various examples of monocyclic systems are worked out: for
    all of their periodic orbits is defined the amount of heat $dQ$ that
    the system receives in a transformation (``work to increase the
    internal motion'' or ``infinitesimal direct
    increment''\phantomsection\label{directincrement} of the internal
    motion, \ie the heat acquired by a warm body), and the amount of work
    $dW$ that the system does against external forces as well as the
    average kinetic energy $L$; in all examples it is shown that
    $\frac{dQ}L$ is an exact differential.
    \\ This leads Boltzmann to generalize the Helmoltz' conception of state
    of a monocyclic system, \ie of a priodic orbit with an invariant
    density. The generalization is that of ``monode'' for a system whose
    states are imagined as regions of phase space filled with a density
    that is invariant under the time evolution. Thus a monode is a nice
    word to indicate an invariant (normalized) distribution on phase
    space. If the points of a monode are evolved with the equations of
    motion the density invariance (or stationarity) implies that they will
    look the same forever (hence their name). In a discrete phase space a
    monode will consist of points each representing a possible microscopic
    configuration. Of course there are many such monodes and in the
    following Boltzmann will collect them into distinguished families
    generating important definitions (p.129-130):}] %

\* \centerline{\bf\S2,  (p.129)} \*
...\\
I would permit myself to call systems whose motion is stationary in this
sense with the name {\it monodes}. 
\footnote{With the name ``stationary'' Mr. Clausius would denote
  every motion whose coordinates remain always within a bounded region.}
They will therefore be characterized by the property that in every point
persists unaltered a motion, not function of time as long as the external
forces stay unchanged, and also in no point, in no region nor through
any surface mass or kinetic energy enters from the outside or goes out.  If the
kinetic energy is the integrating denominator of the differential $dQ$, which
directly gives the work to increase the internal motion, %
\footnote{\alertr{\sl In an infinitesimal transformation, {\it i.e.} 
    variation of internal energy summed to work done by the system
    on the outside, defines the heat received by the
    system.}}
then I will say that the such systems are {\it orthodes}.
....\\
\phantom{.} \kern-7pt[\alertb{\sl Etymologies:}
    monode=\bgr m'onos\egr +\bgr
    e>~idos\egr=unique+aspect; orthode=\bgr >orj'os\egr + \bgr
    e>~idos\egr=\\ right + aspect.%

  \alertb{\sl The notion of {\it monode} and
      {\it orthode} will be made more clear in the next subsection 3. It is
      already clear that while the monode is just a stationary distribution
      on phase space an orthode must be a collection of monodes indexed by
      parameters that can be varied, to represent processes consisting in
      varying the parameters so it can be checked that the collection (or
      ``ensemble'') is a model of thermodynamics, \ie it is ``orthodic'' or
      generates a thermodynamic analogy.}
]
\*
\centerline{\S3. (p.131)}
\*

After these introductory examples I shall pass to a very general
case. Consider an arbitrary system, whose state is characterized by
arbitrary coordinates $p_1,p_2,\ldots, p_g$; and let the corresponding
momenta be $r_1,r_2,\ldots, r_g$. For brevity we shall denote the
coordinates by $p_{\bf g}$ and the momenta by $r_{\bf g}$. Let the internal
and external forces be also assigned; the first be conservative. Let $\ps$
be the kinetic energy and $\ch$ the potential energy of the system, then also
$\ch$ is a function of the $p_{\bf g}$ and $\ps$ is a homogeneous function
of second degree of the $r_{\bf g}$ whose coefficients can depend on the
$p_{\bf g}$.  The arbitrary constant appearing in $\ch$ will be determined
so that $\ch$ vanishes at infinite distance of all masses of the system or
even at excessive separation of their positions.  We shall not adopt the
restrictive hypothesis that certain coordinates of the system be
constrained to assigned values, hence also the external forces will not be
characterized other than by their almost constancy on slowly varying
parameters. The more so the slow variability of the external forces will
have to be taken into account either because $\ch$ will become an entirely
new function of the coordinates $p_{\bf g}$, or because some constants that
appear in $\ch$, which we shall indicate by $p_{\bf a}$, vary slowly.
\*
{\bf 1.} We now imagine to have a large number  $N$ of such systems,
of exactly identical nature; each system absolutely independent from all
the others.%
\footnote{\alertr{\sl  In modern language this is sometimes an ensemble:
    it is the generalization of the Saturn ring of Sec.1: each
    representative system is like a stone in a Saturn ring. It is a way to
    realize all states of motion of the same system.  Their collection does
    not change in time and keeps the same aspect, if the collection is
    stationary, \ie is a ``monode''.}}
The number of such systems whose coordinates and momenta are
between the limits $p_1$ and $p_1+dp_1$, $p_2$ and
$p_2+dp_2\ldots$, $r_g$ and $r_g+dr_g$ be

$$dN= N e^{-h(\ch+\ps)} \fra{\sqrt\D\, d\s\,d\t}{\int\int
e^{-h(\ch+\ps)}\sqrt\D\, d\s\,d\t},$$
where $d\s=\D^{-\fra12} dp_1dp_2\ldots dp_g, \ d\t=dr_1\,dr_2\ldots dr_g$
(for the meaning of $\D$ see Maxwell {\it loc. cit}
p.556).\footnote{\alertr{In general the kinetic energy is a
    quadratic form in the $r_{\bf g}$ and then $\D$ is its determinant: it
    is the Jacobian of the linear transformation $r_{\bf g}\otto r_{\bf
      g}'$ that brings the kinetic energy in the form $\fra12|r'_{\bf
      g}|^2$.}}

The integral must be extended to all possible values of the coordinates and
momenta. The totality of these systems constitute a {\it monode} in the
sense of the definition given above (see, here, especially Maxwell {\it
  loc. cit.}) and I will call this species of monodes with the name {\it
  holodes},
[\alertb{Etymology:}\\
  \bgr <'olos\egr=''global''+\bgr e>~idos\egr=''aspect''].
\footnote{\alertb{\sl Probably because the canonical distribution deals with all
  possible states of the system and does not select quantities like the
  energy or other constants of motion.}}

Each system I will call an {\it element} of the holode.
\footnote{\alertr{\sl Summarizing: a monode is a probability distribution; the
    monodes so defined are a special class of distributions each of which
    is called a holode. The holodes depend on parameters, $h,V$ in this
    case, and it makes sense to ask whether they provide a thermodynamic
    analogy: \ie if they are ``orthodic''.}}
The total kinetic energy of a holode is%
\footnote{\sl  In the case of a gas the number $g$ must be
    thought as the Avogadro's number times the number of moles,
    while the number $N$ is a number {\it much larger} and equal to the
    number of cells which can be thought to constitute a regular
    discretization of the phase space. Its introduction is not necessary
    for the purpose of defining models of thermodynamics, and Boltzmann
    already in 1868 and 1871 had treated canonical and microcanonical
    distributions with $N=1$: it seems that the introduction of the $N$
    copies, adopted later also by Gibbs, intervenes for ease of comparison
    of the work of v. Helmholtz with the preceding theory of 1871. Remark
    that Boltzmann accurately avoids to say too explicitly that the work of
    v. Helmholz is, as a matter of fact, a different and
    particular version of his preceding work. Perhaps this is explained by
    the caution of Boltzmann who in 1884\sl was thinking to move to Berlin,
    solicited and supported by v. Helmholtz. We also have to say that the
    works of 1884 by v. Helmholtz became an occasion for Boltzmann to review and
    systematize his own works on the heat theorem which, after the present
    work, took up the form and the generality that we still use today as
    ``theory of the statistical ensembles''.}

$$L=\fra{Ng}{2h}.$$
Its potential energy $\F$  equals $N$ times the average value
$\lis\ch$ of $\ch$, \ie:

$$\F=N\fra{\int \ch\, e^{-h\ch}\,d\s}{\int \, e^{-h\ch}\,d\s}.$$
The coordinates $p_{\bf g}$ correspond therefore to the v. Helmholtzian
$p_{\bf b}$, which appear in the kinetic energy $\ps$ and potential energy
$\ch$ of an element. The intensity of the motion of the entire ergode%
\footnote{\alertr{\sl This is a typo as it should be holode: the notion
    of ergode is introduced later in this work.}} and hence also $L$ and
$\F$ depend now on $h$ and on $p_{\bf a}$, as for Mr. v. Helmholtz on
$q_{\bf b}$ and $p_{\bf a}$.

The work provided for a direct increase, see 
p.\pageref{directincrement}, of internal motion is:

$$\d Q=\d \F+\d L-N\fra{\int \d\ch\,e^{-h\ch}\,d\s}{\int\,e^{-h\ch}\,d\s}$$
(see here my work%
\footnote{{\small Wien. Ber., {\bf63}, 1871, formula (17).}}
{\it Analytischer Beweis des zweiten {H}auptsatzes der mechanischen
  {W\"a}rme\-theorie aus den {S\"a}tzen {\"u}ber das {G}leichgewicht des
  lebendigen {K}raft}), \citep[\#19]{Bo871-b}.  The amount of internal
motion generated by the external work, when the parameter $p_a$ varies %
\footnote{\alertr{\sl Here we see that Boltzmann considers
among the parameters $p_a$ coordinates such as the dimensions of the
container: this is not explicitly said but it is often used in
the following.}} 
by $\d p_a$, is therefore $-P\d p_a$, with

$$-P=\fra{N\int\fra{\dpr\ch}{\dpr p_a} e^{-h \ch}\,d\s}{\int e^{-h
\ch}\,d\s}$$
The kinetic energy $L$ is the integrating denominator of $\d Q$: all holodes
are therefore orthodic, and must therefore also provide us with
thermodynamic analogies. Indeed let %
\footnote{\alertr{\sl  Here \phantomsection{\label{olodecanonico}}
      the argument in the original relies to some
    extent on the earlier paragraphs: a self contained check is 
    reported in this footnote for ease of the reader:\\
    $$F\defi-h^{-1}\log
    \int e^{-h(\ch+\f)}\defi -h^{-1}\log Z(\b,p_a),\qquad T=h^{-1}$$
    and
    remark that if $P\defi \frac1h\dpr_{p_a}\log Z$ 
    $$dF=(h^{-2} \log Z+h^{-1} (\F+L))dh-h^{-1}\dpr_{p_a}\log
    Z\, dp_a=(-\frac1h F+\frac1h U)dh
    -Pdp_a$$
    Define $S$ via $F\defi U-TS$ and $U=\F+L$ then
    $$dF=dU-T
    dS-S dT =-\frac{d T}T (-(U-TS)+U)-Pdp_a$$
    hence the factor $T^{-1}=h$ is the integrating factor for $dQ\defi dU+Pdp_a$
    because
    $$dU
    -TdS-SdT=-\frac{dT}T TS - P dp_a \ \tto\  Td S=dU+Pdp_a$$
    see \citep[Eq.(2.2.7)]{Ga000}.  }}
$$\eqalign{s=&\fra{1}{\sqrt{h}}\Big(\int e^{-h\ch}d\s\Big)^{\fra1g}
e^{\fra{h\ch}{g}}=\sqrt{\fra{2L}{N g}}\Big(\int
e^{-h\ch}d\s\Big)^{\fra1g} e^{\fra\F{2L}},\cr
q=&\fra{2L}s,\quad K=\F+L-2L\log s,\quad H=\F-L, \cr}$$
[\alertb{\sl the intermediate expression for $s$ is not right and instead of $\ch$
    in the exponential should have the average $\frac{\F}N$ of $\ch$\,}]

{\bf2.} Let again be given many ($N$) systems of the kind considered at the
beginning of the above sections; let all be constrained by the constraints

$$\f_1=a_1,\ \f_2=a_2,\ \ldots\ ,\f_k=a_k.$$
These relations must also, in any case, be integrals of the equations of
motion. And suppose that there are no other integrals. Let $dN$ be the
number of systems whose coordinates and momenta are between $p_1$ e
$p_1+dp_1$, $p_2$ and $p_2+dp_2$, $\ldots\ \ r_g$ and $r_g+dr_g$. Naturally
here the differentials of the coordinates or momenta that we imagine
determined by the equations $\f_1=a_1,\ldots$ will be missing.  These
coordinates or momenta missing be $p_c,p_d,\ldots,r_f$; their number
be $k$. Then if

$$dN=\fra{\fra{N dp_1 dp_2\ldots dr_g}{\sum\pm \fra{\dpr\f_1}{\dpr
p_c}\ldots\fra{\dpr\f_k}{\dpr
r_f}}}
{\int\int\ldots 
\fra{ dp_1 dp_2\ldots dr_g}{\sum\pm \fra{\dpr\f_1}{\dpr
p_c}\cdot
\fra{\dpr\f_2}{\dpr p_d}\ldots \fra{\dpr\f_k}{\dpr r_f}}}$$
the totality of the $N$ systems will constitute a monode, which is defined
by the relations $\f_1=a_1,\ldots$.  The quantities $a$ can be either
constant or subject to slow variations. The functions $\f$ in general may
change form through the variation of the $p_a$, always slowly. Each single
system is again called element. 

Monodes that are constrained through the only value of the equation of the
kinetic energy%
\footnote{\alertr{\sl  The equation of the kinetic energy is the energy
conservation $\f=a$ with $\f=\psi+\chi$, if the forces are conservative
with potential $\ch$.}} will be called {\it ergodes}, while if also other
quantities are fixed will be called {\it subergodes}.  The ergodes are
therefore defined by

$$dN=\fra{ 
\fra{N\,dp_1dp_2\ldots dp_g dr_1\ldots dr_{g-1}}
{\fra{\dpr \ps}{\dpr r_g}}
}
{ \int \int \fra{\,dp_1dp_2\ldots dp_g dr_1\ldots dr_{g-1}}{\fra{\dpr
\ps}{\dpr r_g}}}$$
Hence for the ergodes there is a $\f$, equal for all the identical systems
and which stays, during the motion, equal to the constant energy of each
system $\ch+\ps= \fra1N (\F+L)$. Let us set again $\D^{-\fra12}dp_1
dp_2\ldots dp_g=d\s$, and then (see the works cited above by me and by
Maxwell):

$$\eqalign{
\F=& N\fra{\int \ch \ps^{\fra{g}2-1} d\s}{\int
\ps^{\fra{g}2-1}d\s},\qquad
L= N\fra{\int  \ps^{\fra{g}2} d\s}{\int
\ps^{\fra{g}2-1}d\s},\cr
\d Q=& N\fra{\int \d\ps \ps^{\fra{g}2-1} d\s}{\int
\ps^{\fra{g}2-1}d\s}=\d(\F+L)-N 
\fra{\int  \d\ch\,\ps^{\fra{g}2-1} d\s}{\int
\ps^{\fra{g}2-1}d\s},\cr}$$
$L$ is again the integrating factor of $\d Q$,%
\footnote{\alertr{\sl  The (elementary) integrations on the variables $r_{\bf
      g}$ with the constraint $\ps+\ch=a$ have been explicitly performed:
    and the factor $\ps^{\fra{g}2-1}$ is obtained, in modern terms,
    performing the integration $\int\d(\ch+\ps-a) dr_{\bf g}$ and in the
    formulae $\ps$ has to be interpreted as ${a-\ch}$, as already in
    \citep[1868-71]{Bo868,Bo871-a}.}}%
and the entropy thus generated is $\log (\int \ps^{\fra{g}2}
d\s)^{\fra2g}$, while it will also be $\d Q=q\,\d s$ if it will be set:

$$s= (\int \ps^{\fra{g}2} d\s)^{\fra1g},\qquad q=\fra{2L}s.$$
Together with the last entropy definition also the characteristic function
$\F-L$ is generated. The external force in the direction of the parameter
$p_a$ is in each system

$$-P= \fra{\int \fra{\dpr\ch}{\dpr p_a} \ps^{\fra{g}2-1} d\s}{\int
\ps^{\fra{g}2-1}d\s}.$$
Among the infinite variety of the subergodes I consider those in which for
all systems not only is fixed the value of the equation of kinetic energy
[\alertb{\sl value of the energy}] but also the three components of the
angular momentum. I will call such systems {\it planodes}. Some property of
such systems has been studied by Maxwell, {\it loc. cit.}. Here I mention
only that in general they are not orthodic.
\\
...
\\
The nature of an element of the ergode is determined by the parameters
$p_{\bf a}$\footnote{\alertr{\sl  In the text, however, there is $p_{\bf
      b}$: typo?}}.  Since every element of the ergode is an aggregate of
point masses and the number of such parameters $p_{\bf a}$ is smaller than
the number of all Cartesian coordinates of all point masses of an element,
so such $p_{\bf a}$ will always be fixed as functions of these Cartesian
coordinates, which during the global motion and the preceding developments
remain valid provided these functions stay constant as the kinetic energy
increases or decreases.
\footnote{\alertr{\sl  Among the $p_{\bf a}$ we must include the container
    dimensions $a,b,c$, for instance: they are functions of the Cartesian
    coordinates which, however, are {\it trivial constant functions}. The
    mention of the variability of the kinetic energy means that the
    quadratic form of the kinetic energy must not depend on the $p_{\bf
      a}$.}}.  \\ ....

\*
\0[\alertb{\sl p.138: Follow more examples. The concluding remark (p.140) in
Sec. 3 is of particular interest as it stresses that the generality of the
analysis of holodes and ergodes is dependent on the ergodic
hypothesis. However the final claim, below, that it applies to polycyclic
systems may seem contradictory}]\\
...

\*
\0(p.140)\\
The general formulae so far used apply naturally both to the monocyclic
systems and to the polycyclic ones, as long as they are ergodic,%
\footnote{\alertr{\sl It is sometimes stated that the word ``ergodic'' is never
  used by Boltzmann: here is such an instance.}}  and
therefore I omit to increase further the number of examples.
\\
...

\*

\setcounter{section}{-1}
\def\SEC{References}
\section{\SEC}
\label{sec:L}
\iniz
\lhead{\small \SEC}

\bibliographystyle{plainnat}



\* \large The {\it Wissenshaftliche Abhandlungen} (Scientific works) of
L. Boltzmann have been digitalized on the initiative of the
E. Schr\"odinger Institut (ESI, Vienna) by the {\it\"Osterreichische
  Zentralbibliothek f\"ur Physik}, and made freely available
at:\\ \url{https://phaidra.univie.ac.at/detail_object/o:63668}
\\
In the same site the {\it Popul\"are Schriften} are also available:\\
\url{https://phaidra.univie.ac.at/detail_object/o:63638}
\*\*

\0{\bf Acknowledgement:} I am grateful to W. Beiglb\"ock for his constant
encouragement to undertake this work. And to E.G.D. Cohen for his essential
contribution to the joint works on nonequilibrium, to J. Lebowitz for
stimulating and inspiring my interest and work on nonequilibrium, to D.
Ruelle for teaching his ideas and results on the theory of chaos.

\end{document}
